\documentstyle[aps,graphicx,floats]{revtex}

\input epsf        

\epsfverbosetrue

\def\epsfig#1#2{\begin{center}\leavevmode
\includegraphics[origin=c, width=#2]{#1}\end{center}}
\def\epsfigrot#1#2{\begin{center}\leavevmode
\includegraphics[origin=c, angle=-90, width=#2]{#1}\end{center}}

\makeatletter
\renewcommand{\section}{\@mainheadtrue
\@startsection {section}{1}{0mm}{0.8cm plus1ex minus
 .2ex}{0.5cm plus1ex minus.2ex}{\reset@font\small\bf\raggedright}}
\renewcommand{\subsection}{\@mainheadfalse
\@startsection{subsection}{2}{\z@}{0.8cm plus1ex minus
 .2ex}{0.5cm plus1ex minus.2ex}{\reset@font\normalsize\it\raggedright}}

\renewcommand{\l@section}[2]{\addpenalty{\@secpenalty}%
\addvspace{0.2em plus\p@}%
\@tempdima 1.5em %3.0
\begingroup
\parindent \z@ \rightskip \@pnumwidth
\parfillskip -\@pnumwidth
\leavevmode %
\advance\leftskip\@tempdima %
\hskip -\leftskip %
#1\nobreak\hfil \nobreak\hbox to\@pnumwidth{\hss #2}\par
\endgroup}
\renewcommand{\l@subsection}[2]{\addpenalty{\@secpenalty}%
\addvspace{0.2em plus\p@}%
\@tempdima 1.5em %3.0
\begingroup
\parindent \@tempdima \rightskip \@pnumwidth
\parfillskip -\@pnumwidth
\leavevmode %
\advance\leftskip\@tempdima %
\hskip -\leftskip %
#1\nobreak\hfil \nobreak\hbox to\@pnumwidth{\hss #2}\par
\endgroup}
\makeatother

\begin{document}

\def\LaSr{La$_{2-x}$Sr$_x$CuO$_4$}
\def\YBax{YBa$_2$Cu$_3$O$_{6+x}$}
\def\YBa{YBa$_2$Cu$_3$O$_{7-\delta}$}
\def\YBa124{YBa$_2$Cu$_4$O$_8$}
\def\etal{{\it et al.\ }}
\def\cm-1{cm$^{-1}$}
\def\sig{$\sigma(\omega,T)$}
\def\prl{Phys. Rev. Lett. }
\def\prb{Phys. Rev. B }

% need to be defined
\def\barM{M}
\def\comment#1{}
\def\mathbold{\bf}
\def\BiSr{Bi$_2$Sr$_2$CaCu$_2$O$_8$}
\vbox{\vskip 25mm}
\begin{flushleft}
{\Large\bfseries The pseudogap in high temperature superconductors: 
an\\ experimental survey}\\
\vskip 8 mm
Tom Timusk$^{\dagger}$ and Bryan  Statt$^{\ddagger}$\\

$^{\dagger}$ Department of Physics and Astronomy, McMaster University, 
Hamilton Ont. Canada L8S 4M1 \\

$^{\ddagger}$ Department of Physics,  University of Toronto,
Toronto Ont. Canada M5S 1A7 
\bigskip
\date{6 July 1998}

\vskip 10 mm
{\bf Abstract }
\bigskip

\noindent%
We present an experimental review of the nature of the pseudogap 
in the cuprate superconductors. Evidence from various 
experimental techniques points to a common phenomenology. The 
pseudogap is seen in all high temperature superconductors and 
there is general agreement on the temperature and doping range 
where it exists. It is also becoming clear that the 
superconducting gap emerges from the normal state pseudogap. The 
d--wave nature of the order parameter holds for both the 
superconducting gap and the pseudogap. Although an extensive 
body of evidence is reviewed, a consensus on the origin of the 
pseudogap is as lacking as it is for the mechanism underlying 
high temperature superconductivity. 

\vskip 45 mm
\tableofcontents
\end{flushleft}
\vfill\eject

\section{Introduction}
% I

High temperature superconductivity was discovered by Bednorz and M\"uller 
(1986) in a complex oxide containing quasi--two dimensional copper oxygen 
planes. Copper is a multivalent ion and by chemical doping, either by 
heating in an oxygen atmosphere or by adjusting the composition in the 
layers away from the CuO$_2$ plane, these oxides can be made conducting. For 
example, by replacing some of the trivalent La in  La$_2$CuO$_4$ (La 214) 
with the divalent Sr, this material, which is an antiferromagnetic insulator 
at zero Sr doping, becomes a superconductor. From the point of view of the 
band theory of solids this is a surprise since by simply counting charges 
one expects the undoped materials to be metals with half filled bands. We 
will see in this review that the approach to the insulating state is 
closely related to the growth of a {\it pseudogap} at the Fermi surface. 
By a pseudogap we mean a partial gap. An example of such a partial gap would 
be a situation where, within the band theory approximation, some regions of 
the Fermi surface become gapped while other parts retain their conducting 
properties and with increased doping the gapped portion diminishes and the 
materials become more metallic. We view the pseudogap as a fundamental 
property of underdoped copper oxides. Our story of the pseudogap 
starts with the search for the superconducting gap in the newly discovered 
high temperature superconductors. 

The energy gap is one of the defining properties of a superconductor, but 
despite considerable effort, early experiments failed to find one in 
high--$T_c$ cuprates. Well understood physical properties such as the 
frequency dependent conductivity, the Raman efficiency or the tunneling 
conductance did not show the familiar signatures of the energy gap, namely 
a zero density of excitations below an energy $2\Delta$ appearing abruptly 
at the superconducting transition temperature $T_c$. Instead, in the 
cuprates, the depression of excitations was incomplete and  often started 
well above $T_c$ in the normal state. We now know  that this behaviour 
was not the result of poor sample quality or faulty experimental technique, 
but a consequence of two basic properties of high temperature 
superconductors --- the d--wave nature of the superconducting gap function 
varying as the cosine function around the Fermi surface with nodes at 
$k_x= \pm k_y$, and the persistence of this gap into the normal state.  It is the aim 
of this review to provide an introduction to the experiments that have 
established the pseudogap state as the normal state from which 
superconductivity emerges as the temperature is lowered through $T_c$. An 
earlier review focussing on transport properties was presented by Batlogg 
\etal (1994). 

A number of families of high temperature superconductors have been 
discovered, all of which have  shown evidence of a pseudogap. Within each 
family the properties vary with the doping level through the control of 
carrier density. Properties can also be changed through substitution  of 
impurities into the copper oxygen planes. The system that has received the 
largest amount of attention is \YBax\  (YBCO 123) with less work done on the 
closely related systems YBa$_2$Cu$_4$O$_8$ (YBCO 124) and 
Pb$_2$Sr$_2$(Y/Ca)Cu$_3$O$_8$ (PSYCCO). These systems are characterized by a 
CuO$_2$ bilayer as well as third layer containing copper, chains in the case 
of YBCO 123, double chains for YBCO 124 and isolated twofold coordinated 
coppers for PSYCCO. In transport, the effect of the 1D chains can be 
separated by doing polarized measurements with electric fields normal to the 
chains. A second important bilayer system is  Bi$_2$Sr$_2$CaCu$_2$O$_8$ 
(Bi 2212) and the more recently discovered thallium and mercury versions (Tl 
2212 and Hg 2212). Single layer materials include the much studied La$_{2-
x}$Sr$_x$CuO$_4$ (LaSr 214) with a low $T_c<40$ K as well as the 90 K 
materials Tl$_2$Sr$_2$CuO$_{6+\delta}$ (Tl 2201). 

There are technical reasons why all the experimental probes have not been 
used with success on all the families of high temperature superconductors. 
For example, magnetic neutron scattering requires very large single crystals 
and has almost exclusively been done on YBCO 123 and LaSr 214. On the other 
hand tunneling spectroscopy and angle resolved photoemission (ARPES) are 
surface sensitive probes where the focus has been on Bi$_2$Sr$_2$CaCu$_2$O$_8$ 
which cleaves easily in vacuum along the BiO planes normal to the 
c--direction yielding a high quality virgin surface. Techniques like the dc 
conductivity, ab--plane transport and nuclear magnetic resonance
(NMR) place less onerous demands on the 
crystal growers and have consequently been been applied with success to a 
larger number of high temperature superconducting (HTSC) systems. 

We have organized our review in terms of different experimental techniques 
starting with ARPES and tunneling. Ideally these 
techniques measure the density of single electronic states as a function of 
energy and momentum, occupied states in the case of ARPES and both occupied 
and unoccupied in the case of tunneling. For various practical reasons both 
techniques yield an average over large regions of momentum space, 
particularly tunneling. 

Other experimental techniques do not yield information about single 
electronic states but involve an excitation of the electronic system where 
a  transition takes place from an initial state to a different final state. 
These techniques include transport properties where in the optical 
conductivity and Raman scattering the initial and final states have the same 
momentum and in the case of the dc conductivity, the same energy. Magnetic 
neutron scattering is a spectroscopic technique where there is a spin flip 
and the experimenter can also choose the momentum transfer between the initial 
and final states. 

        NMR is a low energy technique which is
sensitive to various momentum averages  of the spin excitations. By 
probing various nuclei in the unit cell one can probe different parts of
momentum space.
The electronic specific heat also measures the density of all excitations 
within a broad window of width $k_BT$ centered at the Fermi surface.

While NMR experiments were the first to  show evidence of a  normal state 
gap, recent ARPES data have provided us with the most detailed picture of 
the evolution of the electronic structure in the pseudogap state and forms a 
good starting point. We then turn to transport measurements including the 
optical conductivity where the pseudogap manifests itself through reduced 
scattering below a certain temperature $T^*$. The electronic specific heat 
also shows evidence of a pseudogap through a depression of the specific heat 
coefficient $\gamma$ at low temperature. Finally we review the results of  
two other spectroscopies, electronic Raman scattering and magnetic neutron 
scattering. 

\section{Angle Resolved Photoemission}
% II

\begin{figure}[hb]
\epsfig{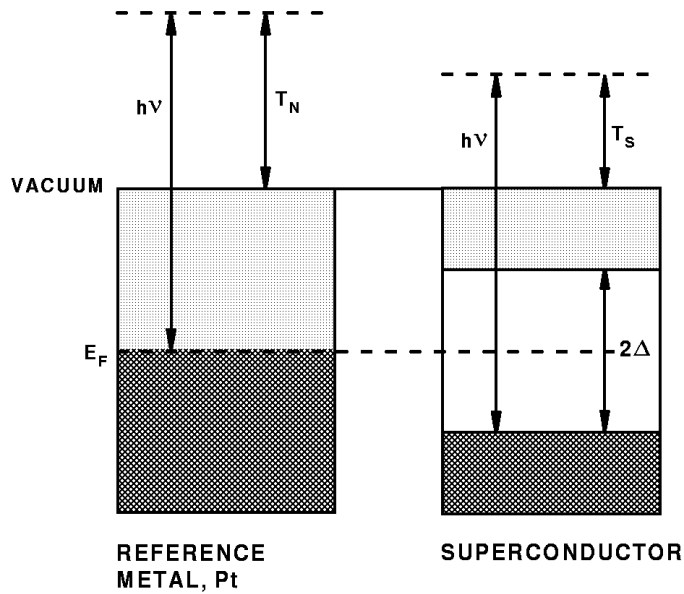}{10cm}

\caption{
Measurement of the superconducting gap by photoemission. The kinetic energies 
of photoelectrons emerging from a normal metal $T_N$ and and a 
superconductor $T_S$ will differ by $\Delta$, the superconducting gap. The 
two systems are electrically connected and  charge will flow until their 
Fermi levels are equal. 
}\label{fgArpes1}
\bigskip
\end{figure}

Angle resolved photoemission spectroscopy is a technique that 
yields the energy and momentum of the filled electronic 
energy states below the Fermi surface. It is a refinement of the 
classical photoelectric effect. In the modern version, a high energy 
photon from a synchrotron, typically $\approx 20 $eV in energy, 
is selected by a monochromator and illuminates the surface of a single 
crystal cleaved in ultra--high vacuum. A photoelectron is ejected from 
the sample at an angle $\theta$ to the surface normal, defined by an 
aperture or a position on a fluorescent screen. The kinetic energy of the 
electron is measured with a electron spectrometer to an energy resolution of 
 20 meV or better. To establish a zero of energy, the sample is connected 
electrically to a reference metal, for example platinum, and the 
photoelectron energies from the reference metal are compared to those of the 
sample under study.  When the two systems are in equilibrium their Fermi 
levels coincide. Fig. \ref{fgArpes1} shows how one can measure a gap 
$\Delta$ below the Fermi level of a metal this way. ARPES experiments in high 
temperature superconductors have been reviewed by Shen and Dessau (1995) and 
more recently by Randeria and Campuzano (1997).

%1 ARPES

Since the photoelectron emerges from a plane surface with
translational symmetry, its momentum component parallel to the surface has 
to be balanced by the momentum of excitations created inside the 
sample. By varying  the angle $\theta$ between the sample normal and the 
direction of the detected photoelectrons, it is possible to map out the 
${\bf k}$ dependence of the filled states below the Fermi level. In this 
oversimplified model an electron of momentum ${\bf k_{\parallel}}$, in 
occupied state of energy $\epsilon_{{\bf k_{\parallel}}}$ below the Fermi 
energy, gives rise to a photoelectron distribution with a sharp peak at 
momentum ${\bf k_{\parallel}}$ and energy $\epsilon_{\bf k}$ measured from 
the Fermi level of the reference sample. The measured energy distribution 
curve of the photoelectrons is given by (Randeria and Campuzano (1997)): 
\begin{equation} 
I({\bf k},\omega) = I_0({\bf k})f(\omega)A({\bf k},\omega) 
\end{equation}
where ${\bf k}$ is the momentum component parallel to the surface and 
$\omega$ its energy relative to the Fermi level. $A({\bf k},\omega)$ is the 
spectral function of the hole:

\begin{equation}
A({\bf k},\omega)={\Sigma ''({\bf k},\omega)/\pi  \over
(\omega - \epsilon_{\bf k}-\Sigma'({\bf k},\omega))^2 + 
\Sigma''({\bf k},\omega)}
\end{equation}
In a non--interacting Fermi liquid $A({\bf 
k},\omega)$ is simply a $\delta$--function in both ${\bf k}$ and $\omega$ 
centered on $\epsilon_{\bf k}$ but in a real materials $A({\bf k},\omega)$  
acquires a width in $\omega$. 
$I_0({\bf k})$ is an intensity factor and $f(\omega)$ the Fermi function.

ARPES spectra in the transition metal oxides have been reviewed by Shen and 
Dessau (1995). They are dominated by a broad background that extends to 
lower kinetic energies from the peak at $\epsilon_{\bf k}$. This background 
is in part caused by energy loss of the photoelectron within the surface layer 
before escaping. However, in strongly correlated materials, such as the 
high--$T_c$ oxides, a large part of the background is due to  electron--electron 
interactions. The electron states acquire an incoherent tail at higher energy 
and the sharp coherent peak at $\epsilon_{\bf k}$ rapidly broadens for 
states below the Fermi level. This loss of coherence is also seen in the optical 
conductivity where the transport scattering rate is found  to equal the 
quasiparticle energy, a sign of very strong electron--electron interaction 
and a breakdown of the Fermi liquid picture. This is described in detail in 
the section on the transport properties below. Quantum Monte Carlo 
calculations, based on the Hubbard model, also show typically a tail extending 
to as far as 0.5 eV from the coherence peak (Rozenberg \etal 
(1996) and Georges \etal (1996)). 

%2
\begin{figure}[htb]
\epsfig{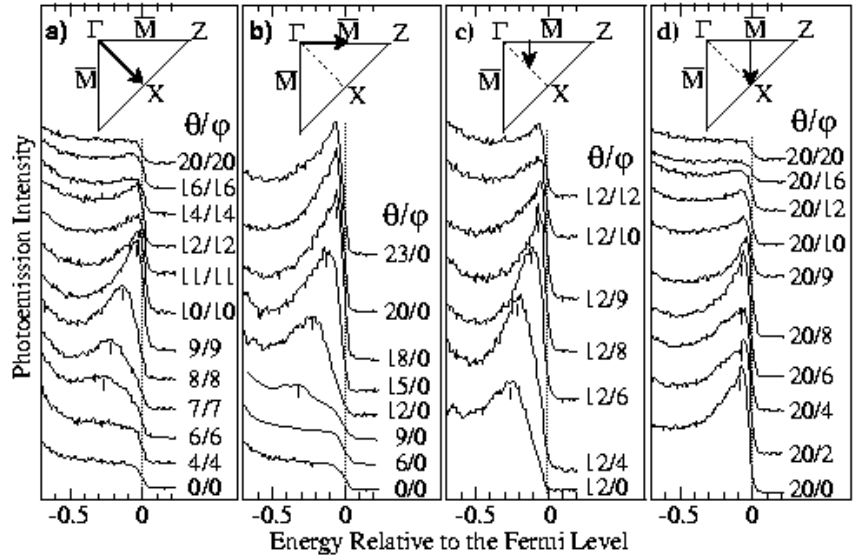}{12cm}
\caption{
Photoelectron spectra at different angles $\theta$ and $\phi$ to the crystal 
normal. In a) $\theta=\phi$ and the vertical line shows the state dispersing 
as it approaches the Fermi energy at $\theta=\phi\approx 12 ^{\circ}$. The peak 
disappears for empty states above the Fermi level at $\theta=\phi\approx > 11 ^{\circ}$.
The other panels show other combinations of $\theta$ and $\phi$. 
}\label{dessau1}
\bigskip
\end{figure}

Nevertheless a well defined quasiparticle peak can be seen in the spectra of 
many HTSC oxides with carefully prepared surfaces under ultra high vacuum 
conditions. An example is shown in Fig. \ref{dessau1} from the work of 
Dessau \etal (1993) where a broad peak, denoted by a short vertical line, 
can be seen to narrow as its energy approaches the Fermi level. The various 
curves are photoelectron spectra taken at different angles relative to the 
sample normal with varying in--plane momentum component of the outgoing 
photoelectron. The left panel scan is from the zone center, the $\Gamma$ 
point, at 45 $\deg$ to the CuO bond direction towards the X point 
($\pi,\pi$). One can clearly see that at $\theta=11 ^{\circ}$ the peak becomes 
very sharp and disappears altogether at steeper angles. This point is taken 
to be the position of the Fermi momentum ${\bf k}_F$.  The second and third 
panels of Fig. \ref{dessau1} are scans in the bond direction and reveal a 
region of flat dispersion where quasiparticle energies are just below the 
Fermi surface for a large region of the Brillouin zone. 

%3
\begin{figure}[ht]
\epsfig{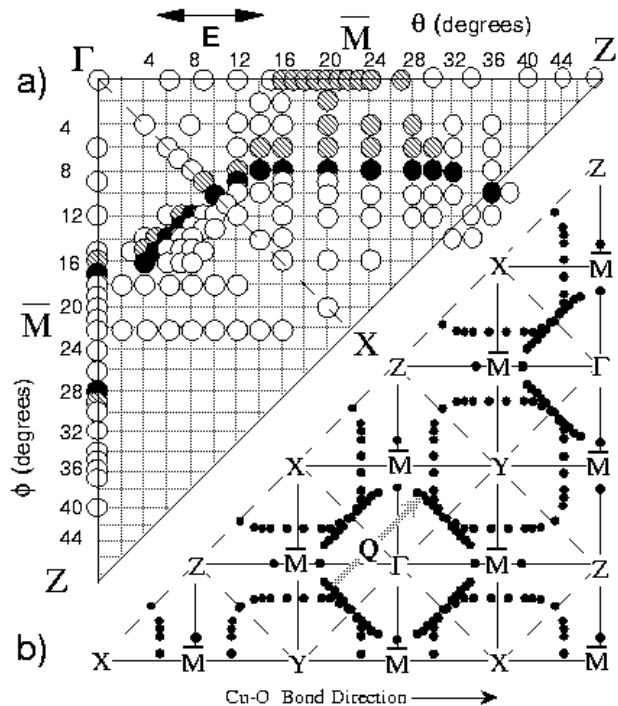}{9cm}
%figure is too large
\caption{
The Fermi surface in Bi 2212 mapped out with photoemission. 
Each measurement is denoted by a circle. The 
filled circles denote points where the dispersion curves cross the Fermi 
surface and the gray circles filled states that lie close to the 
Fermi energy. After Dessau \etal (1993).
}\label{dessau2}
\smallskip
\end{figure}

\begin{figure}[ht]
\epsfig{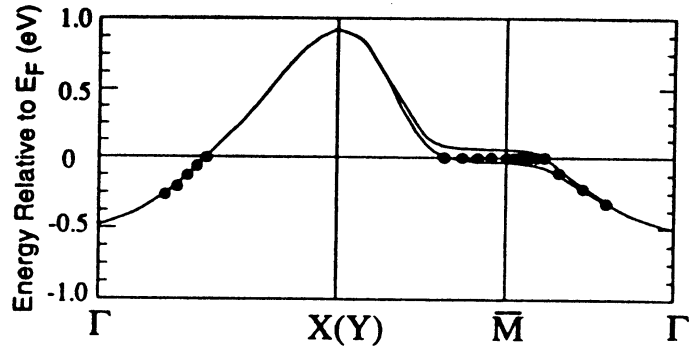}{9cm}
\caption{
$E$ vs ${\bf k}$ relationship along symmetry directions in Bi 2212 obtained 
by photoemission. The solid line is an interpretation of the dispersion 
relationship. There is clear Fermi surface crossing on the zone diagonal and 
a flat region, an extended van Hove singularity, near the zone boundary 
M ($\pi,0$). After Dessau \etal (1993).
}\label{dessau3}
\bigskip
\end{figure}

The whole Fermi surface can be mapped out this way by locating the transverse 
momentum where the quasiparticle peak disappears as it moves towards 
zero energy. This is shown in Fig. \ref{dessau2} where we see a hole--like 
Fermi surface centered on the X point with filled states at the 
$\Gamma$ point ($0,0$) and empty states at the X point. Fig. \ref{dessau3} 
shows the band structure obtained from the ARPES spectra (filled circles) 
and a simple band model compatible with the data. One can see the steep 
Fermi surface crossing on the $\Gamma-X$ line whereas in the vicinity of the 
${\rm \barM}$ ($\pi,0$) points the bands are flat. Detailed 
local density approximation (LDA) calculations 
predict such a band structure but there are serious discrepancies that are 
outside the scope of this review, as discussed in Dessau \etal (1993). In 
particular there is an absence of BiO bands predicted by band theory.

%4

%5
\begin{figure}[!h]
\epsfig{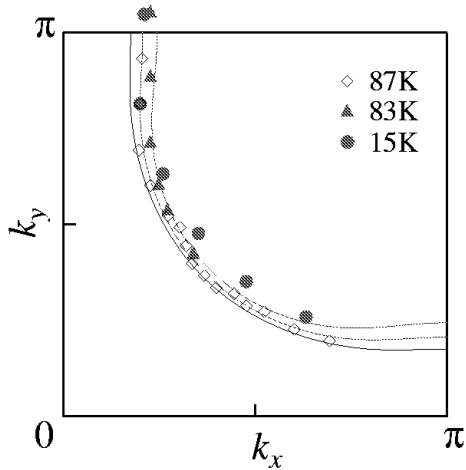}{7cm}
\caption{
Fermi surface of Bi 2212 of samples with different $T_c$'s obtained by 
varying the doping level. Tight binding theory, shown as the solid lines, 
predicts that the hole Fermi surface (centered on $(\pi,\pi)$) expands as 
$1+x$ with doping. Photoemission data, shown as symbols, show a large 
Luttinger type Fermi surface at all doping levels but lack the resolution to 
discern the change in  area with $x$.
}\label{ding3}
\bigskip
\end{figure}

The large Fermi surface has also been seen in YBCO by Campuzano \etal (1990) 
and Liu \etal (1992), in optimally doped Bi2212 by Olsen \etal (1990) and in 
the electron--doped Nd$_{1-x}$Ce$_x$CuO$_4$ by King \etal (1993) and by 
Anderson \etal (1993). Such a large surface is consistent with Luttinger's 
theorem which states that the area enclosed by the Fermi surface is 
independent of interaction and should therefore equal  the free electron 
value. As a function of doping $x$, the Luttinger Fermi surface is expected 
to vary in area as $1-x$. As the doping level $x$ approaches zero the 
Fermi surface should remain large. Photoemission lacks the resolution to 
detect the $1-x$ variation with doping. This is shown in Fig. \ref{ding3} 
where Ding \etal (1997) plot the Fermi surface from samples of Bi 2212 with 
different doping levels. The solid lines are tight binding rigid band 
calculations. The authors note that the error bars are larger than the 
separation between the calculated Fermi surfaces at the different doping 
levels. Furthermore, as discussed below, the underdoped sample shown in Fig. 
\ref{ding3} is in the pseudogap state with a gapped Fermi surface and the 
curve shown is not an actual Fermi surface but a locus of points of minimum 
gap.

The idea of a large $1-x$ Fermi surface is not in accord with 
transport measurements where the Drude spectral weight has been shown to be 
proportional to $x$ (Orenstein \etal (1990), Uchida \etal (1991)) 
and not $1-x$ as expected from 
Luttinger's theorem. This observation has led to  
suggestion that there are small pockets of carriers in the vicinity of the 
d--wave nodes at ($\pi/2,\pi/2$) at low doping. Another possibility within 
the conventional Fermi liquid picture is that a dramatic increase in 
effective mass is responsible for the reduced spectral weight with 
underdoping. This, however, is in disagreement with specific heat data that do 
not show any large changes in the specific heat coefficient $\gamma$ with 
underdoping (Loram \etal (1994a)). As we will see in the next section the 
presence of the pseudogap in the normal state offers a way out of the $x$ vs.
$1-x$ dilemma. 

\subsection{~~~The superconducting gap and the pseudogap} 
% A

The earliest observations of the superconducting gap by photoemission were in 
the angle integrated mode or with low momentum resolution. It was found that 
in optimally doped, vacuum cleaved samples of Bi 2212 there was a shift 
in the leading edge of the electron loss spectrum below $T_c$ (Imer \etal 
(1989), Manzke \etal (1989) and Olsen \etal (1989)). Since the photoelectron 
spectral peak is quite broad, particularly in underdoped samples, it has 
become customary to use the difference between the midpoint of the leading 
edge and the Fermi energy as the "leading edge gap". The true gap is larger 
than the leading edge gap since the linewidth and the experimental 
resolution have to be added. Leading edge gap values of $\Delta$ are 
typically in the 25 meV range in Bi 2212 whereas true gaps, determined by fits, 
are 10 meV larger (Randeria and Campuzano (1997)).   
\vfill\eject

%6
\begin{figure}[!t]
\epsfig{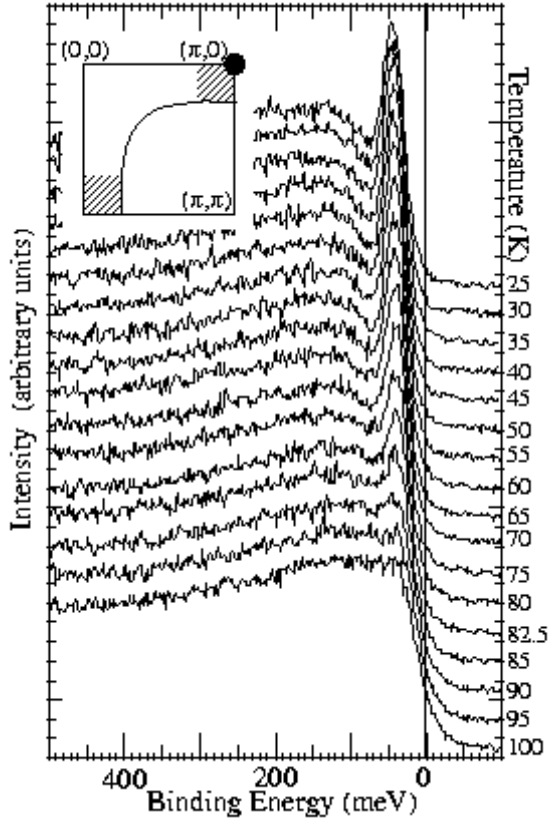}{8cm}
\caption{
Temperature dependence of the photoemission spectra at the ($\pi,0$) point. 
A sharp peak followed by a dip grows below $T_c$. The energy of the peak 
does not change with temperature suggesting that the superconducting gap 
$\Delta$ does not close at $T_c$.
}\label{loeser1}
\bigskip
\end{figure}

A second striking feature seen in the superconducting state is the 
appearance of a strong, resolution limited peak followed by a dip structure 
(Dessau \etal (1991) and Hwu \etal (1991)) as shown in Fig. \ref{loeser1} from 
Loeser \etal (1997). The spectra are taken in the momentum region near the 
($\pi,0$) point as a function of temperature on an underdoped sample of Bi 
2212 ($T_c=79 $K). A sharp peak appears 42 meV below the Fermi level at 
approximately 95 K, about 15 K above the superconducting transition 
temperature. It is followed by a dip in the photoelectron intensity at 
approximately 75 meV. Below the dip is a broad peak centered at 
approximately 100 meV. Comparisons with recent vacuum tunneling spectra 
show that the peak and the dip are symmetric about the Fermi energy whereas 
the broad peak only occurs on the negative energy side, suggesting it may be 
a band structure effect. 

\begin{figure}
\epsfig{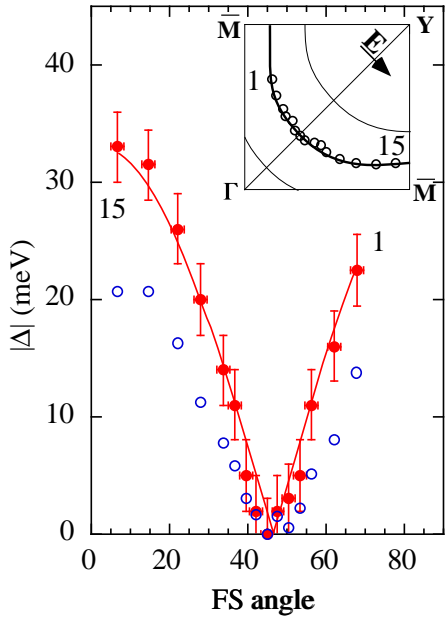}{6cm}
\caption{
The superconducting gap as function of angle around the Fermi surface. The 
gap has d--wave like nodes on the zone diagonals and rises to a maximum value 
of $\approx$ 35 meV at the M point ($\pi,0$). The open circles are the measured 
leading edge shifts of the raw photoemission spectra and the filled circles 
the estimated gap.
}\label{randeria11}
\bigskip
\end{figure}

Olsen \etal found  that the gap in the superconducting state was isotropic 
but later work with better sample quality and improved resolution showed a 
clear anisotropy. Wells \etal (1972) and Shen \etal (1993) found that the gap has 
d--wave symmetry with a minimum in the ($\pi,\pi$) direction. This was 
confirmed with better resolution by Ding \etal (1996). A recent plot of 
the gap as a function of angle around the Fermi surface is shown in Fig. 
\ref{randeria11} from Randeria and Campuzano (1997). The open circles are 
the leading edge shifts while the solid symbols are the results of fits 
based on a BCS model. It is clear that the gap has the form predicted 
for a d--wave superconductor with a minimum gap  along the diagonal at 
45$^{\circ}$ to the CuO bond direction (i.e. in the $\Gamma$--Y direction) and a maximum 
gap in the M direction. 

%7 

\begin{figure}
\epsfig{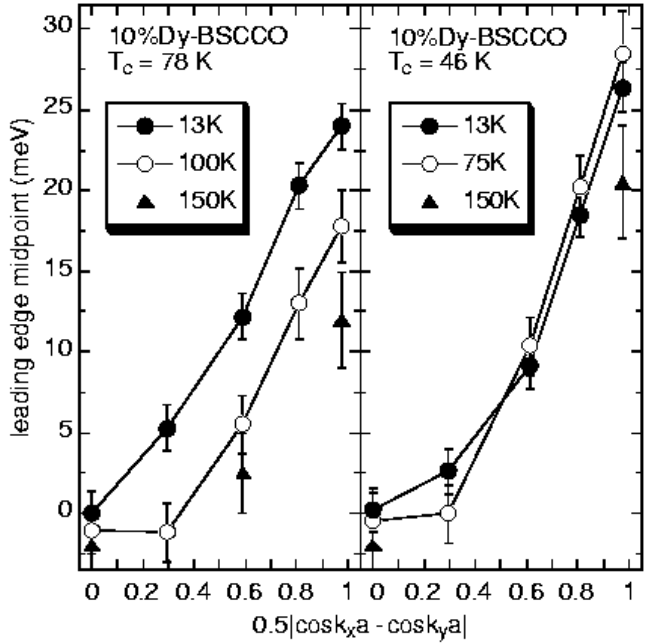}{9cm}
\caption{
Pseudogap in photoemission. Leading edge shifts plotted in such a way as to 
yield a straight line for a d--wave gap which is found in the superconducting 
state at high doping (left panel). The gap fills in near the node as the 
temperature is raised or as the doping level is reduced. The striking 
observation is that a pseudogap remains in the normal state in both samples 
with an overall symmetry and magnitude similar to the superconducting gap. 
}\label{harris2}
\bigskip
\end{figure}

Loeser \etal (1996) and Ding \etal (1996) reported that the leading edge gap 
seen in the superconducting state was also present in the normal state in 
underdoped Bi 2212. Both groups found that the momentum dependence and 
the magnitude of  this normal state gap resembled the gap seen in the 
superconducting state. It had the characteristic d--wave symmetry with a 
node in the $\Gamma$--Y direction and a maximum gap in the $\Gamma$--M 
direction. The gap was only seen in the underdoped samples and its onset 
temperature $T^*$ was found to approach the superconducting transition 
temperature at optimal doping. Fig. \ref{harris2} from Harris \etal (1996) 
shows  the gap for two underdoped samples of Dy--doped Bi 2212 MBE grown 
films. The trivalent Dy substitution on the Ca site reduces the hole 
concentration resulting in underdoping. The leading edge gap is shown as a 
function of $0.5|\cos k_xa - \cos k_ya|$. A d--wave gap would be a straight 
line on this plot.  It can be seen that the d--wave model fits well in the 
superconducting state and that the normal state gap also has clear d--wave 
symmetry but that there is considerable smearing out of the node as expected 
in the ``dirty d--wave''  picture. 
%is thermal smearing also evident?

\begin{figure}
\epsfig{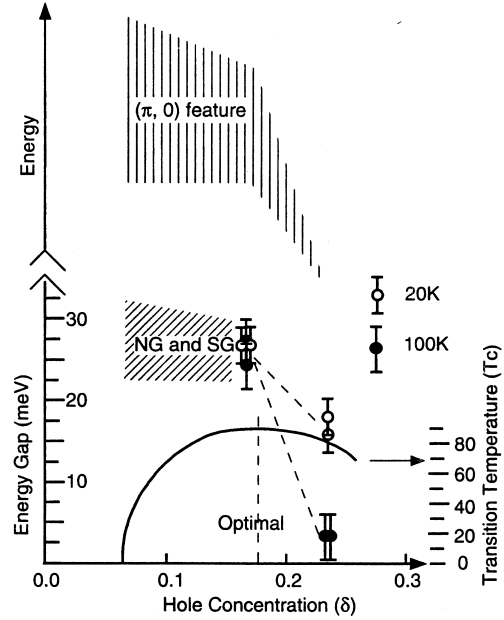}{8cm}
\caption{
Leading edge shifts as function of hole concentration. In the underdoped 
samples there is no difference in the magnitude of the superconducting(SG) and 
normal state(NG) gaps whereas in the overdoped samples the superconducting gap 
is reduced and the normal state gap goes to zero. The shaded area at high 
energy refers to the electronic structure near the ($\pi,0$) point 
discussed in Marshall \etal (1976) 
}\label{white3}
\bigskip
\end{figure}

Harris \etal (1996) plot the size of the gap as a function of doping and find 
that in the underdoped samples the gap magnitude {\it decreases} with 
doping. This is just the opposite of what is expected in BCS theory where 
the ratio of $2\Delta/k_BT_c$ is a constant, in other words the gap is 
proportional to $T_c$. In overdoped samples of 
Bi2212 the gap {\it does} decrease as $T_c$ is reduced (White 
\etal 1996) and 
there is no normal state gap. These results are summarized in Fig. 
\ref{white3} which shows the leading edge gap variation with hole doping. 
In the underdoped samples the normal state gap and the superconducting gap 
are both approximately 25 meV whereas in the overdoped sample the 
superconducting gap has dropped to 20 meV and the normal state gap is 
almost zero. 

%8

\begin{figure}
\epsfig{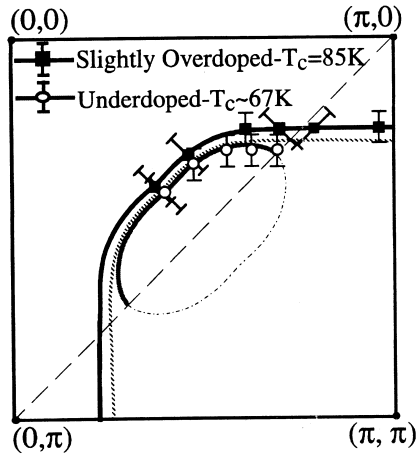}{6cm}
\caption{
Fermi surface in underdoped Bi 2212 shown as open circles in contrast with an 
overdoped sample where a continuous arc can be seen. The disappearance of the 
Fermi surface is due to the formation of the pseudogap. The dot--dashed line 
extends the short Fermi surface arc to form a pocket, an interpretation that 
would require a new zone boundary along the dashed line.
}\label{Marshall4}
\bigskip
\end{figure}

The presence of the pseudogap in the normal state also offers an 
explanation of the loss of spectral weight seen in transport measurements in 
the underdoped samples which, when interpreted in conventional terms, 
suggests the presence of small pockets at ($\pi/2,\pi/2$). The 
pseudogap simply reduces the large Luttinger Fermi surface by gapping  the 
($\pi,0$) region (Marshall \etal (1996)). Fig. \ref{Marshall4} shows the 
large Fermi surface of the overdoped sample as a solid line and the 
partially gapped Fermi surface in the underdoped sample. The dotted line 
that would extend the short Fermi surface arc to form a closed pocket 
centered at ($\pi/2,\pi/2$) corresponds to weak features of uncertain, 
possibly structural 
origin. The authors suggest theoretical scenarios that would fold the zone 
along the dashed line and give rise to a closed pocket.

The issue of the shape of the Fermi surface is addressed in a recent paper by 
Ding \etal (1997). They follow the minimum gap locus in the pseudogap state and 
find it to coincide with the locus of gapless excitations in the ungapped 
state above $T^*$ where a Fermi surface can be seen. These results suggest 
precursor pairing and not the presence of a spin density wave (SDW) gap; 
where the gap is tied to  
the SDW nesting vector and would coincide with the Fermi surface only in the 
case of perfect nesting. The implication of this work to transport 
measurements is that the reduced Drude weight in underdoped samples is due 
to the formation of the pseudogap, away from the ($\pi/2,\pi/2$) points, 
thus restricting the volume of phase space of carriers that are available for 
conduction. 

\begin{figure}
\epsfig{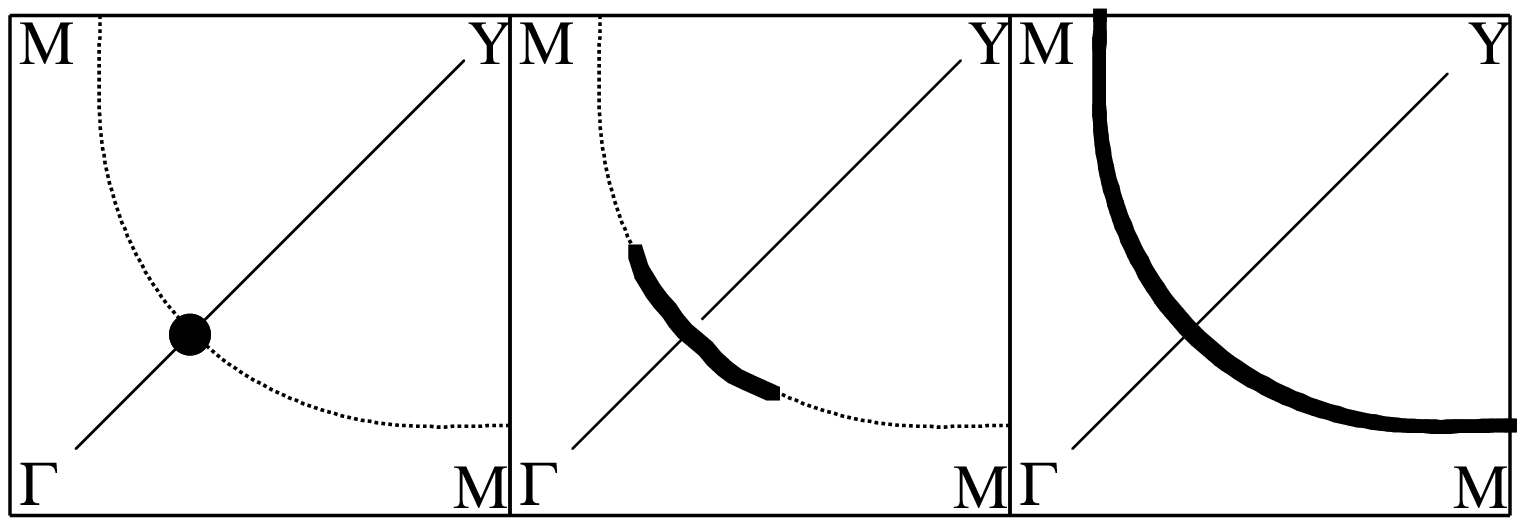}{9cm}
\caption{
Destruction of the Fermi surface by the pseudogap at three doping levels. 
The dotted line shows the large Fermi surface seen in the overdoped 
materials on the left. As the doping is reduced, the Fermi surface becomes 
gapped and only the solid arcs remain. Data show that even at low doping 
(left panel) the minimum gap follows the original dashed Fermi surface.
}\label{norman2}
\bigskip
\end{figure}

\begin{figure}
\epsfig{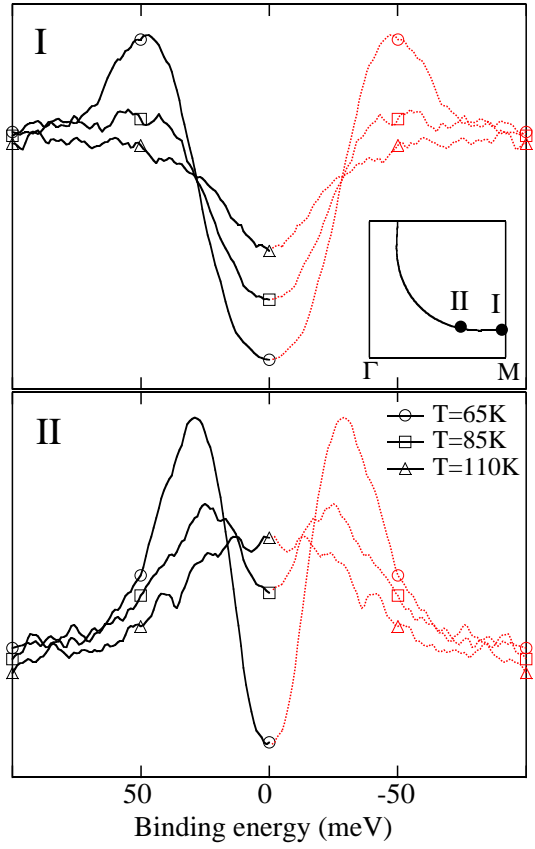}{7cm}
\caption{
Temperature dependence of the gap in underdoped Bi 2212 at two locations in the zone. Symmetrized 
ARPES data show that the gap frequency is temperature independent in the 
region of the gap maximum but closes at points in the zone closer to the 
node. This can also be seen in Fig. 8.
}\label{norman}
\bigskip
\end{figure}

The nature of the  Fermi surface in  underdoped Bi 2212 has been addressed 
in a recent paper by Norman \etal (1997a). They do not find evidence for a 
large Luttinger type Fermi surface in the underdoped material but neither is 
there evidence for pockets in the $\Gamma$--Y direction. Instead, 
the Fermi surface consists of arcs that, in the optimally doped material, 
coincide with the Luttinger Fermi surface but shrink in size to approach 
point nodes of the $d_{x^2-y^2}$ superconducting gap
as the temperature is decreased (Fig \ref{norman2}). They 
find that in underdoped materials, as the temperature is lowered, the 
pseudogap first opens up at ($\pi,0$) and progressively gaps larger portions 
of the Fermi ``contour'' (the arc in $k$ space that would be the Luttinger 
Fermi surface) leading to gapless arcs.  In the 
overdoped case the behavior is more conventional: the gap opens at the same 
temperature at all points of the Fermi surface. The authors find that
in the underdoped case near the 
($\pi,0$) point, the gap ``fills in'' (its frequency does not change but 
spectral weight at $E_F$ drops). In contrast, near the d--wave 
node, the gap ``closes'' (its frequency decreases as it fills in) in a 
mean field fashion. Fig. \ref{norman} illustrates this where the ARPES data 
has been artificially symmetrized to remove the Fermi function.

%9

\begin{figure}
\epsfig{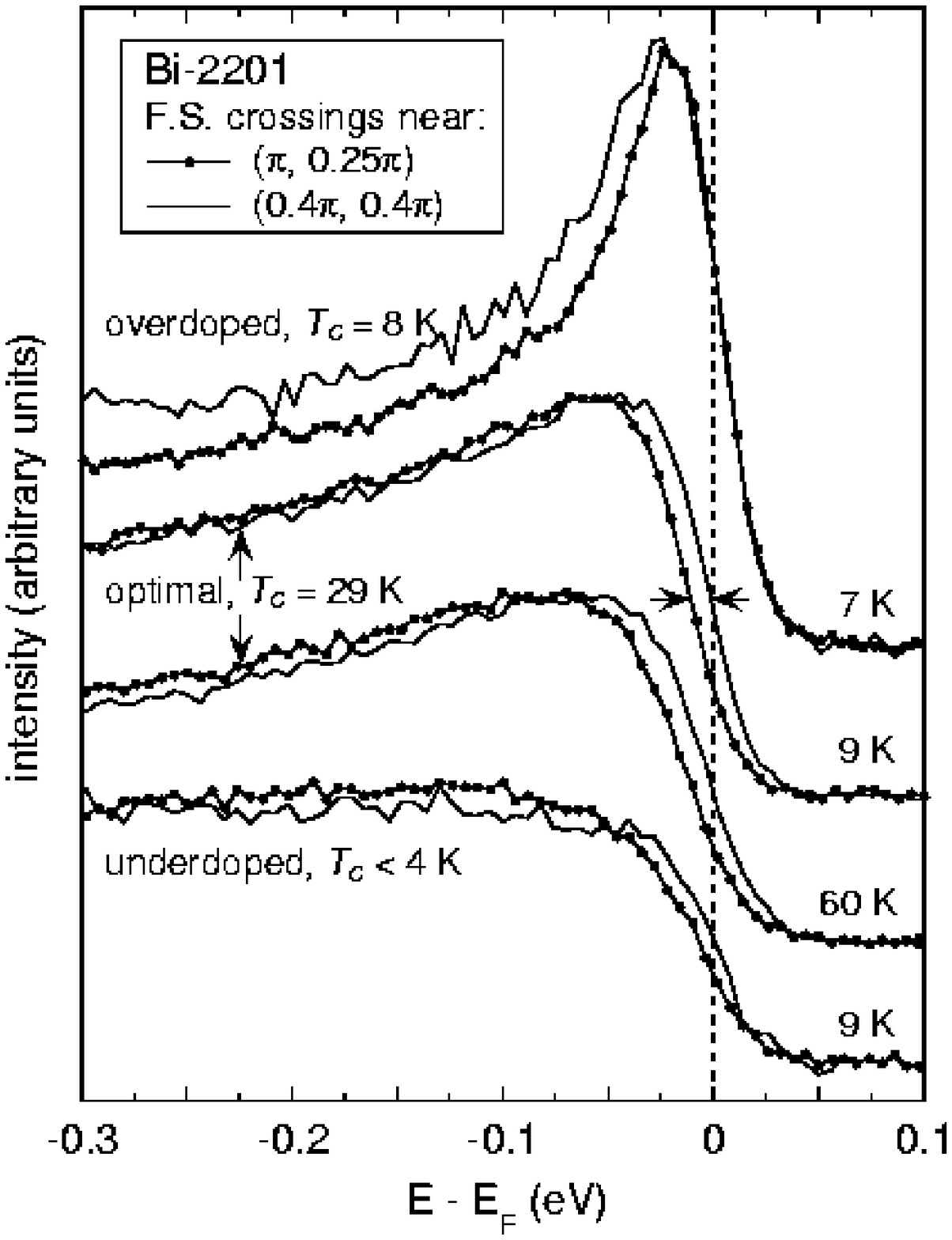}{8cm}
\caption{
The pseudogap in single layer Bi 2201. There is a clear leading edge normal 
state gap in the optimal and underdoped samples but no gap in the highly 
overdoped sample.
}\label{harris2_}
\bigskip
\end{figure}

Harris \etal (1997) have studied the single layer Bi 2201  compound \break
Bi$_{2+x}$Sr$_{2-(x+y)}$La$_y$CuO$_{6-\delta}$ which has been underdoped by the 
substitution of the trivalent La in place of the divalent Sr or Bi. With 
$y$=0.35 an optimally doped sample with a $T_c=29$ K is obtained, while by 
varying the Bi/Sr ratio an  underdoped sample with $T_c<4$ K and an 
overdoped one with $T_c=8$ K was made. Fig. \ref{harris2_} shows that the 
leading edge shifts have a pseudogap of 10$\pm$ 2 meV in the underdoped and 
optimally doped samples but no gap of any kind in the overdoped sample. 

The one--plane material Bi 2201 differs from its two--plane counterpart Bi 
2212 in the line shape in the superconducting state: it lacks the 
sharp peak followed by a dip. As figure \ref{harris2_} shows the spectra 
for the optimally doped sample ($T_c$ =29K) look very similar in the normal 
and superconducting states.  In contrast,  the two plane material has 
a striking peak that appears in the superconducting state followed by a 
dip. Harris \etal argue that the lack of the coherent quasiparticle peak in 
the superconducting state could be the result of either an overall lower 
energy scale, by a factor of 3, or perhaps enhanced impurity scattering.

%10

%11

In conclusion, Harris \etal (1997) note that the main effect in going from a 
two layer to a single layer material is the reduction of the maximum gap 
value by a factor of three which is quite different from the effect of 
underdoping, which has the effect of reducing $T_c$ and {\it 
increasing} the gap.

We next discuss some of the models for the narrowing of the ARPES peak and 
the formation of the peak and dip structure. The most common interpretation  
has been in terms of strong coupling effects in analogy with BCS 
superconductors (Arnold \etal (1991), Coffey and Coffey (1993), Shen and 
Schrieffer (1997) and Norman \etal (1997)). In BCS superconductors the 
phonons produce minima in the tunneling conductance at $eV=\Delta+\Omega$ 
where $\Omega$ is a phonon frequency. The narrowing is 
explained in terms of the formation of a gap $\Delta$ 
at the Fermi surface (Coffey and Coffey (1993)). If the scattering is 
electronic, there will be no scattering for photoelectron energies less than 
$3\Delta$ (in the s--wave case)--- $\Delta$ to overcome the gap energy and 
$2\Delta$ to create a pair of quasiparticles. Coffey and Coffey (1993) show 
that in a d--wave superconductor this dip occurs at $2\Delta$. (The 
corresponding energy for the conductivity is $4\Delta$ since the photon 
creates an electron--hole pair, each particle of which can create a second electron--hole 
pair). This reduction in scattering can be seen most dramatically in the 
reduction of microwave surface impedance at $T_c$ (Bonn \etal (1992)). 
It is also responsible for the absence of an Hebel--Slichter coherence peak in
the NMR spin--lattice relaxation rate just below $T_c$. This 
effect does give an explanation of the peak and dip structure in both ARPES 
and tunneling spectra (see section III). 

%12

Shen and Schrieffer (1997) suggest the coupling is to a strong collective 
excitation centered at ($\pi,\pi$) arguing that, because of Fermi surface 
geometry, a vector of length {\bf Q}=($\pi,\pi$) matches the two high density 
of states regions ($0,\pi$) to ($\pi,0$) but not ($\pi,\pi$) to ($-\pi,-
\pi$). They point out that while the neutron peak is 
proportional to the susceptibility $\chi({\bf q}, \omega)$ ARPES and 
tunneling give the product of the coupling constant and the susceptibility 
$g^2_{\bf kq}\chi({\bf q}, \omega)$. 

Norman \etal (1997) also interpret the peak and hump structure of the 
ARPES spectra at ($\pi,0$) in the superconducting state in terms of a 
collective bosonic mode that appears in the superconducting state. They 
argue that the sharpness of the peak--dip structure is only consistent with a 
resonance peak and not just a step--like change in the lifetime of  
the carriers at $2\Delta$ as suggested by Coffey and Coffey (1991). 
Norman \etal find that by taking the gap $\Delta=32$ meV and $\Omega=42$ meV 
they can account for the ARPES lineshape in the superconducting state. They 
argue that the collective mode is related to the 41 meV resonance seen in 
neutron scattering in the superconducting state of \YBax\ and suggest that a 
mode similar to the 41 meV resonance would also exist in Bi 2212. Related to 
these arguments is the discussion of the width of the 41 meV peak seen in 
neutron scattering by Morr and Pines (1998) who argue that heavy 
damping in the normal state is turned off when the gap in the excitations 
forms in the superconducting state. 

%13

\begin{figure}
\epsfig{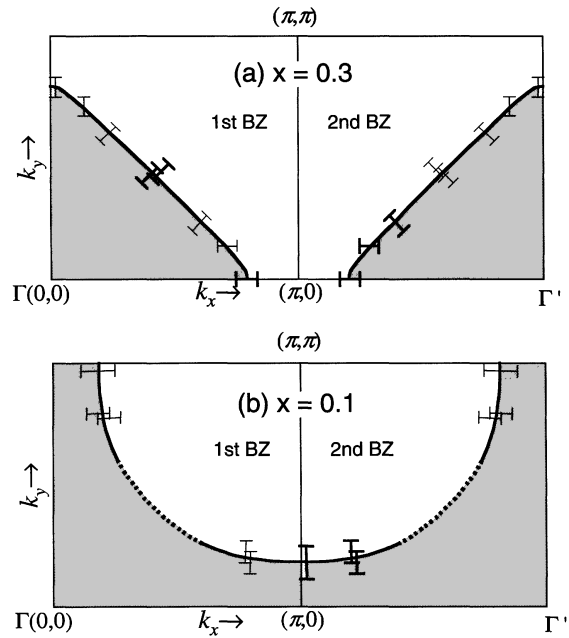}{8cm}
\caption{
Fermi surface in LSCO. This material shows an electron--like Fermi surface in 
the underdoped state where the states at the ($\pi,0$) region lie above the 
Fermi surface. With reduced doping states at ($\pi,0$) drop 100 meV below 
the Fermi level, become filled, yielding a hole--like Fermi surface centered on 
($\pi,\pi$).
}\label{Ino3}
\bigskip
\end{figure}

Ino \etal (1997) report on ARPES spectra in \LaSr\ for both 
underdoped and overdoped samples.  They find a remarkable change from an 
electron--like Fermi surface centered on  $\Gamma$ in the overdoped 
samples to a hole--like one in the underdoped materials. This occurs through 
a decrease in the energy of the states near ($\pi,0$) as the doping level 
is reduced. In the $x$=0.3 samples the ($\pi,0$) states are {\it above} the Fermi 
surface whereas in the underdoped sample they form an extended saddle point 
about 100 meV {\it below} the Fermi surface. As Fig. \ref{Ino3} shows, 
this results in a transformation of the Fermi surface from electron--like 
centered on ($0,0$) to hole--like centered at ($\pi,\pi$). The  authors contrast 
this ``high energy pseudogap''  of 100 meV with the low energy pseudogap seen 
at the same ($\pi,0$) point in Bi 2212 system. The authors also observe a 
superconducting gap  at ($\pi,0.2\pi$) as a leading edge shift 
of 10 -- 15 meV and compare that to the Bi 2212 values of $\approx$ 25 meV. 
They suggest this gap may be consistent with d--wave symmetry. They 
observe no gap in the Fermi surface crossing along the ($0,0$)--($\pi,\pi$) line.

Fujimori \etal (1998) discuss the change of chemical potential $\mu$ as 
determined by the shift of the La 3$d$ and O 1$s$ core levels in BIS 
spectroscopy. In a Fermi liquid the chemical potential shifts with doping in 
a characteristic way as the states near the Fermi level are filled. Fujimori 
\etal find that this indeed happens in the overdoped case in \LaSr\  but for 
$x<0.15 $ they find the chemical potential is pinned at the value for the 
insulating parent La$_2$CuO$_4$. The authors interpret this as evidence for 
the opening of pseudogap that moves with $\mu$ as the doping proceeds. 

%14

\section{Tunneling Spectroscopy}
% III

In the conventional BCS superconductors tunneling spectroscopy has been 
perhaps the most powerful tool used to investigate the electronic 
density of states near the Fermi level. Discovered  by Giaver in 1960, it was 
used first to study the gap, and after further refinements, subtle changes 
in the density of states due to the electron--phonon interaction (Scalapino 
(1969)) found by inverting the tunneling data to yield the spectrum of  
excitations responsible for superconducting pairing. For a recent review see 
Carbotte (1990).

\begin{figure}
\epsfig{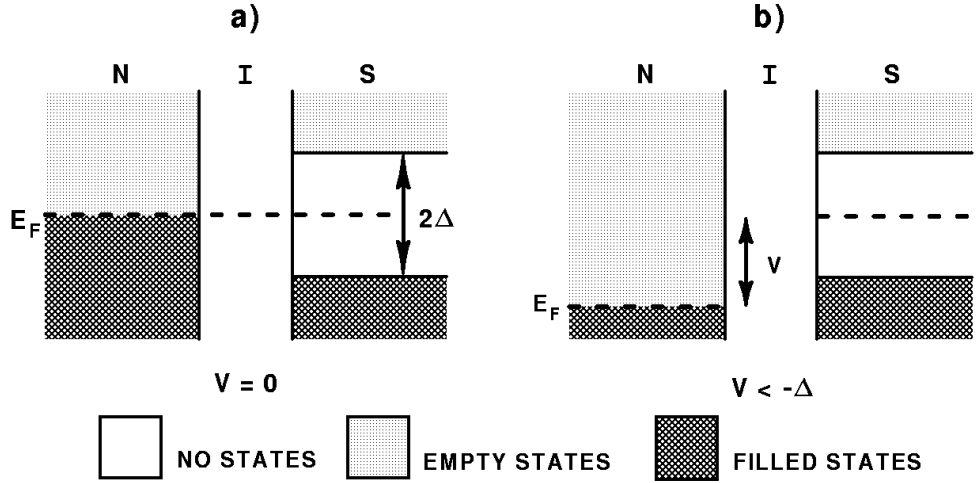}{13cm}
\caption{
Tunneling spectroscopy. In SNS tunneling no current flows at zero voltage 
difference between the normal metal and the superconductor (left panel). 
When the bias $V<-\Delta$ electrons from the filled states in the 
superconductor flow into the empty states of the metal above the Fermi 
level giving rise to a current proportional to the difference between $V$ and 
$\Delta$. 
}\label{tunneling1}
\bigskip
\end{figure}

The physics behind tunneling spectroscopy is simple (Tinkham (1975)). In SIN 
(superconductor--insulator--normal metal) tunneling an oxide layer is grown on a 
superconductor which is  then covered with a normal metal layer, Fig. 
\ref{tunneling1}. In thermodynamic equilibrium at zero degrees, the Fermi 
levels of the two systems are equal and no  current flows through an external 
circuit connected between the metal and the superconductor. If an external 
positive voltage exceeding $\Delta/e$ is applied to the metal, electrons 
tunnel from the metal through the insulator to the unoccupied states of the 
superconductor. In a BCS superconductor there is a peak in the density of 
states just above the gap edges. Therefore there will be a large peak in the 
tunneling conductance as the bias voltage approaches the gap value, 
$eV=\Delta$ .  Conversely with negative bias, when $eV=-\Delta$, electrons will 
tunnel from the occupied states of the superconductor to the normal metal. 
Thus there will be a gap in the conductance of width $2\Delta$ centered on the 
Fermi level at zero bias. In SIS tunneling there is a superconductor on 
both sides of the barrier and the tunneling current starts to flow when the 
bias $eV=\Delta_1+\Delta_2$ where $\Delta_1$ and $\Delta_2$ are the gaps of 
the two superconductors. 

%15 TUNNELING

In high temperature superconductors, because of the short coherence length, 
the tunneling electrons sample the density of states within a few atomic 
layers of the surface and the experiments are extremely sensitive to 
surface quality. The most reliable results have come from 
measurements where the tunneling barrier has been prepared in situ by 
vacuum cleaving. The measurements are then carried out either by using a 
scanning  tunneling microscope tip at some distance above the surface 
(SIN vacuum tunneling, Renner \etal (1996, 1998)) or by break junction 
tunnelling where a crystal is broken in ultra high vacuum to form a fresh 
surface and then the two pieces are allowed to come close enough for a 
tunneling current to flow (SIS tunneling, Mandrus \etal (1991, 1993)).

One of the earliest reports of a pseudogap in tunneling spectroscopy is that 
of Tao \etal (1997) who find a gap--like depression in the tunneling 
conductance of Bi 2212 junctions in the normal state. The junctions were 
prepared by evaporating lead electrodes on cleaved faces on top of a natural 
insulating barrier.  In the superconducting state they find results similar 
to what has been reported previously on this material, a depression of 
tunneling conductance at low bias to 0.4 to 0.6 of the high bias value and a 
pileup of conductance to peaks at $\approx \pm 35 $ meV. The new result in 
this work is the observation that the depression in the tunneling 
conductance remains in the normal state and, at least in one sample, can be 
traced all the way up to room temperature. 

\begin{figure}
\epsfig{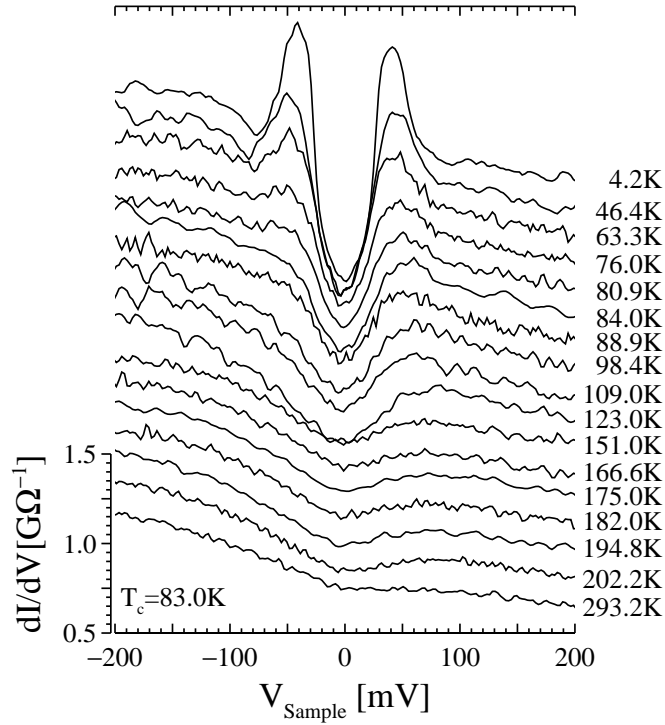}{9cm}
\caption{
Tunneling conductance for underdoped Bi2212. A gap--like feature at zero bias 
is seen to persist in the normal state which is direct evidence of a 
pseudogap in the tunneling conductance. In the superconducting state a peak 
develops at $\pm 45$ meV followed by a dip and a broad maximum. The gap 
frequency does not seem to be temperature dependent.  
}\label{renner2}
\bigskip
\end{figure}

Renner \etal (1998) report tunneling into underdoped Bi2212 crystals 
cleaved in vacuum using a scanning tunneling microscope (STM). They find 
reproducible voltage--current curves as the tip is scanned across the sample 
and suggest that many features previously observed were due to spatial 
inhomogeneity. In regions that are uniform the tunneling spectra are 
independent of bias voltage as the tip--to--sample distance is varied. As Fig 
\ref{renner2} shows, below the superconducting transition temperature they 
find a depression in the conductivity at zero bias and two symmetrically 
placed conductance peaks previously observed by several investigators in 
this material. The figure also shows that while the peaks disappear at $T_c$ 
an inverted bell--like conductance depression  remains in the normal state up 
to room temperature. The authors identify this conductance dip with the 
pseudogap. Finally there is an asymmetry with respect to the sign of the  
bias. When the bias is negative, corresponding to the removal of electrons 
from the sample (equivalent to ARPES where photoelectrons are removed, 
leaving a hole), a broad peak is seen in the spectrum at approximately 100 
meV. This peak is not seen with positive bias. Also there is some asymmetry 
in the disappearance of the sharp peaks: the positive bias peak persists into 
the normal state while the negative bias one vanishes at $T_c$. 

%16

As the temperature is raised, the magnitude of the gap in the  
superconducting state, as measured by the separation between the peaks, does 
not change. Renner \etal confirm this by modelling the high temperature 
spectrum by applying thermal broadening due to the Fermi function to the 
low temperature spectrum (Renner \etal (1996)). Thus, unlike a BCS gap which 
closes as the temperature is raised, the gap magnitude is {\it temperature 
independent}. This is also true of the magnitude of the pseudogap which 
fills in with temperature but retains its frequency width up to room 
temperature. 

\begin{figure}
\epsfig{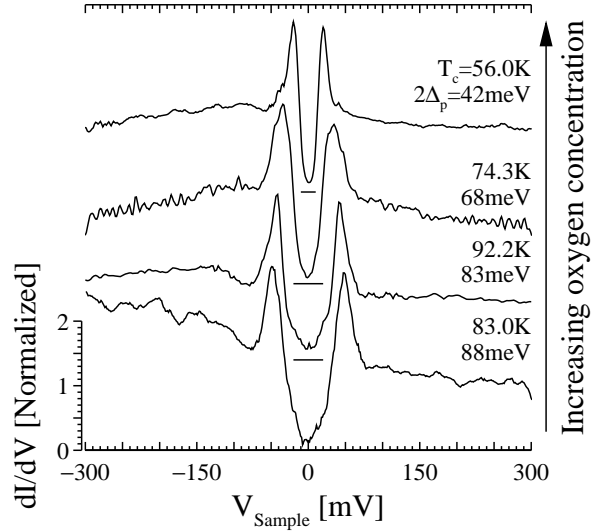}{8cm}
\caption{
Doping dependence of tunneling spectra measured at 4.2 K. The 
superconducting gap, as measured by the separation between the peaks, 
decreases in magnitude in the overdoped region (top 
two curves).
}\label{renner1}
\bigskip
\end{figure}

While there is no change in peak separation with temperature, 
the magnitude of the gap does decreases as the {\it doping level} 
is increased, 
which is just the opposite to what one might expect for a 
superconductors where the ratio $2\Delta/kT_c$ would remain 
constant and the gap would grow in proportion to $T_c$. As Fig. 
\ref{renner1} shows, $2\Delta$ falls from 88 meV (710 \cm-1) in the 
most underdoped sample with a $T_c$ of 83.0 K to 42 meV (340 \cm-1) in the 
overdoped sample.

%17

%18
\begin{figure}[!h]
\epsfig{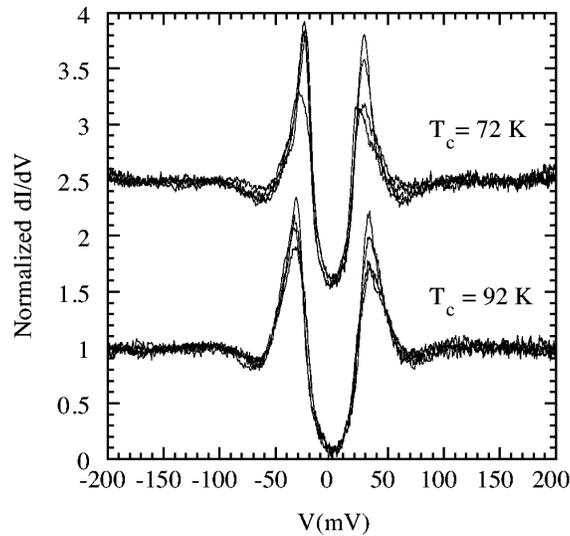}{8cm}
\caption{
Tunneling spectra in optimal and overdoped Bi 2212. The 
gap measured at 4.2 K  increases as the doping level is 
reduced from  overdoped (top curve) to  underdoped (bottom curve). The dip 
is approximately symmetric, suggesting it is not a band structure effect but 
may instead be due to strong coupling effects.
}\label{DeWilde3}
\bigskip
\end{figure}

There is a close relationship between the superconducting gap, as represented 
by the conductance peaks, and the pseudogap as described by the 
depression of conductivity in the normal state. Both features are centered 
at zero bias {\it i.e.} on the Fermi surface. The energy scales are similar 
and both scales are temperature independent.  Furthermore, as the 
doping level is changed, both gaps scale in the same way in that their 
magnitude decreases with doping. It is important to note here that Ding 
\etal (1997) report that in the ARPES spectra a similar parallel can be seen 
in the $k$ dependence of the locus of minimum gap in the pseudogap and the 
superconducting gap --- both seem to be tied to the Fermi surface. 

Another parallel between the ARPES and the tunneling results is the dip in 
the spectra seen in the superconducting state at a frequency of $2\Delta$ 
above zero bias in SIN tunneling. Fig. \ref{DeWilde3} shows recent 
measurements in overdoped Bi2212 by DeWilde \etal (1998). The authors argue 
that  since the peak is symmetrically placed and, as 
a function of doping, the dip scales with $\Delta$, the superconducting gap 
determined from the conductance peaks and therefore the dip is not a band structure 
effect but  
is  intimately associated with superconductivity. 
They note that $\Omega$ 
coincides with $2\Delta$,  which is also seen in Pb where there is a feedback 
effect of a $2\Delta$ electronic singularity on the phonon density of states, 
forming a singularity at $\Omega=2\Delta$. This $4\Delta$ feature was first 
observed in the optical conductivity of Pb by Farnworth and Timusk 
(1975) and has been discussed in detail in a high--$T_c$ context recently by 
Coffey and Coffey (1993). It would occur at $eV=2\Delta$ in SIN tunneling. 
This view of the dip is related to the early interpretation of the dip 
in ARPES spectra discussed in section II and is in accord with the 
results of Renner 
\etal (1998) who also note that, as a function of doping, the peak--to--dip distance 
tracks the gap, decreasing as the gap decreases with doping.  The alternate 
explanation of the dip in terms of coupling to the ($\pi,\pi$) mode at 41 
meV (whose frequency increases with doping ) 
would predict an {\it increase} of the dip--to--peak distance with doping.

\section{Nuclear Magnetic Resonance}

Magnetic resonance experiments were the first used to observe the pseudogap 
in underdoped YBCO. These experiments probe the spin channel as opposed to 
say optical conductivity which probes the charge channel. In a Fermi liquid 
one expects to observe the temperature independent Pauli susceptibility. 
Instead, in the high temperature superconductors a decrease in susceptibility 
with temperature is observed. This has been used to support the hypothesis of a 
spin gap (Warren  \etal (1989)). Subsequent experiments have revealed that 
the pseudogap exists in both the spin and charge channels.

NMR is a versatile tool which allows one to probe  the electronic 
state at different nuclear sites in the lattice as well as probe 
different parts of {\bf q} space.  As  nuclear dipole moments
are several orders of magnitude smaller than the Bohr magneton,
nuclei are ideal probes of the electronic state.
The Knight shift $K_s$ is proportional 
to the real part of the susceptibility $\chi^{\prime}({\bf q} = 0, \omega)$, 
measuring the polarization of  electrons by the applied magnetic 
field. The spin--lattice relaxation rate in the cuprates is dominated 
by antiferromagnetic (AF) spin fluctuations. For example, the 
spin--lattice relaxation rate for copper nuclei in the CuO$_2$ planes 
$1/^{63}T_1$ is enhanced by one to two orders of magnitude by 
these spin fluctuations. On the other hand, the relaxation rate 
$1/^{17}T_1$ seen by 
oxygen atoms in the plane, situated half--way  between the copper 
sites, is barely enhanced. The spin--lattice relaxation rate is related 
to the susceptibility through 
\begin{equation}
1/T_1 = {k_BT \over 4\mu_B^2\hbar^2}\sum_{\bf q} \mid F({\bf q})\mid^2
                \chi^{\prime\prime}({\bf q},\omega)/\omega
                                        \label{eq1}
\end{equation}
where $F({\bf q})$ is the form factor for the particular nuclear site. 
As $1/^{63}T_1$ is dominated by AF spin fluctuations it predominately 
probes ${\bf q} = {\bf Q} \equiv (\pi,\pi)$. Finally the spin--spin 
relaxation rate $1/T_{2G}$ is related to $\chi^{\prime}({\bf q},\omega)$. 

\begin{figure}[t]
\begin{minipage}[c]{7.5cm}
\epsfigrot{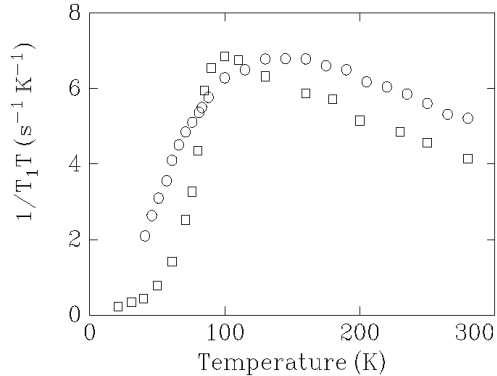}{7cm}
\end{minipage}
\hfill
\begin{minipage}[c]{7.5cm}
\begin{centering}
\caption{
Planar $^{63}$Cu spin--lattice relaxation rate in optimally doped 
YBa$_2$Cu$_3$O$_{6.95}$ (squares) and underdoped YBa$_2$Cu$_3$O$_{6.64}$ 
(circles). The pseudogap causes a suppression in the relaxation rate well above
$T_c$.
}\label{fignmr1}
\end{centering}
\end{minipage}
\bigskip
\end{figure}

In the context of a Fermi liquid the Knight shift is proportional 
to the density of states at the Fermi surface. The spin--lattice relaxation 
rate is dominated by electronic excitations. The pseudogap is observed in 
both the Knight shift and spin--lattice relaxation rate. 
Warren \etal (1989a) were the first to see the pseudogap in 
$1/^{63}T_1$ of underdoped YBCO. They suggested the possibility of spin pairing 
above $T_c$. Figure \ref{fignmr1}  displays the spin--relaxation rate in 
optimally doped YBCO and YBCO 6.7. As temperature decreases $1/T_1$ increases, 
as the AF spin coherence increases, 
until just above $T_c$ in the optimally doped compound. On the other hand, 
for the underdoped compound, $1/T_1$ starts to decrease well above $T_c$. 
This is attributed to the presence of the pseudogap.

%19 NMR

%20
\begin{figure}[!h]
\begin{minipage}[c]{7.5cm}
\epsfigrot{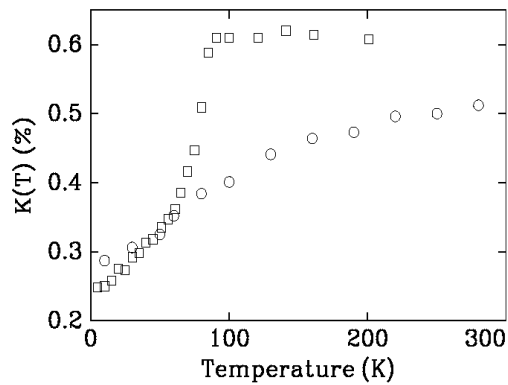}{7cm}
\end{minipage}
\hfill
\begin{minipage}[c]{7.5cm}
\begin{centering}
\caption{
Planar $^{63}$Cu Knight shift in \break YBa$_2$Cu$_3$O$_{6.95}$ (squares) 
and underdoped \break YBa$_2$Cu$_3$O$_{6.64}$ (circles). The normal state
susceptibility is temperature independent in the optimally doped compound
but decreases with temperature in the underdoped compound.
}\label{fignmr2}
\end{centering}
\end{minipage}
\bigskip
\end{figure}

Walstedt \etal (1990) have observed the pseudogap in 
$^{63}K_s$ of YBCO 6.7. As can be seen in Fig. \ref{fignmr2} $K_s$ drops to 
about 20\% of its room temperature value at $T_c$, leaving a much smaller 
decrease in the superconducting state than for the optimally doped compound. 
Entry into the superconducting state is dramatic for YBCO 7.0 as the spins 
condense out into spin singlet pairs. YBCO 6.7 on the other hand, barely 
provides any hint of the superconducting transition in the Knight shift.

It is worth noting that if one views the pseudogap as a loss in 
density of states in the CuO$_2$ layers, then one would naturally expect it 
to be accompanied by a reduction in the AF spin fluctuations. This is 
particularly the case in a one band model where the local moments on the 
copper sites are hybridized with the itinerant holes on the oxygen sites. 
From this perspective one should expect the temperature at which the 
pseudogap develops to be similar for the Knight shift and the relaxation 
rates. 

Observation of the pseudogap is not constrained to  copper nuclei.
Alloul's group (Alloul \etal (1989)) have performed $^{89}$Y NMR studies on 
underdoped YBCO which display the pseudogap. $^{17}$O measurements on the 
planar oxygen sites by Takigawa \etal (1991a)  also show the 
pseudogap. These authors point out that the Knight shift for copper and 
oxygen have the same temperature dependence in the normal state. This 
suggests that a single band model may apply where the Cu, O and Y nuclei all 
couple to the same $\chi({\bf q}, \omega)$.

Many other underdoped compounds exhibit the pseudogap in their 
magnetic resonance properties. Some examples are the naturally underdoped 
and stoichiometric YBa$_2$Cu$_4$O$_8$ (Bankay \etal (1994)) and the related  
Y$_2$Ba$_4$Cu$_7$O$_{15}$ (Stern \etal (1994)), the two layer bismuth 
compound (Walstedt \etal (1991)) and the mercury compounds (Julien \etal (1996) and 
Bobroff \etal (1997)). 
The pseudogap is observed in the Knight shift and relaxation rates of the 
planar copper and oxygen nuclei as well as in the nuclei situated between 
the CuO$_2$ planes in multilayer compounds.

\begin{figure}
\begin{minipage}[b]{0.45\linewidth}
\epsfigrot{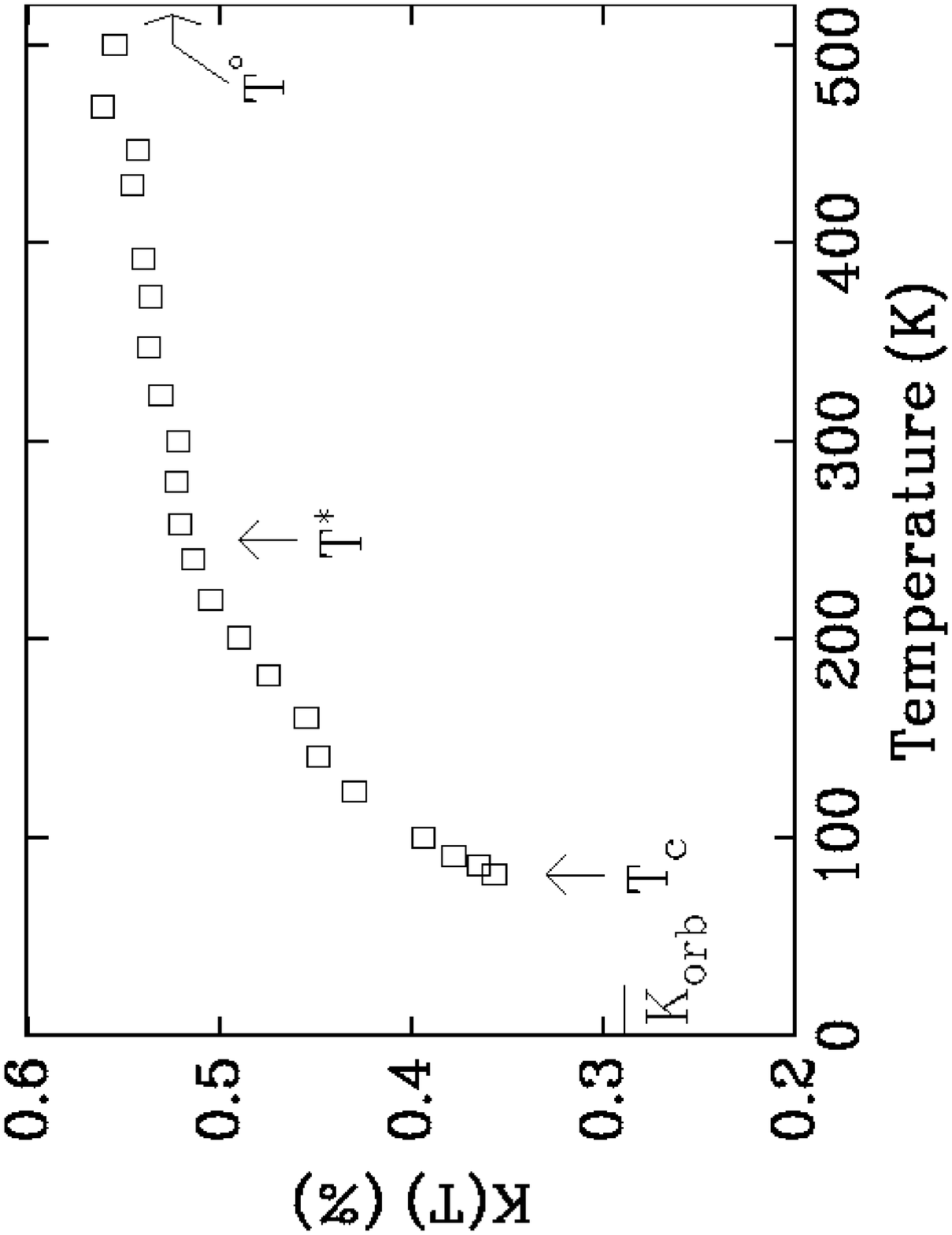}{7cm}
\caption{
Crossover displayed  in planar $^{63}$Cu Knight shift in YBa$_2$Cu$_4$O$_8$.
The upper crossover is taken from Fig. 23.
}\label{fignmr3}
\end{minipage}\hfill%
\begin{minipage}[b]{0.45\linewidth}
\epsfig{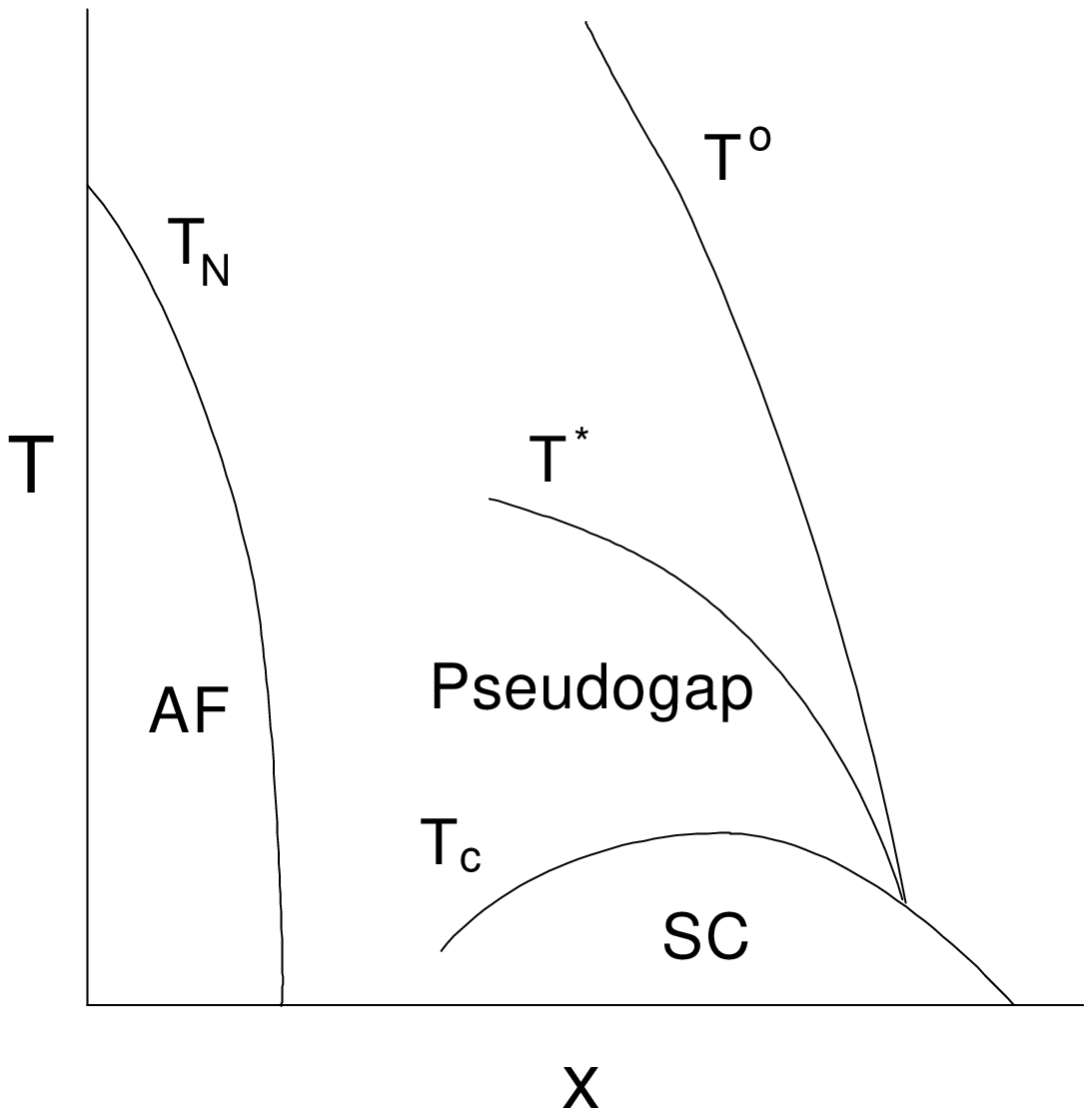}{6cm}
\caption{
Pseudogap crossover temperatures(K) suggested  from NMR data.
Note that the crossover temperatures merge into $T_c$ slightly
into the overdoped region of the phase diagram.
}\label{phasediag}
\end{minipage}
\bigskip
\end{figure}

Certain compounds exhibit crossovers in the pseudogap as a function of 
temperature. This is in contrast to others in which a smooth evolution with 
temperature is observed. The latter is typified by underdoped YBCO,  
illustrated in Fig. \ref{fignmr2}, whereas crossover temperatures can be  
discerned, for example, 
in YBCO 124 (Bankay \etal (1994)) displayed in 
Fig. \ref{fignmr3}. 
At the upper crossover temperature  $T^{\circ}$ the Knight shift 
changes behaviour from being temperature independent to decreasing 
linearly with temperature below $T^{\circ}$. Below the lower crossover 
temperature $T^*$ the Knight shift decreases faster than linear with 
temperature. As  we will discuss below, these two crossover temperatures may
have their origin in different physical phenomena. Indeed, different 
scaling relations are observed in the relaxation rates delineated 
by $T^{\circ}$ and $T^*$.

%21

The magnitude of the crossover temperatures increases with decreasing doping 
and is discussed below. Several slightly underdoped compounds 
have been observed with crossover temperatures much lower than those of 
YBCO 124. For example, the crossover temperatures of slightly underdoped 
Pb$_2$Sr$_2$(Y,Ca)Cu$_3$O$_{8+\delta}$ (Hsueh \etal (1997)) are about a 
factor of 2.5 lower than those of YBCO 124. Two other examples are slightly 
underdoped Bi2212 (Walstedt \etal (1991)) and 
Y$_{0.95}$Pr$_{0.05}$Ba$_2$Cu$_3$O$_7$ (Reyes \etal (1991)). Optimally doped 
YBCO also exhibits  crossover  temperatures (Martindale \etal (1996)) not 
too far above $T_c$. This is consistent with the phase diagram in Fig. 
\ref{phasediag} where the crossover temperatures approach the 
superconductivity phase boundary on the overdoped side.

A common feature to the pseudogap observed in the 
slightly underdoped compounds is that the drop in the 
susceptibility in the normal state, as measured by the Knight  
shift, is less than that in the strongly underdoped compounds. 
This drop is smallest for the optimally doped YBCO. 
Thus as the doping is increased, the  drop in the 
susceptibility in the normal state relative to that in the superconducting 
state decreases. In the overdoped region the pseudogap disappears; 
in this case the susceptibility drops only in the superconducting state. 

Measurements of the pseudogap in the underdoped, three layer mercury 
compound HgBa$_2$Ca$_2$Cu$_3$O$_{8+\delta}$ have been  
performed by Julien  \etal (1996). The lower crossover 
temperature $T^*$ is detected in both the Knight shift and 
$1/^{63}T_1T$. As the authors point out, the value of $T^*$ 
= 250K is the largest reported to date and may indeed be related 
to the large value of $T_c$ = 115K. 

A table summarizing the crossover 
temperatures of various underdoped cuprates appears below.  
Values of $T^*$ and $T^{\circ}$ are presented from both 
Knight shift and relaxation data. Where available, the 
ratio $^{63}T_1T/^{63}T^n_{2G}$ is used, otherwise 
the point at which $^{63}T_1T$ breaks away from 
linearity is used to extract $T^*$. For comparison, 
$T_c$ is also tabulated.

\begin{table}
\hskip2cm%
\begin{minipage}{10.6cm}
\begin{center}
\caption{Pseudogap crossover temperatures(K) from NMR data.}
\smallskip
\begin{tabular}{p{4cm}p{1.3cm}p{1.3cm}p{0.5cm}p{0.8cm}p{1.3cm}p{0.5cm}}
&&Knight shift &&& Relaxation rates\\
\cline{3-4}\cline{6-7}
& $T_c$ & $T^*$ & $T^{\circ}$ && $T^*$ & $T^{\circ}$ \\
\hline
YBa$_2$Cu$_3$O$_{6.95}$ $^a$       & 92 & 110 & 150 &&      &      \\
YBa$_2$Cu$_4$O$_8$ $^b$                  & 81 & 240 &      && 200 & 500 \\
Y$_2$Ba$_4$Cu$_7$O$_{15}$ $^c$  & 93 &  190  & 250 && 130 & 250 \\
Pb$_2$Sr$_2$(Y,Ca)Cu$_3$O$_{8+\delta}$ $^d$ & 80 & 100 & 180 && 140 & \\
Bi$_2$Sr$_2$CaCu$_2$O$_8$ $^e$    & 90 & 110 & 170 && 200 & \\
HgBa$_2$Ca$_2$Cu$_3$O$_{8+\delta}$ $^f$ & 115& 250&     && 250 & \\  
\end{tabular}
\end{center}
\vskip -\baselineskip
$^a$Martindale(1996)\\
$^b$Corey(1996)\\
$^c$Stern(1994,1995)\\
$^d$Hsueh(1997)\\
$^e$Walstedt(1991)\\
$^f$Julien(1996)\\
\end{minipage}\hfill
\label{table1}
\end{table}

Comparing the  Knight shift values of $T^*$ with those obtained 
from the relaxation rates one sees that although they are comparable, 
they are nevertheless distinct. Thus the lower crossover temperature 
is not necessarily the same for the {\bf q} = 0 dynamic susceptibility 
as for the {\bf q} = {\bf Q} dynamic susceptibility. This difference in 
values of $T^*$ has been noted 
explicitly by Hsueh  \etal (1997). Many others have warned 
that one must be careful to distinguish between the {\bf q} = 0 
and {\bf q} = {\bf Q} regions of the susceptibility. 
Julien  \etal (1996) have noted that $T^{\circ}$ depends 
strongly on hole doping but is insensitive to Zn doping, whereas 
$T^*$ is weakly dependent on hole doping but significantly 
suppressed by Zn doping. This suggests that the lower crossover 
temperature is related to magnetic phenomenon (eg. spin gap) 
and that the upper crossover may be associated with a feature 
in the density of states. As we have noted in section II, ARPES
results also display a {\bf q} dependent behaviour of the pseudogap.

There appears to be no strong connection between the presence of a pseudogap 
and the number of adjacent CuO$_2$ layers in the compound.  
Winzek  \etal (1993) have observed the 
pseudogap in the single layer Hg--compound. More recently 
Alloul's group (Bobroff \etal (1997)) has measured the $^{17}$O 
Knight shift on the single layer Hg--compound to find a  
pseudogap in the underdoped compound. The upper crossover 
temperature $T^{\circ}$ was observed with the expected 
linear temperature dependence below $T^{\circ}$. 
Itoh  \etal (1996) observed the pseudogap in 
the $^{63}$Cu relaxation rates of HgBa$_2$CuO$_{4+\delta}$ as well. 
Another distinct case to consider is a three layer compound 
where a pure bilayer coupling model (Millis \etal (1993)) 
would have a different dynamic susceptibility in the inner 
and sandwiching layers. Measurements of $1/^{63}T_1T$ 
in HgBa$_2$Ca$_2$Cu$_3$O$_{8+\delta}$ reveal that the 
two relaxation rates are identical. 
Thus it would seem that models of the pseudogap 
based exclusively on bilayer coupling are ruled out. 

The peculiar behaviour of the lanthanum compound 
warrants special mention. The temperature dependence 
of the spin susceptibility suggests that a pseudogap may 
be present in the underdoped lanthanum compound. 
On the other hand the copper spin--lattice relaxation 
rate (Ohsugi \etal (1991)) shows no sign of a pseudogap for any strontium 
concentration between $0.075 \le x \le 0.15$. Further discussion on
this compound appears below in section V and section XI. 

\begin{figure}[t]
(a)
\epsfig{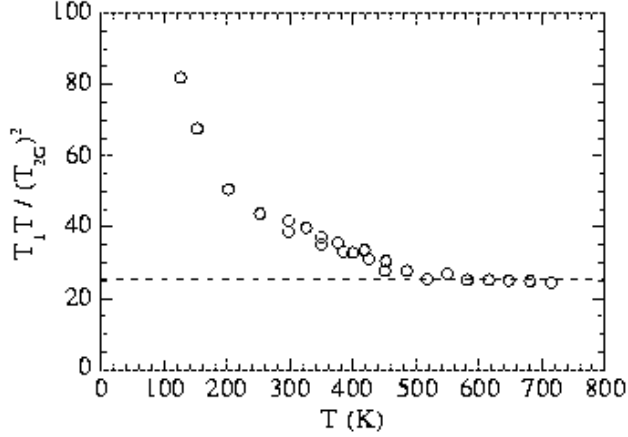}{9cm}
(b)
\epsfig{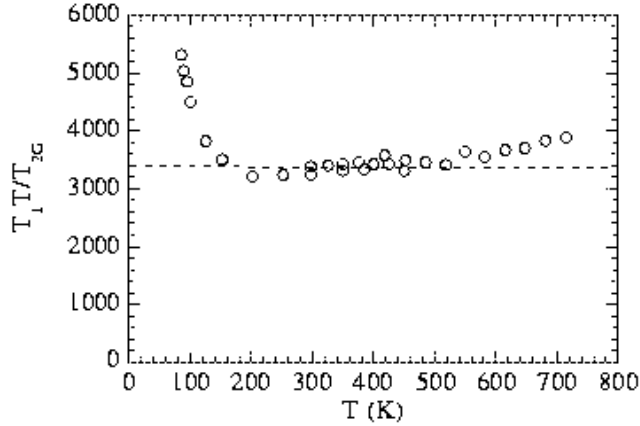}{9cm}
\caption{
Relaxation rate ratios in YBa$_2$Cu$_4$O$_8$. 
        (a) $^{63}T_1T/^{63}T^2_{2G}$, (b) $^{63}T_1T/^{63}T_{2G}$. 
These two ratios clearly indicate the presence of two distinct 
crossover temperatures.
}\label{fignmr4}
\bigskip
\end{figure}

Chakravarty, Halperin and Nelson (1989) have shown that the $S$=1/2 2D 
Heisenberg antiferromagnet crosses over from a mean field  state with a 
dynamical critical exponent of $z$ = 2 to a quantum critical state with $z$ 
= 1.  Barzykin and Pines (1995) have extended this framework into the doped 
region of the phase diagram. Here one expects a crossover from mean field 
behaviour with $z$ = 2 to a pseudoscaling regime with $z$ = 1 at $T^{\circ}$ 
followed by a crossover to a quantum--disordered state at $T^*$.  These 
scaling relations have been observed in YBCO 124 by Corey  \etal (1996). Above 
$T^{\circ}$ the ratio $^{63}T_1T/^{63}T^2_{2G}$, which is dominated by {\bf 
q} = {\bf Q}, is found to be temperature independent (Fig. \ref{fignmr4}a), 
the same as for $z$ = 2 scaling. Between $T^{\circ}$ and $T^*$, on the other 
hand, the ratio $^{63}T_1T/^{63}T_{2G}$ is temperature independent (Fig. 
\ref{fignmr4}b), consistent with $z$ = 1 scaling. This temperature 
independence of $^{63}T_1T/^{63}T_{2G}$ above $T^*$ has also been observed 
in HgBa$_2$Ca$_2$Cu$_3$O$_{8+\delta}$\  (Julien \etal 1996). A more thorough 
description of these scaling relations is found in the theory section, 
section X. 

Williams \etal (1997) have fit planar oxygen Knight shift data of several
compounds to discern whether an s or d--wave gap was applicable
in the normal and superconducting states. The authors found that a
d--wave model for both the superconducting gap and the pseudogap
fit better than an s--wave model. Although not as direct a measurement,
this is consistent with the ARPES results on the symmetry of the gaps.

Impurities not only affect superconductivity but also the pseudogap.   Many 
studies have revealed Zn to be more effective at suppressing $T_c$ than Ni. 
This may be related to the local  moment associated with the Zn impurity in 
underdoped materials (Alloul \etal (1991)  and Mahajan \etal (1994)). Zheng 
\etal (1993,1996) have investigated the Zn  substituted YBCO 124 compound. They 
have monitored the pseudogap with both the Knight shift and spin--lattice 
relaxation rate of the planar $^{63}$Cu nuclei. The crossover temperature 
$T^*$ in $1/^{63}T_1T$ decreases as the Zn concentration is increased. With 
1\% Zn the effect of the pseudogap is completely suppressed, i.e. $T^* < 
T_c$. Further increasing Zn impurities to 2\% increases $1/^{63}T_1T$ above 
the rate observed in pure YBCO 124 just above $T_c$. Most striking is the 
observation that the Knight shift is {\it unaffected} by Zn doping. Thus the 
pseudogap in the {\bf q} = {\bf Q} region of  the susceptibility is strongly 
affected in sharp contrast to the {\bf q} = 0 region which is not 
discernibly affected by Zn impurities. 

Suppression of the pseudogap in YBCO 124 by Zn impurities has been observed with 
other techniques. The deviation from T--linear dependence of the resistivity 
is also a signature of the pseudogap. This deviation is suppressed upon Zn 
doping of YBCO 124 (Miyatake \etal (1991)). Optical conductivity measurements 
also show a suppression of the pseudogap with Zn doping(Puchkov \etal 
(1996a)). As 
antiparamagnon scattering is largely responsible for the conductivity  
(Statt and Griffin 1992, 1993), one would expect the behaviour 
near {\bf q} = {\bf Q} to dominate these measurements. Thus the suppression 
of the pseudogap observed with these various measurements is consistent. 

Zheng  \etal (1996) have studied the effects of doping YBCO 124 and YBCO 6.7 with 
Zn and Ni impurities. They find that Zn is much more effective at 
suppressing the pseudogap than Ni. A plot of $T_c$ vs. impurity 
concentration $x$ in YBCO 124 shows that Zn is about 3 times more effective in 
depressing $T_c$ than Ni. A similar plot of $T^*$ vs. $x$ illustrates the 
suppression of the pseudogap with doping. In this case Zn is found to be 4 
times as effective in suppressing $T^*$ as Ni. Thus it appears that  Zn 
induces local moments in the CuO$_2$ plane which strongly affects the local 
antiferromagnetic correlations. This in turn suppresses $T_c$ and the 
pseudogap at {\bf q} = {\bf Q}. On the other hand, doping has  little or no 
effect on the {\bf q} = 0 susceptibility. 

%23 a,b

In summary, magnetic resonance experiments have been used to probe the 
nature of the pseudogap in the spin channel. Two crossover temperatures have 
been observed. The upper crossover temperature $T^{\circ}$ may very well be 
associated with a phenomenon affecting the density of states whereas the lower 
crossover temperature $T^*$ seems to be a result of a gap in the 
antiferromagnetic excitations. These conclusions are based, in part, on 
differences noted in the behaviour of the dynamic susceptibility at {\bf q} 
= 0 and {\bf q} = {\bf Q}.  They are also consistent with the ARPES results 
showing that the pseudogap opens up in the ($\pi,0$) region of the Fermi 
surface leaving an ungapped arc centred about ($\pi/2,\pi/2$).

\section{Transport Properties}
% V

Within the Fermi liquid picture a gap or a partial gap in the density of 
states near the Fermi level can affect the electrical conductivity in two 
ways: through the reduction of the number of current carrying states 
as a gap forms, and secondly, if the current carriers are scattered by 
electronic excitations, through the reduction in the density of such 
excitations.  The standard formula for the current ${\bf J}$ in the presence 
of an electric field ${\bf E}$ (Ashcroft and Mermin (1976)): 

\begin{equation}
{\bf J}={\bf \sigma} {\bf {\cdot}}{\bf E}={e^2 \over 4\pi^3}\int {\tau_{\bf 
k}{\bf v_k v_k}
\over 1-i\tau_{\bf k}\omega} {dS_F \over v_{\bf k}}{\bf  \cdot  
E_0} \label{ziman}
\end{equation}
where ${\bf \sigma}$ is the conductivity tensor, ${\bf E_0}$ the applied 
field, $\tau_{\bf k}$ the lifetime of state ${\bf k}$ and ${\bf v_k}$ the 
velocity of state ${\bf k}$. The integration is carried out over the Fermi 
surface. The  conductivity is proportional to the component of the velocity 
in the direction of the applied field ${\bf E}$ averaged over the Fermi surface 
and it is also proportional to the factor $\tau_{\bf k}$ which can vary over 
the Fermi surface independently.      

If one further assumes that $\tau$ is a constant one gets the classical 
Drude formula for the conductivity:

\begin{equation}
\sigma(\omega)= {1 \over {4\pi}}
{{\omega_p^2} \over {1/\tau-i\omega}}
\end{equation}
which yields a Lorentzian peak centered at zero frequency with an oscillator 
strength $\omega_p^2/8$, where $\omega_p^2=e^2/(3\pi^2\hbar){\int}{\bf v} 
{\cdot} d{\bf S}_F$ and ${\bf v}$ is the electron velocity and ${\bf S}_F$ 
is an element of Fermi surface. For a spherical Fermi surface 
$\omega_p^2=4{\pi}ne^2/m_e$, where $n$ is the free--carrier density and $m_e$ 
is the electronic band mass. The imaginary part of $\sigma(\omega)$ is just 
the real part multiplied by $\omega\tau$. 

The simple Drude formula is widely used to describe the optical conductivity 
of metals but it is  only valid for impurity scattering where $\tau$ 
is a frequency independent constant. If the processes that limit the 
lifetime $\tau$ are inelastic $\tau$ becomes frequency dependent, 
and to preserve the causal nature of \sig, it is necessary to generalize the 
Drude formula. This is usually done through the extended Drude model or 
memory function technique (Mori (1965), G\"otze and W\"olfle (1972), Allen 
and Mikkelsen (1976), Puchkov \etal (1996a)). The extended Drude formula is 
written as: 

\begin{equation}
\sigma(\omega,T)=
{1 \over {4\pi}} {{\omega_p^2}
\over {1/\tau(\omega,T)-i\omega[1+\lambda(\omega,T)]}}
\end{equation}
where $1/\tau(\omega,T)$ describes the frequency--dependent scattering rate 
and $\lambda(\omega,T)$ can be viewed as a mass enhancement of the electronic 
excitations due to the interactions. 

One can solve for $1/\tau(\omega)$ and $1+\lambda(\omega)$ in terms of the 
experimentally determined optical conductivity to find

\begin{equation}
1/\tau(\omega)={\omega_p^2 \over {4\pi}} Re({1 \over \sigma(\omega)}). 
\label{tausig}
\end{equation}
The dc resistivity is the zero frequency limit 
$\rho_{dc}(T)=1/\sigma_{dc}(T)=m_e/(\tau(T)ne^2)$ since \sig\ is real in the 
zero frequency limit.

The mass enhancement factor $\lambda(\omega)$ is given as the
imaginary part of $1/\sigma(\omega)$:

\begin{equation}
1+\lambda(\omega)=-{\omega_p^2 \over {4\pi}} {1 \over \omega} Im({1
\over \sigma(\omega)}).
\end{equation}
The total plasma frequency $\omega_p^2$ can be can be
found from the sum rule
$\int_0^{\infty}\sigma_1(\omega)d\omega=\omega_p^2/8$. 

Memory--function analysis has been widely used to describe 
electron--phonon scattering (Allen (1971)). Shulga \etal (1991) 
give the following expression for $1/\tau(\omega,T)$: 

\begin{eqnarray}
{1 \over \tau}(\omega,T)={\pi \over \omega} \int_0^{\infty} d\Omega
\alpha_{tr}^2(\Omega)F(\Omega)[2{\omega}{\coth}({\Omega \over {2T}})-
(\omega+\Omega){\coth}({{\omega+\Omega} \over {2T}})+ \nonumber \\ +
(\omega-\Omega){\coth}({{\omega-\Omega} \over {2T}})] +
{1 \over \tau_{imp}}. \label{Shulga}
\end{eqnarray}
Here $\alpha_{tr}^2(\Omega)F(\Omega)$ is a weighted phonon density of states
and $T$ is the temperature measured in frequency units. The last term 
in eqn. \ref{Shulga} represents impurity scattering. 

The quantity $\alpha_{tr}^2(\Omega)F(\Omega)$ is closely related to 
$\alpha^2(\Omega)F(\Omega)$ obtained from the inversion of tunneling spectra 
in BCS superconductors (Allen (1971)) and  can be found by inverting the 
optical conductivity spectra at $T=0$ (Allen (1971), Farnworth and Timusk 
(1974), Marsiglio \etal (1998)) to yield: 

\begin{equation}
\alpha^2(\Omega)F(\Omega) = {1 \over 2\pi\omega}{\partial \over 
\partial\omega}\Bigg[\omega^2{\partial \over \partial\omega}{1 \over 
\tau(\omega)}\Bigg]. \label{a2F}
\end{equation}
The presence of the second derivatives in the expression places severe demands 
on the signal to noise ratio of the $1/\tau(\omega)$ experimental spectrum. 
The method has been applied with considerable success for BCS superconductors to 
extract $\alpha^2(\Omega)F(\Omega)$ and  most recently to K$_3$C$_{60}$ 
(Marsiglio \etal (1998)). It has not been explored to date for the 
high--$T_c$ cuprates. 

We will first discuss the coherent ab--plane conductivity. The 
conductivity normal to the copper oxygen planes, the c--axis conductivity, 
will be discussed separately below. 

\subsection{~~~dc resistivity}

\begin{figure}[!ht]
\epsfig{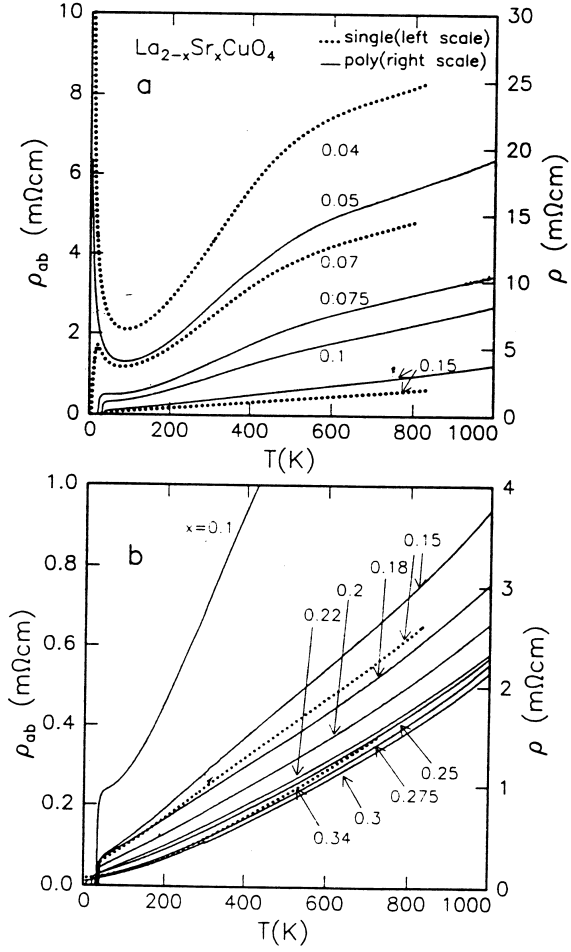}{9cm}
\caption{
Temperature dependence of the resistivity of \LaSr\  at various doping levels: 
underdoped (top panel) and optimal and overdoped (lower panel). Both single 
crystals and polycrystalline films are shown. There is a drop in resistivity 
at a temperature $T^*$ due to reduced scattering as the result of the 
formation of the pseudogap.
}\label{Takagi1}
\bigskip
\end{figure}

One of the striking properties of the high temperature superconductors is 
the remarkable temperature dependence of the normal state resistivity. 
Seen most clearly in optimally doped materials, the variation is linear with 
temperature, extending from 10 K to 1000 K and extrapolating to zero 
resistance at zero degrees. This singular behaviour (Gurvich and Fiory 
(1987),Martin \etal (1990)) is to be contrasted with the behaviour in 
conventional, good metals, where the resistivity is linear only over a 
limited range of temperatures, has an intercept on the temperature axis at 
some fraction of the Debye temperature and saturates at high temperature. 
The saturation occurs when the inelastic mean free path approaches the 
lattice spacing, the Ioffe--Rigel criterion, which in 2 dimensions reads 
$\rho=\rho_M/k_Fl$ where $\rho_M=(h/e^2)d$, $d$ the interplane distance and 
$l$ must be larger than the CuO bond length $a_0/2$. For a Fermi surface 
with $k_F\approx \pi/a$ this yields a resistivity that saturates at 
$\approx$ 1 m$\Omega$cm. It is clearly illustrated in Fig. \ref{Takagi1}, from the 
work of Takagi \etal (1992), that the 
resistivity of optimally doped \LaSr\ shows no sign of saturation at this 
value and in the underdoped materials appears to be much higher than what is 
allowed by the Ioffe-Rigel limit. 

\begin{figure}
\epsfig{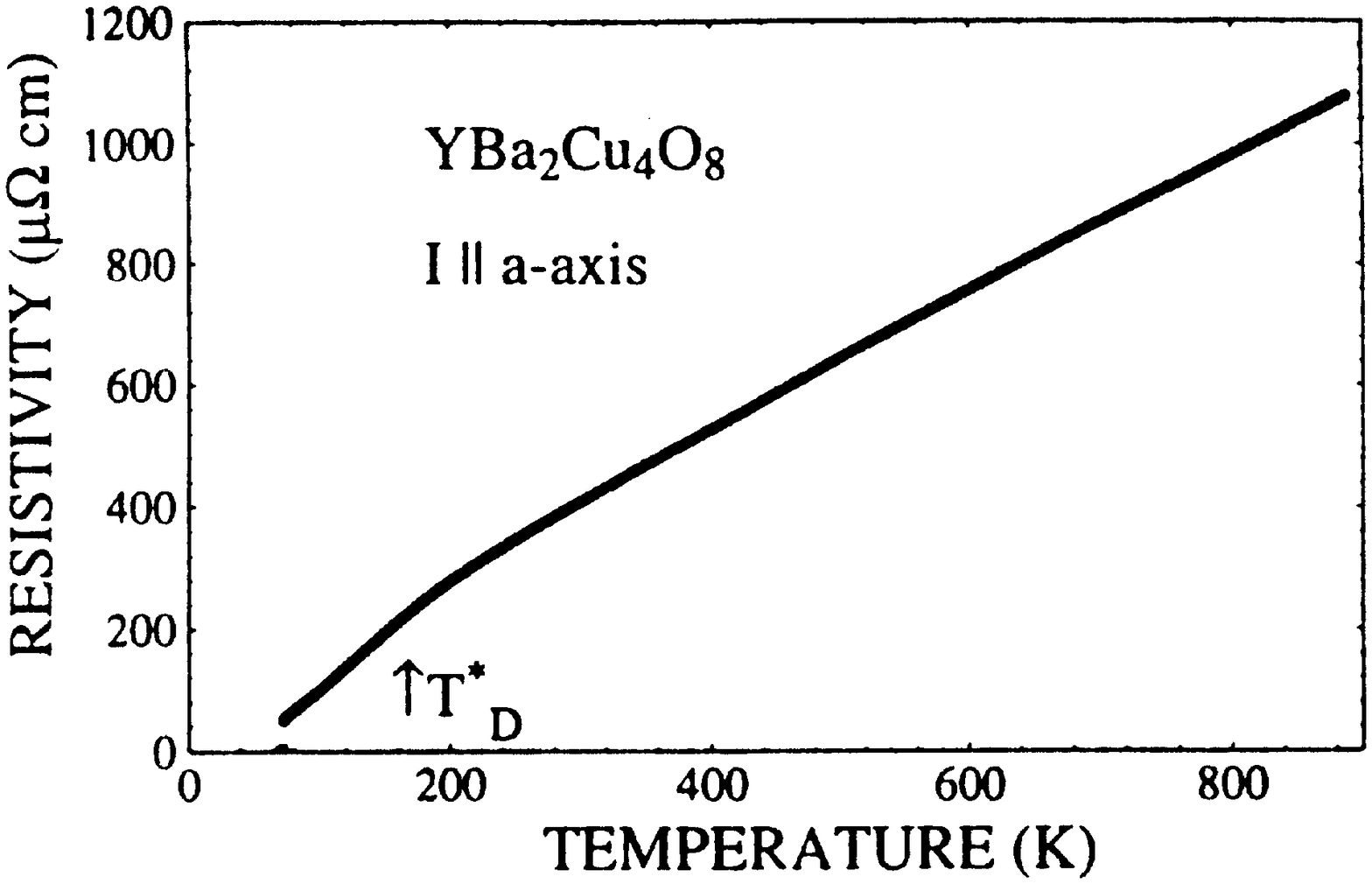}{10cm}
\caption{
Formation of the pseudogap state in the resistivity of \YBa124. This 
naturally underdoped material shows the presence of a pseudogap  below 200 
K. Here the dc resistivity drops below the linear dependence set at higher 
temperatures.
}\label{bucher2}
\bigskip
\end{figure}

In underdoped materials, the presence of the pseudogap leads to striking 
deviations from this simple linear behavior. Bucher \etal (1993) observe a 
reduced resistivity in \YBa124\ below $T^*$, near the temperature where the  
pseudogap, as seen by NMR, opened. The double chain YBCO 124 is an almost ideal 
system for the study of the pseudogap phenomenon. It is simultaneously 
underdoped, stoichiometric with a fixed doping level and naturally untwinned 
(Bucher \etal (1993)). By measuring the resistivity with currents normal to 
the chains it is, to a large extent, possible to suppress the chain response. 
As Fig. \ref{bucher2} shows the resistivity of \YBa124\  is linear at 
high temperature with a slight positive intercept but has a clear break in 
slope below 200 K. The authors point out that the temperature where this 
happens is correlated with the NMR pseudogap. The Knight shift has its
lower crossover temperature below $\approx$ 250 K and the spin--lattice relaxation 
time $1/T_1T$ below $\approx$ 200 K. Thus the resistivity anomaly coincides 
roughly with these two temperatures and Bucher \etal suggest that the 
decrease in resistivity below 180 K is caused by the reduced scattering by 
spin fluctuations resulting from the opening of the gap in the spin 
excitation spectrum. 

%24 TRANSPORT

\begin{figure}
\epsfigrot{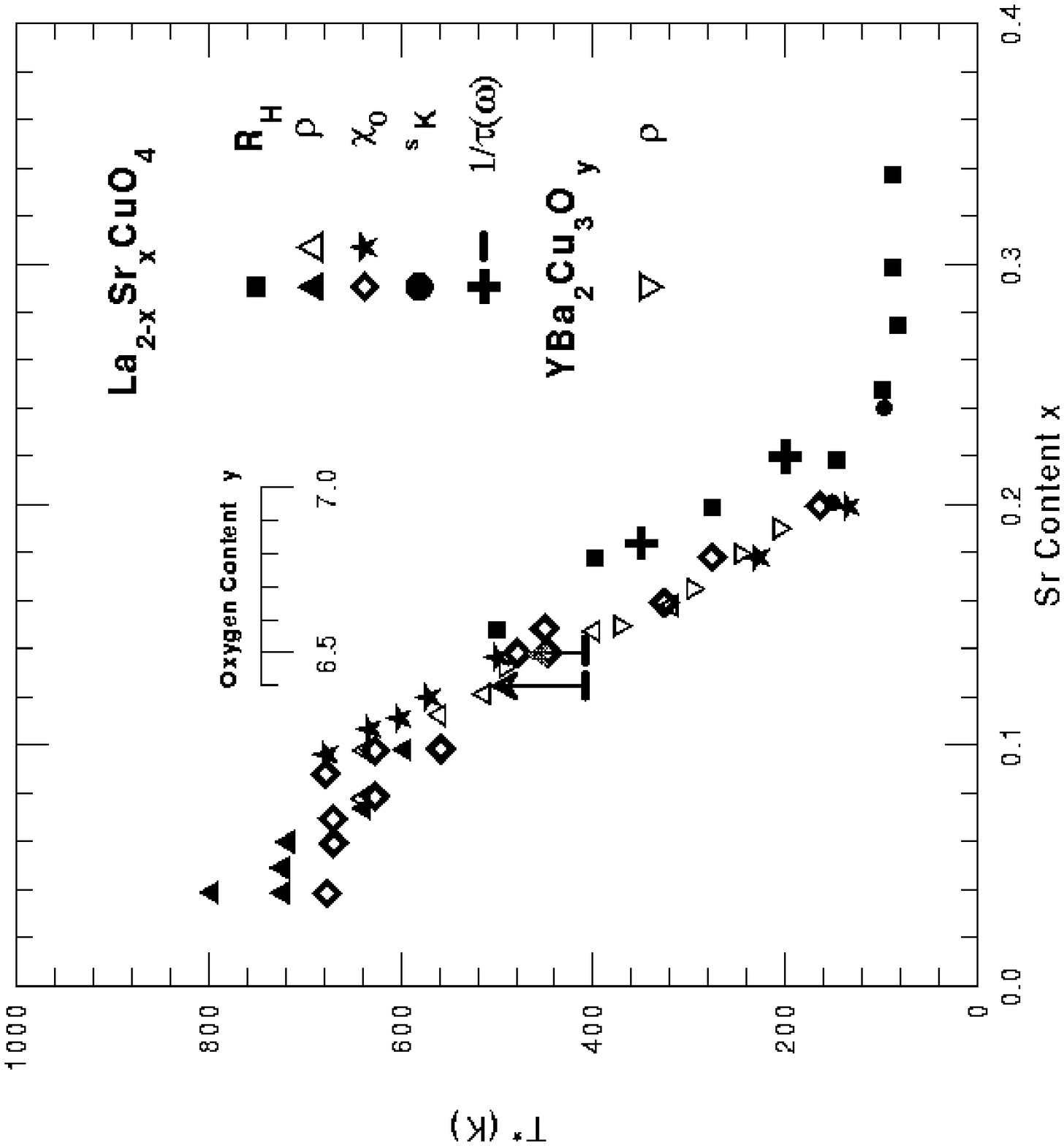}{12.5cm}
\caption{
Variation of $T^*$ with doping for \LaSr\  as measured by various probes. The 
filled squares denote the temperature below which the Hall coefficient has a 
rapid temperature dependence. The open circles refer to maxima in the static 
susceptibility $\chi(T)$ and the solid circles the temperature where the 
Knight shift starts to decrease. The triangles refer to  the 
temperature where there is a slope change in the dc resistivity, 
the crosses infrared measurements of $1/\tau$ suppression and the 
horizontal lines to lower limits of infrared data. 
}\label{Batlogg4} 
\bigskip \end{figure}

A complete study of the doping dependence of the characteristic temperature 
$T^*$ was done in \LaSr\  (Takagi \etal (1992) and Batlogg \etal (1994)). 
Fig. \ref{Takagi1} (upper panel) shows the resistivity data for a series of 
Sr compositions. At the lowest temperatures $\rho_{ab}(T)$ shows 
``semiconducting'' behavior with the resistivity decreasing with temperature, a 
signature of localization. At intermediate temperatures, the resistivity 
increases superlinearly up to a temperature $T^*$ where there is a clear 
break in slope and the temperature dependence becomes approximately linear. 
Batlogg \etal use the variation of $T^*$ with $x$ to define a boundary of 
the low temperature ``pseudogap'' phase. This phase diagram is shown in Fig. 
\ref{Batlogg4} from Batlogg \etal (1994) with $1/\tau$ data 
from Startseva \etal (1998a,1998b). 

%25 

\begin{figure}
\epsfig{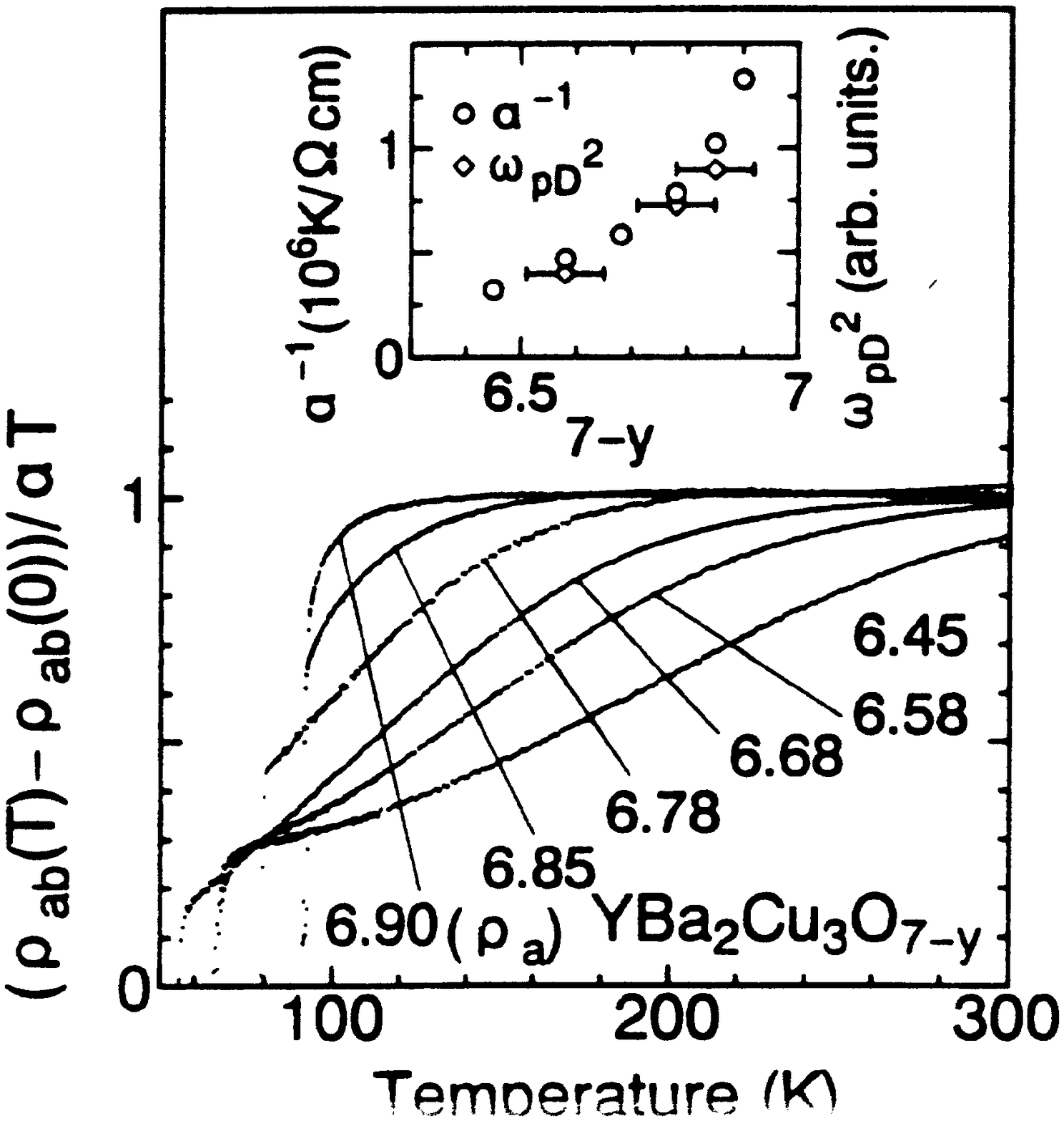}{7cm}
\caption{
Dc resistivity of \YBax\  at various doping levels plotted to show the deviation 
from the high temperature linear variation, unity on this plot. A pseudogap 
onset temperature can be defined from this plot. The inset shows the 
variation of the Drude plasma frequency with doping along with the slope of 
the resistivity. 
}\label{ito2}
\bigskip
\end{figure}

Deviations from $T$--linear resistivity in \YBax\  have been investigated by 
Ito \etal (1993), Carrington \etal (1993) and Takenaka \etal (1994). 
A systematic depression of $\rho(T)$ is found below the 
linear trend set at high temperature. The crossover temperature $T^*$ 
decreases as the doping level rises much in the same way as it does in the 
\LaSr\  system. Figure \ref{ito2} from Ito \etal (1993) shows a plot 
of $(\rho_{ab}(T) - \rho_{ab}(0))/\alpha T$ where $\rho_{ab}(0))/\alpha T$ 
is the $T=0$ intercept of the extrapolated $T$--linear high temperature curve 
and $\alpha$ the slope of the linear part of the resistivity. It is clear 
from the figure that a temperature $T^*$ can be defined from the deviations 
from the linear variation as the curves fall below unity. Ito \etal suggest 
that the reduced scattering below $T^*$ is the result of the development of 
the pseudogap as seen in the NMR Knight shift and spin--relaxation rate
as discussed in section IV.

The inset in Fig. \ref{ito2} shows that both the inverse slope and the 
Drude weight vary linearly with the doping level (Orenstein \etal (1990)). The 
authors suggest that since $\rho_{ab}=(4\pi/\omega^2_{pD})\tau^{-1}$, where 
$\omega^2_{pD}$ is the Drude weight and $\tau^{-1}$ the scattering rate, 
that the scattering rate is independent of doping level in the underdoped 
region and all the variation of the in--plane resistivity arises from the 
changing Drude weight. This suggestion is in approximate agreement with the 
optical results reviewed by Puchkov \etal (1996a) where the 300 K scattering 
rate for a large variety of systems with different doping levels was found 
to be $\approx 1050 \pm 200$ cm$^{-1}$ ($130 \pm 25$ meV). 

The measurements of Ito \etal were done mainly on untwinned single 
crystals of \YBax\  with some checks on detwinned samples to show that there 
was relatively little ab contribution from the chains. This is a little 
surprising since it has been shown by Basov \etal (Basov \etal 
(1995a) that 
in very pure materials where the copper oxygen chains are intact, the  chain 
contribution is very large. This suggests that the crystals of 
Ito \etal  
contained a substantial concentration of chain--breaking defects which acts to 
localize the chain conductivity at low frequencies. However, from a comparison 
with YBCO 124 untwinned measurements of Bucher \etal and the infrared work 
of Basov \etal, there is good reason to conclude that the data of Ito \etal 
do indeed represent the properties of the copper oxygen plane. 

%26

Underdoped Hg 1223 (Fukuoka \etal (1997)) shows very similar deviations 
from the $T$--linear law. The optimally doped sample, the one with the maximum 
$T_c$,   has a deviation from the linear relationship at $T^*\approx 
200 $ K. For the most underdoped sample  $T^* = 280 $ K. 

Another remarkable property of the normal state in high 
temperature superconductors is the temperature dependent Hall 
coefficient.  In a normal metal the Hall coefficient 
$R_H=E_y/J_xB=-1/ne$ is a constant since it only depends on 
the carrier concentration, which in a metal is temperature 
independent. Hwang \etal (1994) studied the in--plane Hall coefficient as 
a function of doping in \LaSr and found that this anomalous 
temperature dependence is confined to the low temperature, low 
doping region of the phase diagram where the NMR spin gap 
exists. Fig. \ref{Batlogg4} shows (solid squares) the 
temperatures $T^*$ below which the Hall coefficient is 
temperature dependent. Above $T^*$ the Hall coefficient $R_H$ 
approaches a temperature independent constant value. Similar 
effects in the Hall angle at $T^*$ were seen in YBCO 124 by 
Bucher \etal (1993).   

\subsection{~~~The ab--plane optical conductivity}

We now turn to the evidence for the pseudogap from an analysis of the 
optical conductivity. Much of this work has been reviewed by Puchkov \etal 
(1996a) and Basov and Timusk (1998). Here we will briefly summarize this 
work and  update it with recent results. 

%27 

The optical conductivity \sig\ offers several advantages over the dc 
resistivity which is the zero--frequency limit of $1/\sigma(\omega,T)$. By 
extending the measurements to finite frequencies one can separate the 
factors $\Omega_p^2$ and $\tau$ in the Drude formula and investigate their 
frequency dependence. Experimental problems with electrical contacts are 
eliminated and accurate absolute values are easily obtained without the need 
for precision measurements of the geometries of submillimeter size samples. 
However, when the crystals lack natural shiny growth faces they have to be 
polished which introduces the possibility of damage. 

It was clear from the earliest work on the optical properties of cuprates 
that their infrared reflectance spectra did not resemble those of simple 
metals which are basically flat from dc to the plasma frequency.  Instead, 
that there was a knee--like onset of strong absorption in the 500 \cm-1 region 
which was present in the normal state spectra in all the cuprates. Some of 
this work is reviewed by Puchkov \etal (1996a) and by Basov and Timusk 
(1998).  When the NMR measurements began to be interpreted in terms of a 
spin gap it was also conjectured that this feature was associated with the 
spin gap (Orenstein \etal (1990), Rotter \etal (1991)) This suggestion was 
based on the observation that the 500 \cm-1 structure was seen in the 
temperature and doping range that matched   $T^*$, the cross--over 
temperature where the spin gap appeared. 

Most of the early analysis of the optical properties was focussed on a study of 
the optical conductivity \sig\  with an aim to find a conductivity gap such as 
the one seen in dirty limit BCS superconductors (Mattis and Bardeen (1961)).
However, in a coherent system dominated by inelastic scattering, one does 
not expect to see a gap in the conductivity if there is a pseudogap in the 
density of states. In this clean limit scenario the Drude peak narrows and 
there will be shift of spectral weight from the Drude peak to an incoherent 
sideband at the frequency of the excitations. These processes result in 
changes to the simple Drude lines shape and are best described by the 
extended Drude model, first used to discuss single crystal high--$T_c$ 
spectra by Thomas \etal (1988). 

\begin{figure}[!ht]
\epsfig{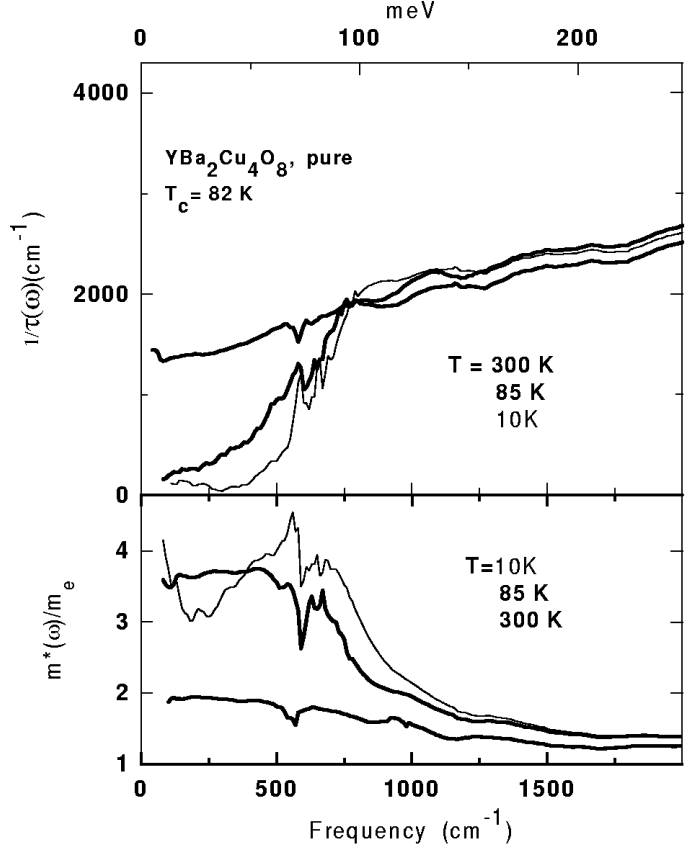}{10cm}
\caption{
The frequency dependent scattering rate and the effective mass of \YBa124. 
The scattering rate varies linearly at room temperature but develops a gap 
like depression in the normal state. At the same time the effective mass of 
the carriers develops a resonance peak at $\approx$ 600 \cm-1 (75 meV).
}\label{Puchkov12}
\bigskip
\end{figure}

Basov \etal (1996) pointed out that the scattering rate $1/\tau$, as 
determined from the extended Drude analysis of reflectance spectra, 
provided clear evidence of the pseudogap in the transport properties 
of underdoped YBCO 6.6 and the naturally underdoped double chain material 
YBCO 124 (Basov \etal (1996)).  Fig. \ref{Puchkov12} shows the scattering 
rate in YBCO 124 as a function of frequency at three temperatures. At 300 K, 
well above $T^*$, the scattering rate has a characteristic linear 
frequency dependence with a rather large zero frequency intercept. At 85 K, 
still in the normal state, a gap like depression in the scattering is seen 
below 750 \cm-1. At 10 K, in the superconducting state, there is a further 
sharpening of the gap--like spectrum. The effective mass, shown in the lower 
panel of Fig. \ref{Puchkov12}, develops a resonance in the superconducting 
state at 550 \cm-1. 

%28

These observations of a pseudogap in the scattering rate of YBCO 124 in  the 
normal state at 85 K are in agreement with other evidence for a pseudogap in 
this material. Bucher \etal (1993) find a break in the dc resistivity at 160 
K  and Zimmermann \etal (1989,1990) find a spin gap by NMR opening at $\approx 160$K in 
polycrystalline c--axis oriented YBCO 124. 
\comment{Tom - which Zimmermann ref is this, I would like to make it consistent
with the NMR section, Bryan see new references to Zimmeerman}

\begin{figure}
\epsfig{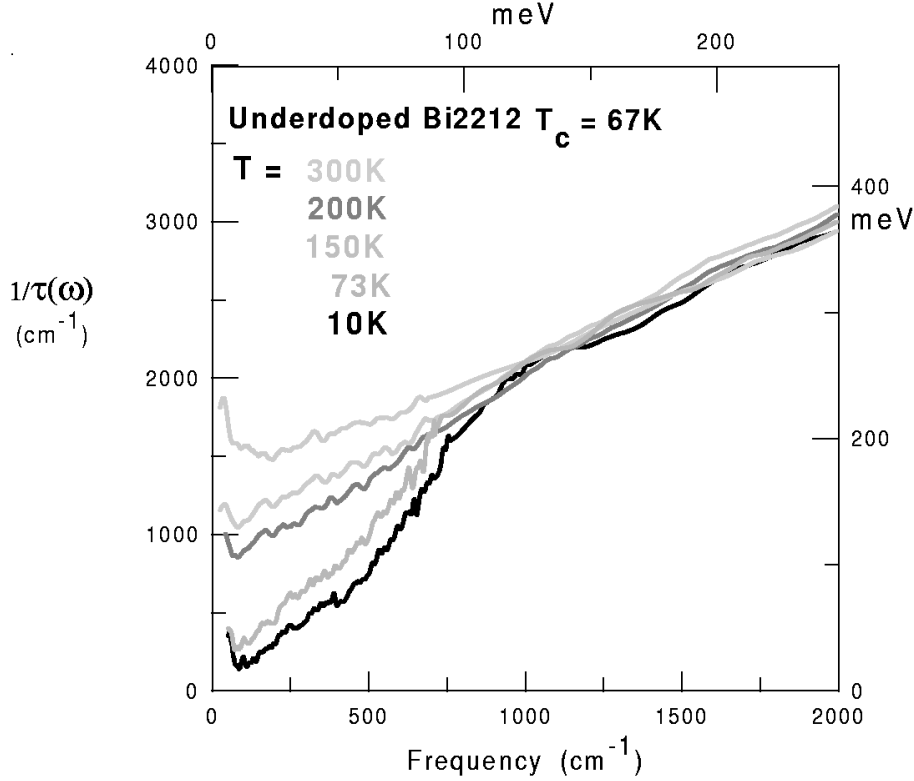}{12.5cm}
\caption{
The frequency dependent scattering rate and the effective mass of 
underdoped Bi 2212. 
The linear scattering rate develops a gap below 150 K. The frequency scale 
of this gap is only slightly enhanced in the superconducting state. There is 
no temperature dependence of the scattering at high frequency.
}\label{puchkov_bi_under_doped}
\bigskip
\end{figure}

The pseudogap in Bi2212 has received the most attention through the 
extensive angle resolved photoemission work discussed in detail in 
section II.  Puchkov \etal (1996c) reported on \sig\ in crystals 
from the same source as those used in the some of the photoemission work. 
They found strong evidence for a  pseudogap in several underdoped samples of 
this material as well as in some of the slightly overdoped ones. 
Fig. \ref{puchkov_bi_under_doped} shows the frequency dependent scattering 
rate at a series of temperatures.  At high frequencies 
$\omega>700$ \cm-1 the frequency dependence is approximately linear but 
there is no temperature dependence. At low frequencies for $\omega<700$ 
\cm-1 there is a pronounced temperature dependence in the form of a 
gaplike depression that starts below 150 K, well above the superconducting 
transition temperature of 67 K. 

\begin{figure}[!t]
\epsfig{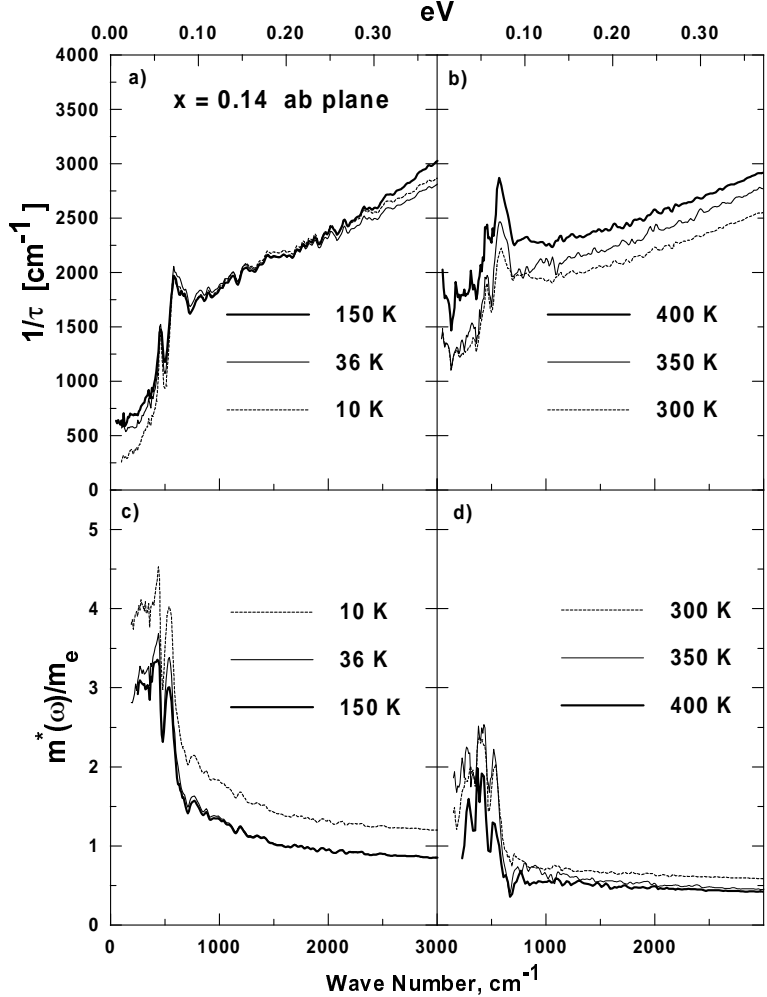}{12cm}
\caption{
Frequency dependent scattering rate in underdoped \LaSr\  (top panels) and 
effective mass (bottom panels). There is a marked depression of scattering 
below a frequency scale 700 \cm-1 (85 meV) in the normal state and 
temperature independent scattering at higher frequencies. The high frequency 
scattering becomes temperature dependent at $T>300$ K and the pseudogap 
vanishes at $\approx 450$ K.
}\label{startseva1}
\smallskip
\end{figure}

Recently Startseva \etal (1998a,b) have shown that a pseudogap can also be 
seen below a temperature $T^*$ which is well above room temperature in 
several underdoped and slightly overdoped samples of LSCO. The gap is 
defined as a region of depressed scattering below the high frequency 
linear behavior. As in the case of the two plane materials 
$1/\tau(\omega)$ is temperature independent in the pseudogap temperature 
region {\it i.e}. $T<T^*$. Startseva \etal find that above $T^*$ a 
temperature dependence {\it does} appear at all frequencies and the linear 
curves move parallel to one another as the temperature is raised. This is 
illustrated in Fig. \ref{startseva1}. 

\begin{figure}[!t]
\epsfigrot{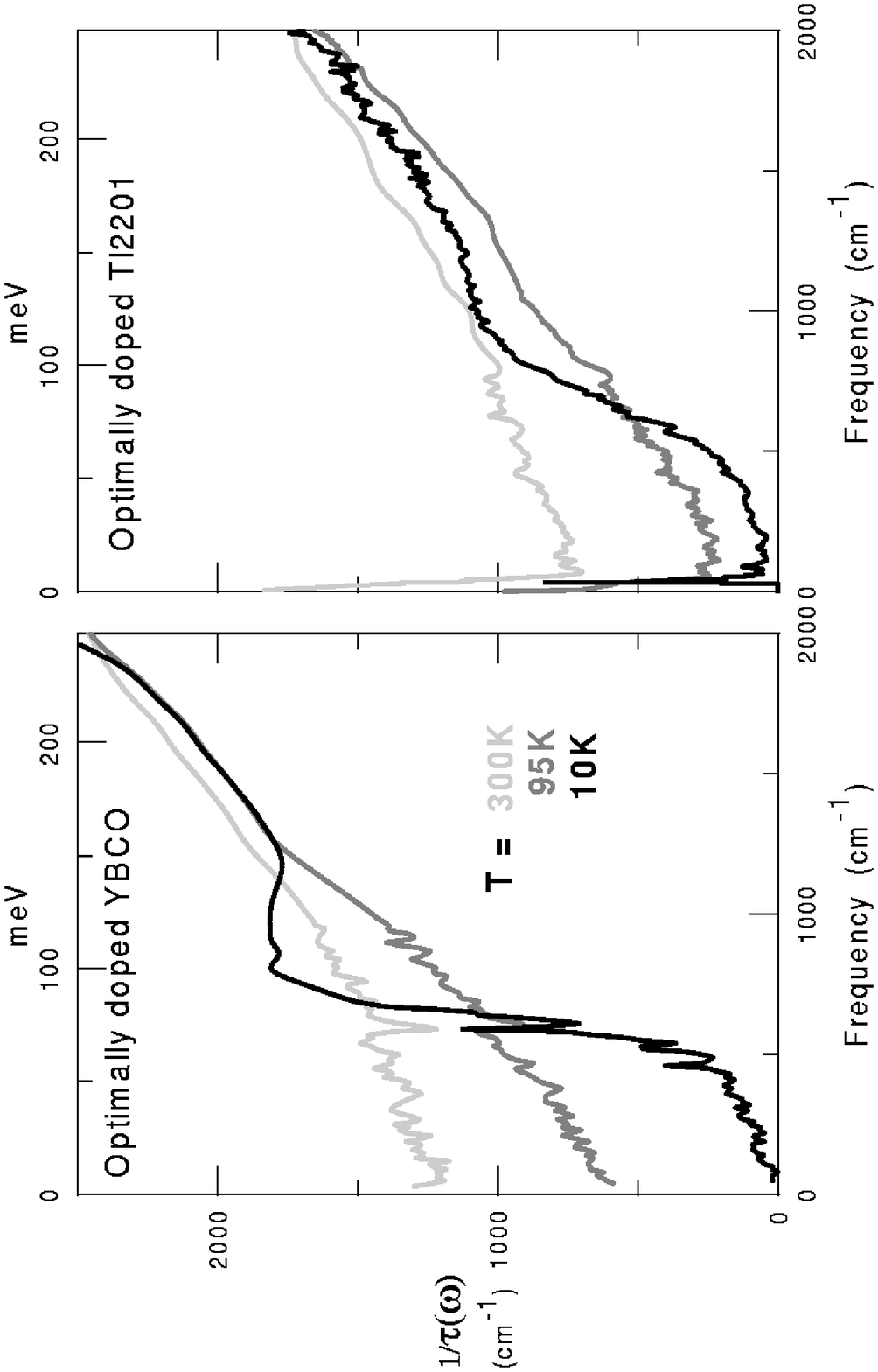}{13cm}
%\caption{Scatttering rate in optimally doped YBCO (left panel) 
%and underdoped YBCO (right panel). The scattering rate spectrum 
%is similar in the superconducting state of the optimally doped 
%and the pseudogap state of the underdoped sample (at 65 K).
\caption{Scattering rate in optimally doped YBCO (left panel) 
and Tl 2201 (right panel). There is no evidence of a pseudogap
in the normal state.
}\label{puchkov_9}
\bigskip
\end{figure}

The scattering rates of the optimally doped YBCO and Tl 2201 do not show 
any evidence of a pseudogap in the normal state. However, below the 
superconducting transition temperature a gap like depression appears 
which is very similar to what is seen in the normal state in the underdoped 
materials. There are some differences however between the spectra at optimal 
doping below $T_c$ and the spectra of  underdoped samples below $T^*$.  
First there is an overshoot in the $1/\tau$ spectra as shown in Fig. 
\ref{puchkov_9}. Secondly there is a slight reduction of slope as the 
temperature is raised above $T_c$. 

One should note that the YBCO 123 and Tl 2201 spectra in Fig. 
\ref{puchkov_9} are very similar. Yet they refer to HTSC materials that are 
very different in many ways. YBCO 123 is a two--plane material with a 
copper--oxygen chain layer while Tl 2201 is single plane material with 
thallium oxide layers. The almost identical scattering rate spectra suggest 
that the properties displayed here are those of the copper oxygen plane 
without significant contamination from interplane interactions or other 
constituents in the unit cell such as the charge  transfer layers. However, 
there is an overall change in the magnitude of the scattering rates: in the 
one plane material the scattering is a factor of two weaker at all 
temperatures and frequencies. In contrast, the Bi 2212 system shows vestiges 
of the pseudogap at optimal doping in the form of a clear break in the 
scattering at 90 K in the normal state.

\begin{figure}[t]
\epsfigrot{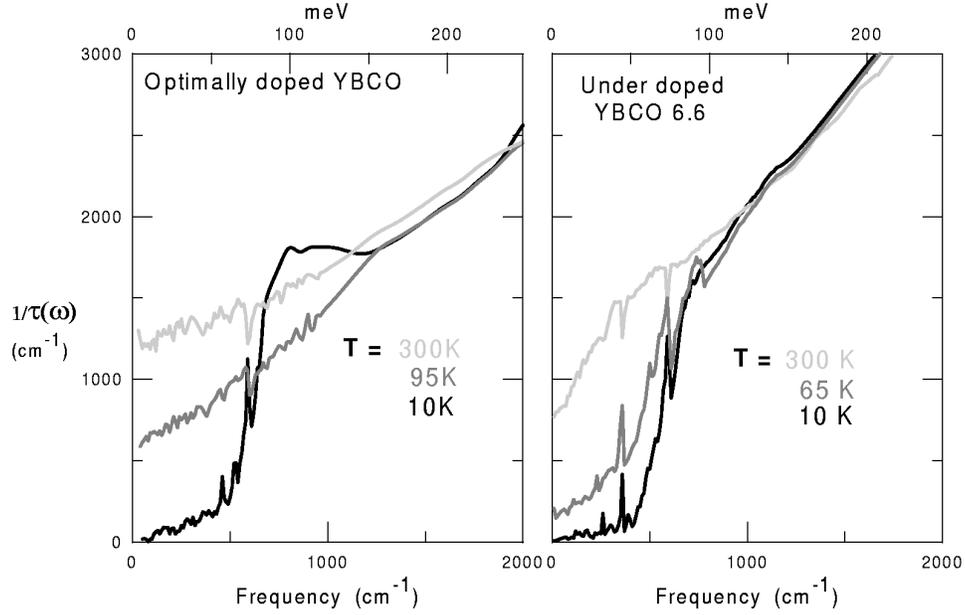}{13cm}
\caption{
Comparison of the pseudogap in the ab--plane scattering rate in optimally 
doped YBCO (left panel) and underdoped YBCO (right panel). As in the two 
chain material the pseudogap and the superconducting gap have  very similar 
frequency dependences. The middle curve in each panel is taken at a 
temperature just above the superconducting transition.
}\label{puchkov7}
\end{figure}

Finally, we address the question of the relationship between the pseudogap 
seen in the underdoped materials in the normal state well above $T_c$ and 
the superconducting gap that seems to appear at $T_c$ in the optimally doped 
samples.  Fig. \ref{puchkov7} shows the scattering rate spectrum of an 
underdoped YBCO 6.6 crystal and for comparison an optimally 
doped crystal of the same material YBCO 6.95. It is clear that the general 
shape of the scattering rate suppression is similar in the pseudogap state 
and in the superconducting state of the optimally doped crystal. There are 
however important differences. There is an overshoot of the scattering rate 
in the optimally doped material whereas in the underdoped sample there is 
a monotonic increase in scattering with frequency.  It is not possible to get  
such a steep decrease of $1/\tau(\omega)$ 
from any spectrum of $\alpha^2F(\Omega)$ using eqn.~\ref{Shulga} and one 
must conclude that the simplified analysis derived from eqn. \ref{tausig} is 
not appropriate in the presence of a superconducting gap. The second 
feature, seen in the optimally doped samples but not in the underdoped 
samples, is a slight temperature dependence in the slope of the scattering 
rate in the normal state.

%29

%30

%31

We will summarize the salient features of the scattering rate spectra as 
they relate to the issue of the pseudogap, neglecting for the moment the small 
differences that are seen between the different materials. First there is 
in {\it all materials} a suppression of scattering below the linear trend 
seen at high temperatures or high frequencies at all temperatures. 
Second there is the similarity in the spectra in the underdoped state to those in 
the  optimally doped materials in the superconducting state.  This leads us 
to notion that the pseudogap in the normal state and the superconducting gap 
are closely related. One should add, however, that in the superconducting 
state there is a slight increase in the over--all frequency scale of the gap of 
the order of 125 \cm-1 (16 meV). 
%\vfill\eject

Table II shows the approximate width of the pseudogap in 
$1/\tau(\omega)$ in various 
samples that have been studied (Puchkov \etal (1996a)). It seems from the 
table that the pseudogap is relatively constant at $\approx 700$ \cm-1 with 
perhaps a slight tendency to increase slightly with doping.

%Table {pgap}

\begin{table}[!h]
\hskip2cm%
\begin{minipage}{9.5cm}
\begin{center}
\caption{Maximum pseudogap $2\Delta$ from $ab$--plane \sig.}
\smallskip
\begin{tabular}{llr}
Material&$T_c$(K)&$2\Delta$(\cm-1)\\
\hline
YBCO 6.6 &58&660\\ 
YBCO 124 &82& 770\\
Bi 2212 &67& 660\\
Bi 2212 &82& 660\\
Bi 2212 &90& 740\\
Bi 2212 &82  (OD)& 810\\
LSCO  x=0.13&32     &$\approx$700\\      
\end{tabular} 
\end{center}
\end{minipage}\hfill
\label{table2}
\end{table}

\begin{figure}[!h]
\epsfig{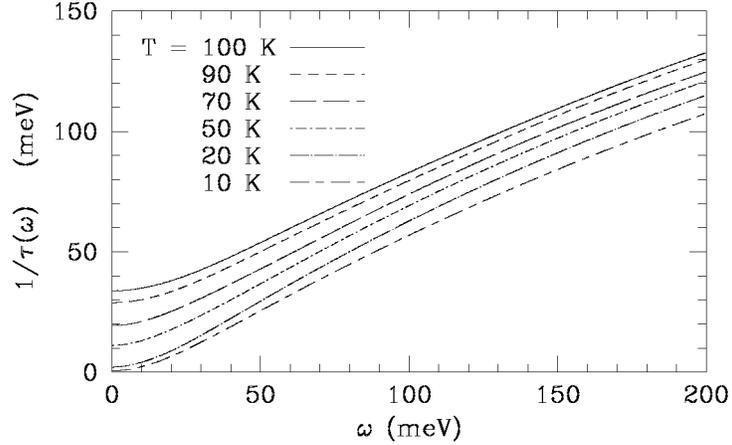}{10cm}
\caption{
Calculated scattering rate based on the nearly antiferromagnetic Fermi 
liquid model. 
}\label{Branch13b}
\bigskip
\end{figure}

Model calculations of the pseudogap in the ab--plane scattering rate have been 
done by Stojkovi\'c and Pines (1997) and by Branch (1998). Based 
on the nearly antiferromagnetic Fermi liquid model of spin fluctuation 
scattering these authors use formulas similar to eqn.~\ref{ziman}  
to 
calculate an accurately weighted conductivity as a function of frequency and 
then use eqn.~\ref{tausig} to calculate an ``effective'' scattering rate in the 
same way the experiments are analyzed. An example of such a calculation is 
shown in Fig.~\ref{Branch13b} from the work of Branch and 
Carbotte (1998). A calculation of $1/\tau$ for the $t-J$ model by Plakida 
(1997) gives similar results.
While the general trend and magnitude of the scattering rate variation is 
reproduced well by these models, there is no sharp break in the 
scattering as seen in the experiments in the 700 \cm-1 region. 
Also these calculations fail to reproduce the temperature 
independent scattering seen above 700 \cm-1. 

\subsection{~~~The c--axis pseudogap}

The earliest spectroscopic measurement of the pseudogap in charge 
fluctuations was the observation by Homes \etal (1993) of a region of 
depressed conductivity in the c--axis optical conductivity of underdoped YBCO 
6.6. Since most cuprate single crystals grow in the form of thin plates, 
measurements with ${\bf E}\parallel c$ are difficult because of the small areas 
of their ac faces. The earliest data on $\sigma_c$ 
showed a very low conductivity in that direction, much lower than the 
Ioffe--Rigel limit. Consistent with this low absolute value, the frequency 
dependence is flat and there is no temperature dependence. Thus the c--axis 
transport can be described as incoherent hopping (Cooper 
(1993a,1993b), Leggett(1994) and Timusk (1996)). 

%32

%33

\begin{figure}[!t]
\begin{minipage}[c]{9.5cm}
\epsfig{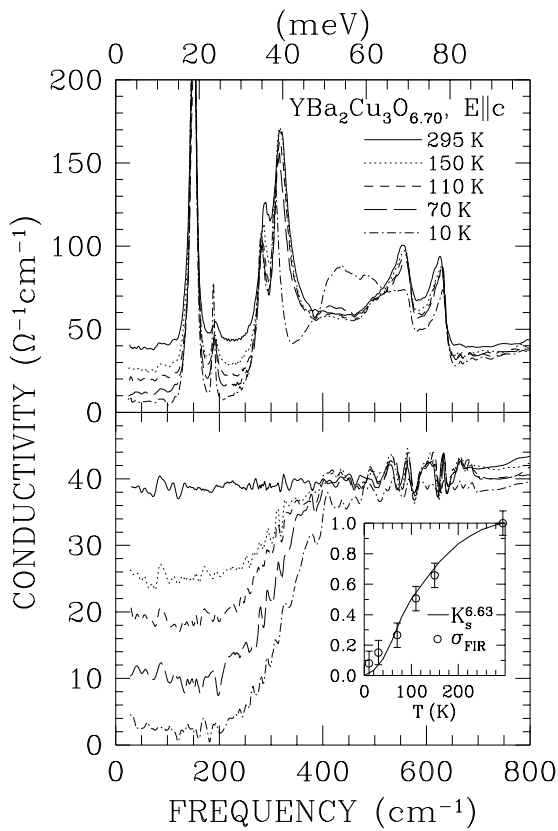}{9cm}
\end{minipage}
\hfill
\begin{minipage}[c]{7.5cm}
\caption{
Optical conductivity measured along the $c$--axis, normal to the planes for 
underdoped \YBax\  (top panel). There is marked depression of conductivity at 
low frequency. There are considerable changes in the phonon spectra between 
250 and 650 \cm-1 including the appearance of a broad peak at 420 \cm-1. 
These features have been subtracted in the lower panel which displays a 
pseudogap in the normal state. The actual magnitude of the gap frequency 
depends on assumptions made when the phonons are subtracted. The inset in 
the bottom panel compares the low frequency conductivity (open circles) with 
the Knight shift. 
}\label{homes2}
\end{minipage}
\smallskip
\end{figure}

Fig.\ref{homes2} shows the c--axis conductivity of YCBO 6.6 at a series of 
temperatures obtained from a Kramers Kronig analysis of  the reflectance 
(Homes \etal (1993)).  The top panel shows the raw conductivity which is 
dominated by several strong optic phonons. In the lower panel the phonons have 
been subtracted and the electronic background is shown more clearly. At room 
temperature the background is flat and frequency independent. As the 
temperature is lowered below $T^*$, which in this material from NMR data is 
of the order of 250 K, a pseudogap develops in the conductivity. The inset 
in the lower panel compares the Knight shift with the conductivity in the 
low frequency, pseudogap region. 

%34

The striking feature of the pseudogap as seen in the c--axis conductivity  
is its unusual temperature dependence. The gap frequency appears to be 
temperature independent and the gap fills in as the temperature increases. 
This behavior of the pseudogap is consistent with what has been seen with 
other spectroscopic probes such as ARPES, vacuum tunnelling spectroscopy 
and the ab--plane scattering rate. It should also be noted that at the lowest 
temperatures there is still a residual conductivity at low frequency and no 
true gap exists in the optical conductivity. One should also note that the 
conductivity in the c--axis pseudogap region is flat and frequency 
independent in contrast to what is seen in tunneling where the tunneling 
conductance rises uniformly in the gap region (Mandrus \etal (1993) and 
Renner \etal (1998)). 

The magnitude of the c--axis conductivity, when extrapolated to zero 
frequency, agrees with the dc value quite well (Homes \etal (1993)).  This 
suggests that there is no low lying coherent peak in the conductivity below 
the lowest frequency of the infrared measurements. Recent microwave 
measurements of \sig\  (Hosseini \etal (1998)) have confirmed this up to a 
frequency of 25 GHz. In accord with these observations Takenaka 
\etal (1994) noted that the temperature region where the
ab--plane dc resistivity showed its characteristic deviation from the 
high temperature linear behaviour coincided with the temperature 
and doping region where the c--axis resistivity became 
"semiconducting" {\it i.e.} acquires a positive temperature 
coefficient of resistivity. 

%35 
\begin{figure}[!ht]
\epsfig{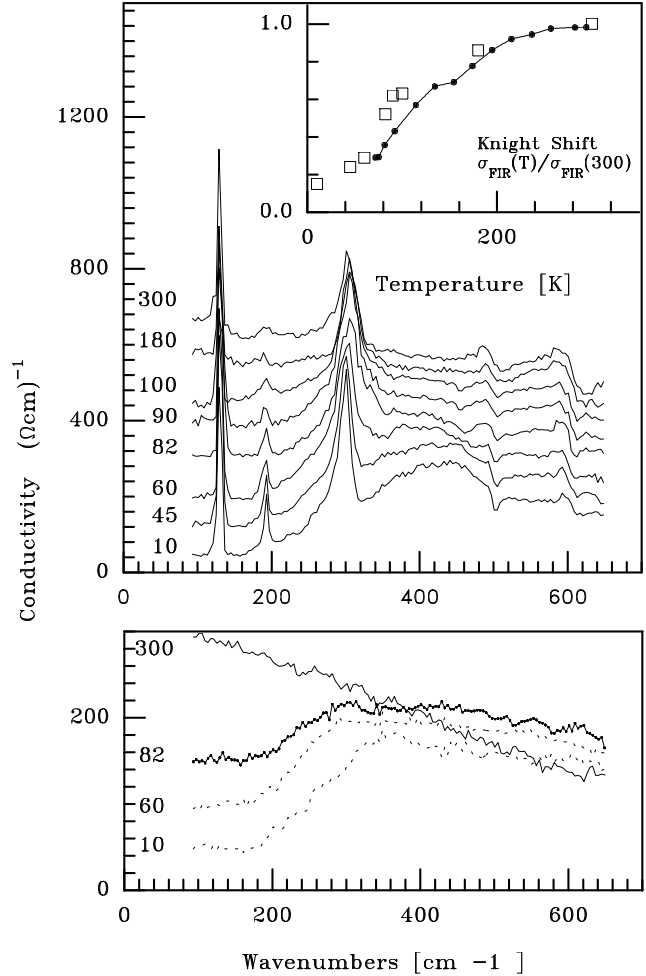}{11cm}
\caption{
Pseudogap in the $c$--axis conductivity of \YBa124. There is a marked 
depression of conductivity below 300 \cm-1 in the normal state and a further 
depression at $T_c$ extending to higher frequencies. The inset compares the 
low frequency conductivity with the Knight shift.
}\label{basov2}
\bigskip
\end{figure}

When the YBCO materials become superconducting a further reduction in the 
c--axis conductivity takes place but over a wider  spectral region than the 
normal state pseudogap scale. The total spectral weight transferred to the 
delta function is shown as the hatched box at low frequency in Fig. 
\ref{homes2}. This effect is even clearer in the YBCO 124 case, shown in 
Fig. \ref{basov2}, where there is a uniform depression in conductivity up to 
as high as 80 meV (Basov \etal (1994a)) in the superconducting state. It 
appears that at least in these two compounds the c--axis pseudogap scale is 
lower than the characteristic energies where the spectral weight of the 
superconducting condensate originates but it should be recognized  that 
because of uncertainties in the phonon subtraction process having to do with 
assumed baselines, there is great uncertainty in the energy scale of the 
c--axis pseudogap. A pseudogap in the c--axis conductivity 
has also been seen  by Tajima \etal (1995) in YBCO and  Reedyk \etal 
(1997) in underdoped Pb$_2$Sr$_2$(Y/Ca)Cu$_3$O$_8$.

The overall evidence of the presence of the pseudogap in the single layer 
\LaSr\ is controversial particularly since the NMR Knight shift reduction is 
weak and no sign of a gap is seen in the spin--lattice relaxation rate.
In the same way, the c--axis conductivity depression does not show the 
strong gap signature seen in the YBCO systems. Basov \etal 
(1995b) found 
depressed conductivity from 10 $\Omega$\cm-1 to 5 $\Omega$\cm-1 
in the normal state for a slightly underdoped crystal with x=0.15 over a 
very large frequency range of 1000 \cm-1 at the expense of increased 
conductivity in the 2000 to 4000 \cm-1 range. More recently Uchida \etal 
(1996) and Startseva \etal (1998) have found reduced normal state 
conductivity in the frequency below $\approx 500$ to 600 \cm-1 (60 to 75 
meV) in underdoped LSCO x=0.12 and x=0.13 respectively. 

Any discussion of the pseudogap phenomenon in the c--axis conductivity must 
address the anomalous broad peak at 400 \cm-1 shown in Fig. \ref{homes2} 
which appears below 150 K in YBCO. It grows in intensity as the temperature 
is lowered with no discontinuity at the superconducting transition. We will 
call it the c--axis resonance in analogy with the 41 meV resonance seen in 
neutron scattering with which it has many similarities. 

Like the neutron resonance, the c--axis resonance frequency $\Omega_0$ 
increases with doping with $\Omega_0\propto T_c$ in the underdoped range and 
it is not seen in the \LaSr\  material. The temperature dependence of the 
resonance intensity is similar as well in the underdoped case. A similar 
mode is seen in the c--axis spectrum of YBCO 124 (Basov \etal 
(1994a)) shown 
in Fig. \ref{basov2} as well as in Pb$_2$Sr$_2$RCu$_3$O$_8$ where R is a 
rare earth (Reedyk \etal (1997)). It is not seen in the c--axis spectra of 
\LaSr\ (Basov \etal (1995b)) or in Bi 2212 (Tajima \etal (1993)). In the 
former case the peak in the ($\pi,\pi$) scattering occurs at 25 meV which 
coincides with an anomalously strong phonon at this frequency and may well 
hide the infrared resonance. 

Homes \etal (1993) originally included the resonance as a part of the 
electronic conductivity but in a later publication (Homes \etal (1995b)) they 
noted that the spectral weight in the 200 \cm-1 to 700 \cm-1 region was 
constant in the temperature range where the peak grew suggesting that the 
new peak was growing at the expense of the phonons, particularly the plane 
buckling mode at   319 \cm-1. This suggested that the resonance had 
a strong phonon component. However, in view of many similarities between the 
c--axis resonance and the magnetic mode seen in neutron scattering, it is clear 
that  the mode is coupled to the electronic degrees of freedom.  More work 
is needed to establish the exact nature of these resonances. 

\section{Specific Heat}

The electronic specific heat is a fundamental property that provides 
thermodynamic evidence for a gap in the normal state of high 
temperature superconductors (Loram \etal (1994a)). The measurements are 
difficult since, at high temperatures where the normal state gap can 
be seen, the specific heat of a typical high temperature 
superconductor is dominated by phonons (Junod (1989)). Typically the 
phonon contribution is a hundred times stronger than that of the 
charge carriers. Nevertheless, using a sensitive differential 
technique, Loram \etal have succeeded in accurately measuring 
the electronic specific heat of several high temperature 
superconductors in the pseudogap temperature range. 

In the differential technique a reference sample is compared with an 
undoped  standard sample, both in the form of fine powders containing 
exactly the same number of moles of material (Loram \etal (1993)). The 
undoped reference is assumed to have zero electronic specific heat. 
If the same quantity of heat $dQ$ is injected into the two samples, 
then according to $dQ=CdT$ where $C$ is the specific heat and $dT$ 
the temperature change, the specific heat difference is $\delta 
C=C_0\delta T/dT_r$, where $dT_r$ is the temperature rise of the 
reference and $\delta T$ is the temperature difference between the 
reference and the sample due to the extra electronic specific 
heat. 

%36 SPECIFIC HEAT
\begin{figure}[!ht]
\epsfig{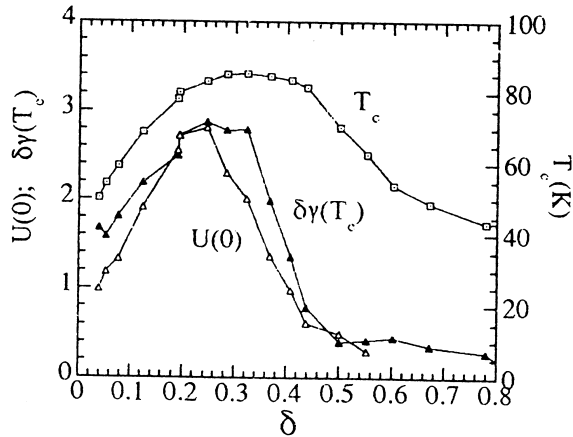}{8cm}
\caption{
$T_c$, $\delta\gamma(T_c)$ the specific heat jump (mJ/g-at.K$^2$) at $T_c$ 
and the condensation energy $U(0)$ (J/g-at.K) {\it vs.} doping level 
$\delta$ in Y$_{0.8}$Ca$_{0.2}$Ba$_2$Cu$_3$O$_{7-\delta}$. Both the 
condensation energy and the specific heat jump drop on the underdoped side 
(right hand  side of the diagram).
}\label{fgLoram1}
\bigskip
\end{figure}

A number of corrections have to be made, the largest being the change 
in the phonon spectrum due to doping which alters the phonon 
contribution of the specific heat (Loram \etal (1993)).  For example, 
in \YBax\ the additional oxygens that enter the chains not only give 
rise to  new modes involving the chain oxygens, but also 
shift the frequency of the bridging oxygen from its twofold 
coordinated copper position at 615 \cm-1 to the 560 \cm-1 of the 
fourfold coordinated site as the chain is built up (Homes \etal (1995b)). 
The corresponding changes to the phonon specific heat are, at their 
maximum at $T=40$ K, an order of magnitude larger than the electronic 
term. Loram \etal use a Zn doped sample as a reference to estimate 
these changes by noting that 7 \% Zn destroys superconductivity but 
leaves a constant, temperature independent metallic $\gamma$ term. An 
independent estimate of doping induced changes to the phonon density 
of states comes from inelastic neutron scattering. Loram \etal find 
that for an estimated uncertainty of  1--2\% in the residual change in 
the phonon contribution and error of 10 \% is propagated to the 
electronic specific heat. 

\begin{figure}
\epsfig{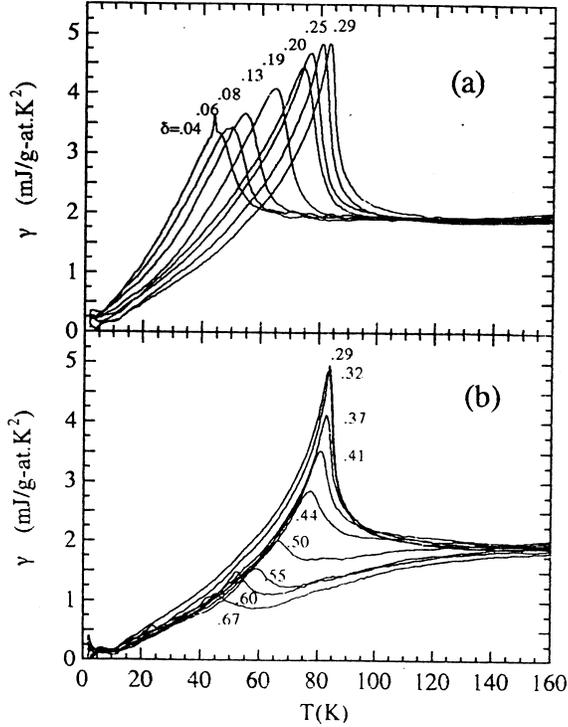}{8cm}
\caption{
Specific heat coefficient $\gamma$ for (a) overdoped and (b) underdoped 
Y$_{0.8}$Ca$_{0.2}$Ba$_2$Cu$_3$O$_{7-\delta}$. In the overdoped material 
a gap, signaled by a depression in $\gamma$, opens up below $T_c$. In the 
underdoped samples a gap starts to form in a the normal state below 140 K. 
}\label{fgLoram2}
\bigskip
\end{figure}

In a recent paper Loram \etal (1997) investigate the electronic specific 
heat of Y$_{0.8}$Ca$_{0.2}$Ba$_2$Cu$_3$O$_{7-\delta}$ as a function 
of oxygen doping. The replacement of some of the yttrium by calcium 
shifts the maximum $T_c$ from O$_{6.92}$ for 0 \% Ca to 
O$_{6.68}$ for 20 \% Ca. By varying the oxygen level from 
O$_{6.20}$ to O$_{6.96}$ the same sample can be studied from the 
underdoped region with $T_c=44$ K to the substantially overdoped region with 
$T_c=53$ K as shown in Fig. \ref{fgLoram1} from Loram \etal 
(1997). The specific heat 
coefficient $\gamma$, where $\gamma=C/T$, of both the overdoped ($\delta 
< 0.32$) and underdoped ($\delta > 0.32$) samples is shown in Fig. 
\ref{fgLoram2}. There are three regions of interest, the underdoped 
region where a pseudogap is seen, the plateau where $T_c$ 
remains roughly constant and finally the overdoped region where $T_c$ 
decreases as oxygen is added. We will first examine the overdoped 
region.  

%37

\begin{figure}[!t]
\begin{minipage}[c]{9.5cm}
\epsfig{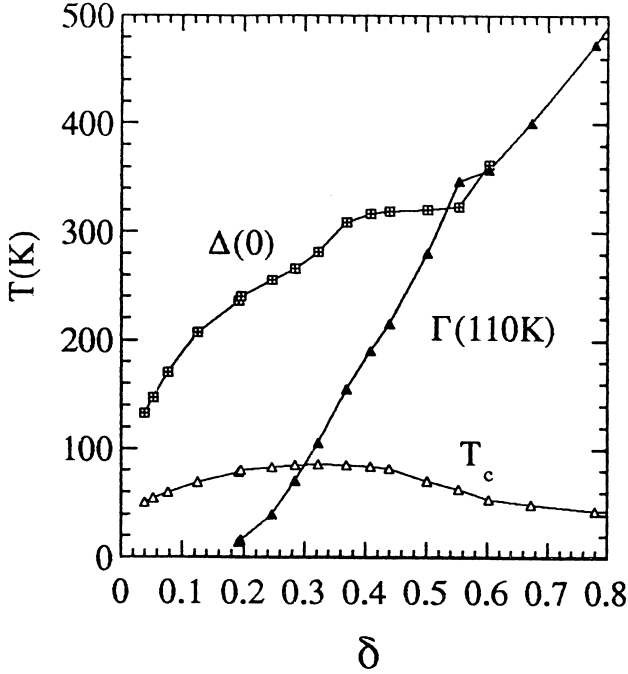}{9cm}
\end{minipage}
\hfill
\begin{minipage}[c]{7.5cm}
\caption{
Magnitude of the superconducting gap $\Delta(0)$ and the normal state gap 
$\Gamma$(110K) as a function of doping $\delta$. At optimal doping the 
normal state gap goes to zero whereas the superconducting gap has a 
magnitude of $\approx$ 300 K (25 meV).
}\label{fgLoram3}
\end{minipage}
\smallskip
\end{figure}

In the overdoped samples, Fig. \ref{fgLoram2}a, the normal state 
$\gamma$ is temperature independent. Since $\gamma$ is proportional 
to the density of states at the Fermi surface this shows that the 
overdoped material resembles a normal metal. Surprisingly $\gamma$ is 
also doping independent. However, a doping independent infrared 
spectral weight was also seen in all overdoped high temperature 
superconductors examined by Puchkov \etal (1996b). The overdoped 
samples show a specific heat jump at $T_c$, larger than the  
$\Delta\gamma/\gamma=1.43$ expected for the weak coupling BCS model 
suggesting possible strong coupling superconductivity. Evidence of 
nodes in the density of states in the overdoped state is shown by the 
approximately linear temperature dependence of $\gamma$ in the 
superconducting state at low temperatures, where $\gamma=\alpha T$. 
The authors use the slope of this linear term to estimate the d--wave 
superconducting gap.  This yields $\alpha=3.29\gamma_n k_B/\Delta_0$ 
where $\gamma_n$ is the normal state specific heat coefficient and 
$\Delta_0$ the zero temperature maximum gap. Fig. \ref{fgLoram3} 
shows the superconducting gap found this way as a function of doping.

%38

%39
\begin{figure}[!h]
\begin{minipage}[c]{9.5cm}
\epsfig{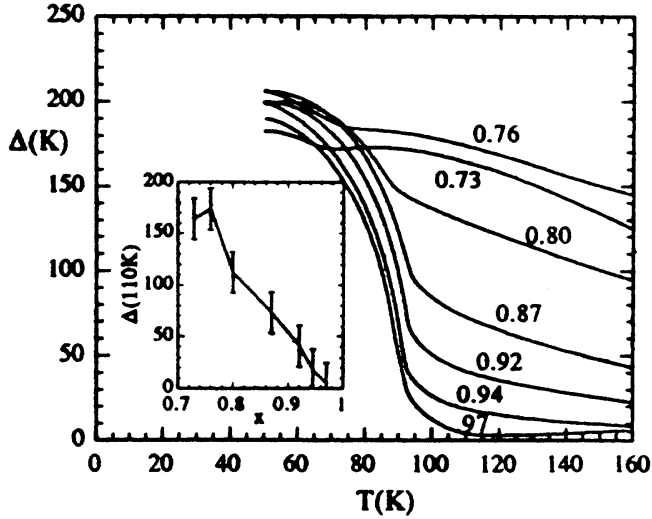}{9cm}
\end{minipage}
\hfill
\begin{minipage}[c]{7.5cm}
\caption{
The energy gap $\Delta$ for YBa$_2$Cu$_3$O$_{7-\delta}$ as a function of temperature $T$. In the 
underdoped state, with $\delta =0.76$, the gap appears well above 160 K and 
its magnitude does not change substantially on entry into the 
superconducting state.
}\label{fgLoram4}
\end{minipage}
\smallskip
\end{figure}

As we move towards the underdoped region by reducing the oxygen content 
below the maximum $T_c$ value, there is a plateau where $T_c$ does not vary 
but the condensation energy collapses rapidly, an effect first seen by Junod 
\etal (1989) in specific heat and by D\"umling \etal (1991) from reversible 
magnetization measurements. This effect can be seen clearly in the sudden 
decrease in the height of the specific heat jump $\delta\gamma$ at $T_c$ 
(the condensation energy is $\propto \delta\gamma(T_c)T_c^2$ ). Fig. 
\ref{fgLoram1} shows that between $\delta=0.34$ and $\delta=0.50$ the 
condensation energy drops rapidly while $T_c$ only drops slowly. 

In the underdoped samples Loram \etal find a pseudogap. Where the 
overdoped samples show a temperature independent $\gamma$ above $T_c$, 
both the 20 \% calcium series (Loram \etal (1997)) and the calcium 
free material (Loram \etal (1994a)) have a depression of the specific 
heat coefficient $\gamma$ in the normal state below a temperature 
$T^*$ as shown in Fig. \ref{fgLoram2} and Fig. \ref{fgLoram4}. As oxygen is 
removed the depression starts at higher temperatures.  

\begin{figure}
\epsfig{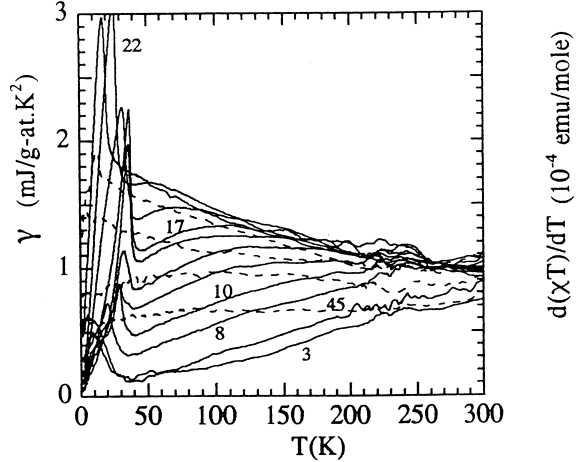}{8cm}
\caption{
Specific heat coefficient $\gamma$ of \LaSr\  for $x=0.08$ to $x=0.30$. There 
is a suppression of $\gamma$ in the normal state for $x<0.22$ signalling the 
presence of a pseudogap.
}\label{fgLoram6}
\bigskip
\end{figure}

In \LaSr\ Loram \etal (1996) also found evidence for a pseudogap in the 
normal state. As Fig. \ref{fgLoram6} shows, there is loss 
of condensation energy below $x=0.15$ and $x=0.135$ but, in 
contrast with the \YBax\  system, the pseudogap appears already in the 
overdoped state. The specific heat coefficient is depressed below 
values seen in the overdoped state already at $x=0.20$ below $\approx 
100$ K. This is accord with measurements with 
other techniques: the pseudogap temperature scale is much higher in 
\LaSr\ than in \YBax\ and the pseudogap state extends well into the overdoped 
region. In the superconducting state the specific heat 
coefficient has a linear dependence on T, $\gamma=\gamma(0) + \alpha 
T$, first observed by Momono \etal (1994).

In both systems Loram \etal also measure the bulk magnetic 
susceptibility $\chi$ of their specific heat samples. They find that 
in the normal state, there is striking agreement between the entropy 
$S$, as determined from an integration of $\gamma$ 
($S(T)=\int^T\gamma(T')dT'$)  and $\chi T$. This is expected for a 
Fermi liquid where the Wilson ratio $a_0=S/\chi T$ is a constant 
$a_0=(\pi k_B/\mu_B)^2/3$. This shows that the spin and charge 
densities of states have similar energy dependences and are equally weighted. 
In spin--charge separation models the Wilson ratio is found to deviate 
from unity.  In one--dimensional Hubbard model simulations Schultz 
(1991) finds that for non--interacting Fermions 
$v_{\sigma}=v_{\rho}=v_{F}$ where $v_{\sigma}$, $v_{\rho}$ and 
$v_{F}$ are the spin and charge and fermion velocities respectively. 
In the   limit of complete spin and charge separation the Wilson 
ratio $a=a_0/2$. For disordered local moments the Wilson ratio is 
$a=a_0/4.75$. Orbital angular momentum affects the Wilson ratio which 
for $g=2.1$ yields about a 10\% reduction. The electron phonon 
interaction enhances the Wilson ratio by $(1+\lambda)$. 

%40

Within the Fermi liquid model, inelastic interactions  enhance the mass of 
the electrons through the mass enhancement factor $\lambda$ given by 
$m^*=m(1+\lambda)$  where $m^*$ is the effective mass and $m$ the band mass. 
Loram \etal find a weak coupling value of $\lambda\approx 0.5$ for both pure 
and zinc doped YBCO (Loram \etal (1994b)). This observation is in 
contradiction with the masses deduced from the frequency dependent 
scattering rate analysis of the optical conductivity where it is found that 
$\lambda\approx 4$. However, this is a high frequency result. At low 
frequencies, at energies that are relevant to the specific heat 
measurements, a smaller $\lambda$ can be inferred; for example from the 
resistivity slope in the context of a two component model  $\lambda=0.3$ 
is obtained. For a review of this issue see Tanner and Timusk 
(1992).

\section{Electronic Raman Scattering}

In metals the Raman effect is difficult to observe. First, the Raman 
process is  an intrinsically weak second order process, much weaker than 
the dipole absorption responsible for the optical conductivity.  Second, 
in conductors electromagnetic radiation only penetrates a few thousand \AA
making the effective interaction volume small. Finally, for free carriers, 
momentum conservation dictates that only carriers up to an energy of $qv_F$ 
can be excited, $\approx$ 50 \cm-1 for the cuprates. Here $v_F$ is the Fermi 
velocity and $q$ the reciprocal skin depth. In addition to this low 
frequency band, free carriers are also expected to show Raman activity at the 
plasma frequency. Such a band has been observed in doped semiconductors 
(Mooradian \etal (1960)). 

In the cuprates the normal state Raman spectrum is dominated by phonons 
which are superimposed on a broad frequency independent continuum extending 
up to 2 eV. The origin of this continuum has been a mystery but  it is 
now clear that it arises from the same incoherent processes that are 
responsible for the strong scattering seen in other probes of the electronic 
excitations (Varma \etal (1989), Shastry and Schraimann (1990), Kostur and 
Eliashberg (1991), Branch (1996)). 

In conventional superconductors excitations across the energy gap are Raman 
active. Predicted by Abrikosov and Fal'kovskii (1961), these transitions 
which break Cooper pairs into pairs of quasiparticles at $\mathbold k$ and 
$\mathbold -k$, thus conserving momentum, result in a threshold in 
scattering at $2\Delta$. They were first observed by Sooryakumar and Klein 
in (1980) in 2H--NbSe$_2$ and in V$_3$Si by Klein and Dierker 
(1984) and Hackl 
\etal (1983). The Raman response of superconductors should be contrasted to 
the infrared response. For the infrared conductivity, type II coherence 
factors cancel out the pairbreaking excitations, and there is no absorption 
at $2\Delta$ in a clean superconductor. For the Raman response (being 
proportional to ${\mathbold A\dot A}$ rather than ${\mathbold p\dot A}$ of the 
infrared) type I coherence factors apply and even in a clean s--wave 
superconductor there is a distinct feature at $2\Delta$. 

It was found early on that in the cuprates there was no clear onset of 
the expected scattering at $2\Delta$. Instead, in the superconducting state, 
there is a redistribution of spectral weight where the broad continuum 
scattering is depressed at low frequencies and the spectral weight lost is 
transferred to higher frequencies, forming a broad peak whose position 
depends on scattering geometry. This peak, often referred to as the 
pairbreaking peak in the high--$T_c$ Raman literature, is not present in all 
experiments. Furthermore, there is no sharp $2\Delta$ threshold. Instead, 
there is scattering intensity to the lowest frequencies. As in the case of 
the optical conductivity some of these deviations from the expected BCS 
behaviour can be understood in terms of a d--wave order 
parameter and inelastic scattering that  increases as a function of 
frequency (Branch (1996)). 

\begin{figure}[!hb]
\epsfig{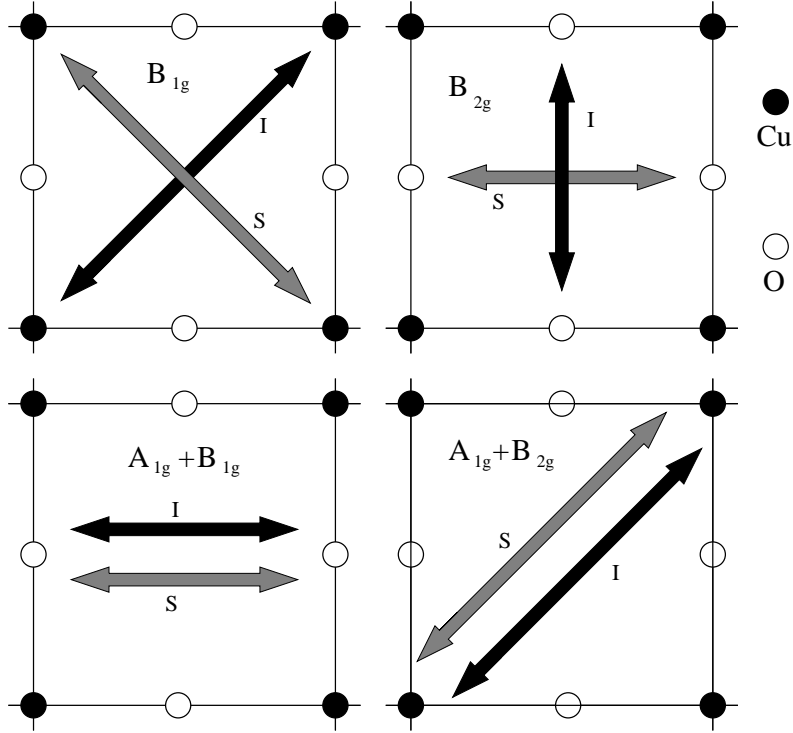}{10.5cm}
\caption{
Relation between Raman symmetries and photon polarization. Incident (I) and 
scattered (S) polarizations are shown. With crossed polarizers B$_{2g}$ is 
measured along the CuO bond axes whereas B$_{1g}$ is measured when the 
polarizers are rotated by 45 $^{\circ}$ to the bond directions.
}\label{Branch4.1}
\end{figure}

Unlike the optical conductivity which, for a tetragonal system, yields 
a single constant for the ab--plane conductivity, Raman scattering can be 
used to give two spectra,  B$_{1g}$  and B$_{2g}$, each 
representing a different average over the Fermi surface. The 
A$_{1g}$ spectrum cannot be obtained independently. The 
different geometries are illustrated in Fig. \ref{Branch4.1} 
from Branch (1996). It can be shown (Chen \etal (1994)) that 
the B$_{1g}$  spectra emphasize processes that involve states in 
the ($\pi, 0$) direction of the maximum gap, whereas the 
B$_{2g}$ spectra involve states near the nodes of the 
$d_{x^2-y^2}$ function at ($\pi$, $\pi$). 

%41 RAMAN 

\begin{figure}[!hb]
\epsfig{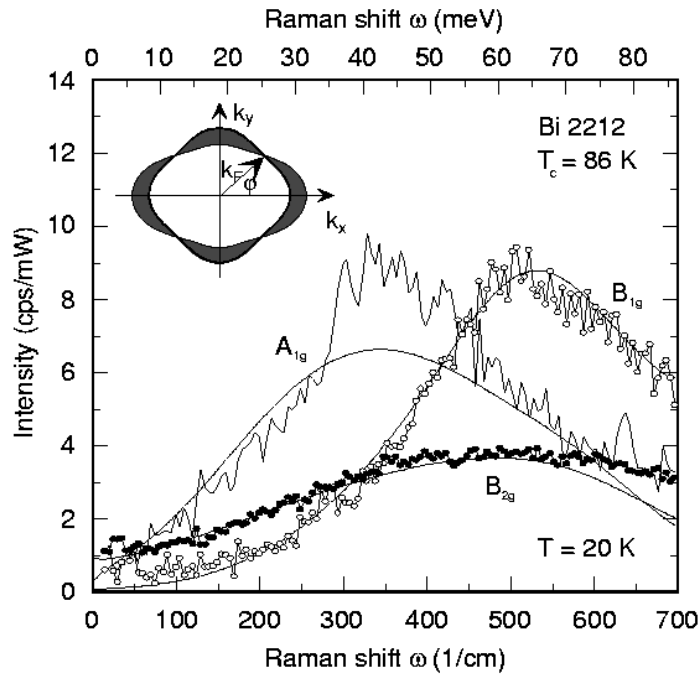}{10cm}
\caption{
Raman spectra of Bi 2212. B$_{1g}$ emphasizes processes in the ($\pi,0$) 
direction whereas B$_{2g}$ is sensitive to the ($\pi,\pi$) direction. The 
solid lines are the result of a $d$--wave model for the Raman tensor shown in 
the inset.
}\label{Hackl6}
\bigskip
\end{figure}

Recent measurements by Hackl \etal (1996) for the Raman intensity in the 
superconducting state for various symmetries in optimally doped Bi 2212 are 
shown in Fig. \ref{Hackl6} from Einzel and Hackl (1996). As this 
figure illustrates, the Raman 
spectra of B$_{1g}$  and B$_{2g}$  symmetry  are quite different. The solid 
curves are theoretical predictions that assume d--wave ($d_{x^2-y^2}$) 
symmetry of the gap function and a maximum gap of $\Delta_0=280$ \cm-1 (34.7 
meV). Kendziora \etal (1996), using similar analysis find that $\Delta_0=264$ 
\cm-1 (32.7 meV). 

The earliest evidence for a pseudogap in the Raman spectra came from the 
observations of Slakey \etal (1990) in underdoped YBCO 123 with $T_c=60$ K. 
They noted that a peak at 500 \cm-1, seen at low temperature in B$_{1g}$ 
spectra, and associated with superconductivity, could be seen well into the 
normal state. They also noted that the frequency of the feature did not 
depend on the carrier concentration, in clear contradiction to the 
expectation from BCS theory that the $2\Delta/k_BT_{c}$ ratio would remain 
constant or even decrease as one moved into the underdoped state. The 
authors suggest that if the 500 \cm-1 energy is a pairing energy then their 
experiments could be understood in terms of models of preformed pairs 
advanced by Randeria \etal (1989) and by Friedberg and Lee (1989). They also 
pointed out that this pairing energy did not show any significant 
temperature dependence below $T_c$.                       

These early observations have been confirmed by recent work. A loss in 
spectral weight below 700 \cm-1 in underdoped Bi 2212 and YBCO 123 has been 
reported by Nemetschek \etal (1997) in the B$_{2g}$ spectrum but not in the 
B$_{1g}$. The overall low frequency spectral weight is reduced by some 10 \% 
starting at 200 K in both materials.

The power law behaviour of the Raman continuum below the gap frequency has 
been the subject of much recent discussion. It is generally found that in 
the normal state the Raman continuum, after corrections for Bose occupation 
factors, varies linearly with frequency.  This can be understood in terms of 
the collision limited or ``Drude'' model of Raman response for a conducting 
system (Zawadowski and Cardona (1990), Hackl \etal  (1996)) where the Raman 
cross--section $\sigma_s$ for a channel $s$ ($s=$ A$_{1g}$, B$_{1g}$, B$_{2g}$) 
is given by: 

\begin{equation} 
{\partial^2\sigma_s \over \partial\omega\partial\Omega} = (1+n_\omega) 
{\omega\Gamma_sB_s \over \omega^2+\Gamma^2_s}
\end{equation}
where $n_\omega=[exp(\hbar\omega/k_BT)-1]^{-1}$ is the Bose--Einstein 
factor and $\Gamma_s$ the electronic scattering rate. The quantity $B_s$ is a 
Raman response, proportional to the reciprocal of the band mass, averaged 
over the Fermi surface with a weight that depends on symmetry $s$ 
(Deveraux and Einzel (1995)). The scattering rate $\Gamma_s$ is also a 
weighted, symmetry dependent, average over the Fermi surface. 

%42 

%43
\begin{figure}[!t]
\epsfig{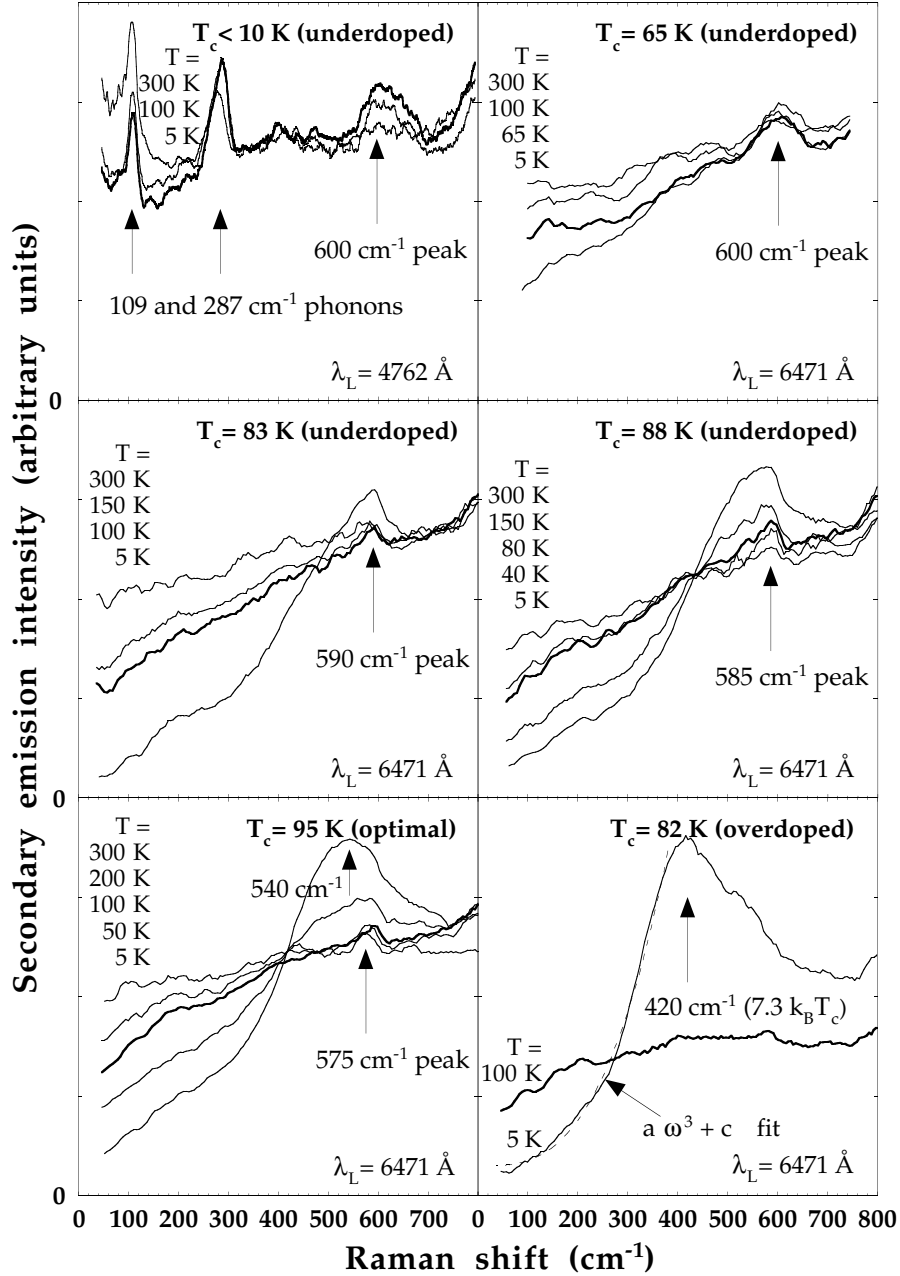}{12cm}
\caption{
Raman intensity for B$_{1g}$ symmetry in Bi 2212 at different doping levels. 
There is a suppression of spectral weight below a frequency of $\approx$ 600 
\cm-1 and a sharp resonance at this frequency. The suppression of scattering 
has been attributed to the formation of a pseudogap in the normal state.
}\label{fgBlum1}
\bigskip
\end{figure}

In the normal state this collision limited model predicts a slope of the 
Raman spectrum that is a direct measure of the magnitude of $\Gamma_s$ the 
scattering rate for the symmetry $s$. Thus if a pseudogap develops at 
($\pi,0$), the slope of the Raman response at low frequencies will be 
proportional to $1/\Gamma_{{\rm B}_{1g}}$. Experimental results in optimally 
doped materials show that the scattering rate determined this way has a 
linear temperature dependence with a zero intercept very similar to what is 
observed in the optical conductivity (Hackl \etal (1996), Naeini \etal 
(1997)). The absolute magnitude of the renormalized Raman scattering rate for 
optimally doped Bi 2212 at 300 K is 450 \cm-1 (Hackl \etal (1996)) in 
reasonable agreement with the infrared conductivity value of 600 \cm-1 
(Puchkov \etal (1996a)). One does not expect the infrared and Raman 
scattering rates to be identical in the presence of anisotropy since they 
result from different averages around  Fermi surface. The comparison between 
Raman and conductivity scattering rates is discussed in detail by Branch 
(1996).

Raman scattering data at various doping levels of \BiSr\ is presented by 
Blumberg \etal (1998). The authors argue that the B$_{1g}$ Raman continuum is 
a measure of both the gapped density of states and the low 
frequency dependent scattering rate.  Fig. \ref{fgBlum1} shows the Raman 
continuum for a series of doping levels for \BiSr samples.  
There is prominent peak in B$_{1g}$ Raman spectra, at 600 \cm-1 in the 
overdoped state.  Blumberg \etal argue that this narrow peak is evidence of a 
bound state between the carriers, such as preformed pairs. 

\section{Magnetic Neutron Scattering}

Magnetic neutron scattering is a powerful spectroscopic tool that provides 
information on both the energy and momentum dependence of excitations that 
involve the change of the spin of an electron. In insulators elastic 
magnetic scattering has been used to map out the static spin structure and 
inelastic scattering to excite spin waves. In metals inelastic scattering is similar  
to Raman and optical spectroscopy in that the final result of the scatting 
process is a hole below the Fermi level and an excited electron above it. 
But unlike optical spectroscopies, where the momentum transfer to the 
electron--hole system is nearly zero, a neutron scattering experiment can be 
set up to transfer any momentum ${\bf Q}$ to the electron--hole pair. Since 
the neutron interacts with the magnetic moment of the electron, the electron 
spin can be flipped in the course of the scattering.

The drawback of this technique is that due to the weak interaction 
between the neutron and the electron spin, very large, centimeter size, 
crystals must be used.  Because of this limitation only two high--$T_c$ 
systems have been investigated in detail by magnetic neutron scattering: 
\LaSr\ and \YBax.

\begin{figure}[t]
\epsfigrot{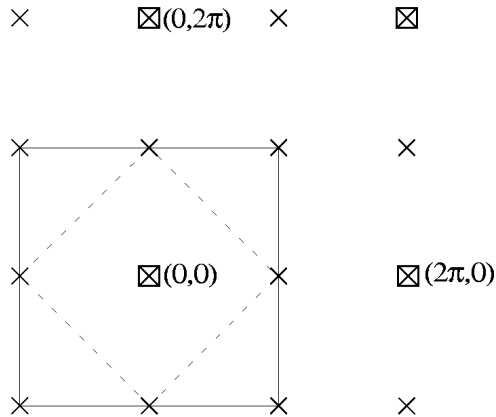}{7cm}
\caption{
Magnetic Brillouin zone shown as dashed lines. The squares denote the 
reciprocal lattice points of the  atomic unit cell. With the onset 
of antiferromagnetic order the unit cell in doubled in both directions and 
the new reciprocal lattice, shown as crosses, is formed. 
}\label{fig1neut_BZ}
\bigskip
\end{figure}

The Bragg diffraction pattern of an antiferromagnetic insulator with a 
square lattice and a lattice spacing $a$ is illustrated in Fig. 
\ref{fig1neut_BZ}. Nuclear scattering occurs at momentum transfers of 
$2\pi/a (h,k,l)$ where $h,k,l$ are integers. Because the alternating spins 
double the unit cell, in a 3D antiferromagnet scattering takes place at 
$2\pi/a(h/2,k/2,l/2)$. (In the neutron scattering literature momentum 
transfer is measured in reciprocal lattice units with $(Q_x,Q_y,Q_z)=(2\pi/a 
h,2\pi/b k,2\pi/c l)$, with $ a \approx b = 1.63$ \AA$^{-1}$ and 
$2\pi/c\approx 0.53$ \AA$^{-1}$ for \YBax). Since the doped cuprates are 
two--dimensional antiferromagnetic insulators, with no coherence between the 2D 
layers,  their Bragg pattern will consist of ``rods'' of magnetic scattering 
located at $(h/2,k/2,l)$ where $h$ and $k$ are integers but $l$ is a 
continuous variable and the scattering intensity is independent 
of $l$. In the limit of zero doping 3D magnetic order is established and 
the rods turn into Bragg  spots at $l/2$

%44

\begin{figure}
\epsfig{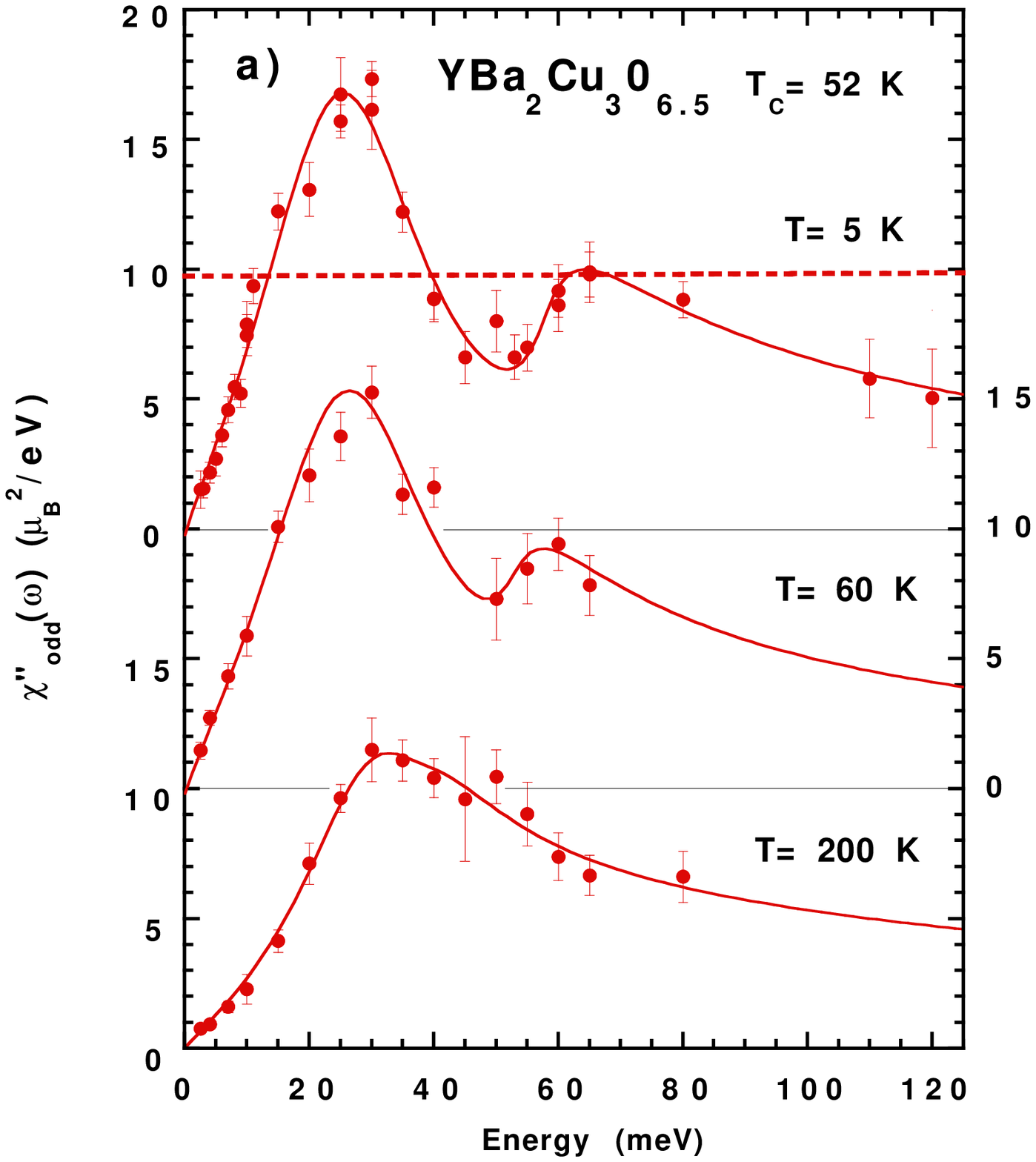}{11.5cm}
\epsfig{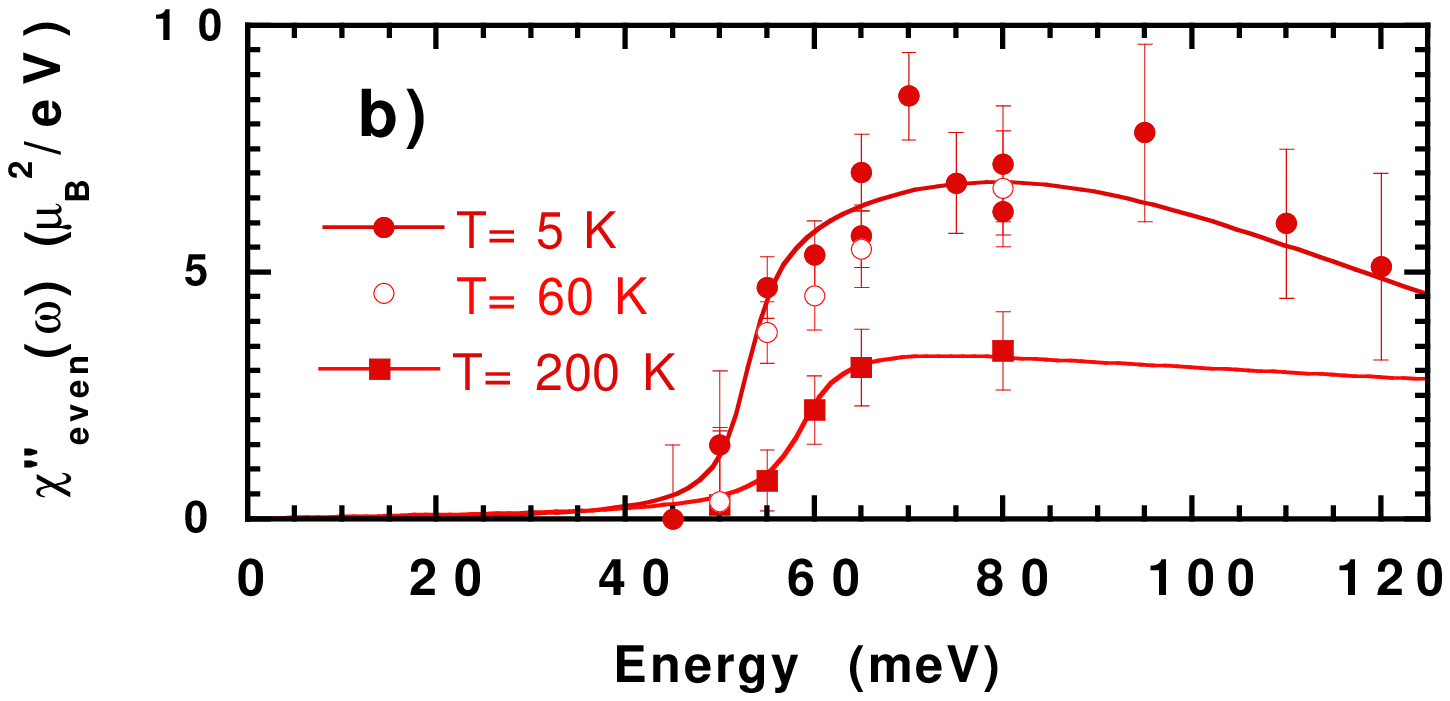}{11.5cm}
\caption{
Magnetic neutron scattering in underdoped YBCO. a) The susceptibility in the 
odd channel where the spins rotate out of phase in the bilayers is  plotted.
The scattering is dominated by a peak at $\approx$ 30 meV. b) Even channel 
magnetic scattering. The even channel is dominated by a gap at 
$\approx$  50 meV.
}\label{bourges3}
\bigskip
\end{figure}

In doped \YBax\ there is a coupling between the spin systems of the two 
CuO$_2$ layers in the unit cell. As a result the rods acquire a sinusoidal 
modulation with a spatial frequency corresponding to the interlayer spacing 
of 3.227 \AA\ and not the $c$--axis spacing $c$=11.83 \AA. The first maximum of 
intensity occurs at {\bf Q}=$(1/2,1/2,1.83)$. Spectra at $l=1.83$ are called odd or 
acoustic since at this $l$ value spins in the two bilayers rotate out of 
phase. Even or ``optic'' spectra can be observed at $(1/2,1/2,0)$. Most of the 
work to date has been done on the odd spectra since the even scattering is 
absent in the normal thermal neutron range. Recent experiments 
show that in the metallic regime 
even scattering has a sharp threshold at 50 meV (Bourges \etal (1997), 
Hayden \etal (1997)) as shown in  Figs. \ref{bourges3} and 
\ref{hayden7}. In the insulating regime there is an optic gap at 
$\approx 70$ meV (Reznik \etal (1996)).

The early work on magnetic neutron scattering in \YBax\ has been reviewed by 
Rossat--Mignod (1993). The undoped material is a two dimensional 
antiferromagnet with the copper spins pointing along the CuO$_2$ planes. 
The N\'eel temperature is 415 K and the in--plane exchange constant 
$J=0.170$ eV. With doping  $T_N$ decreases and drops to 
zero sharply at $x$=0.4 and long range order can no longer be seen at 
$x=0.42$. At  the same time the 3D ordered magnetic moment decreases to 
reach zero abruptly at $x=0.4 \pm 0.02$. The inelastic magnetic 
scattering is mainly confined to  the vicinity of the antiferromagnetic 
point $(1/2,1/2)$ but there is a tendency for the scattering to spread away 
from $(1/2,1/2)$ and recently Dai \etal (1998) reported the presence of four 
incommensurate peaks centered on $(1/2,1/2)$. 

%45 a,b

% 46
\begin{figure}
\epsfig{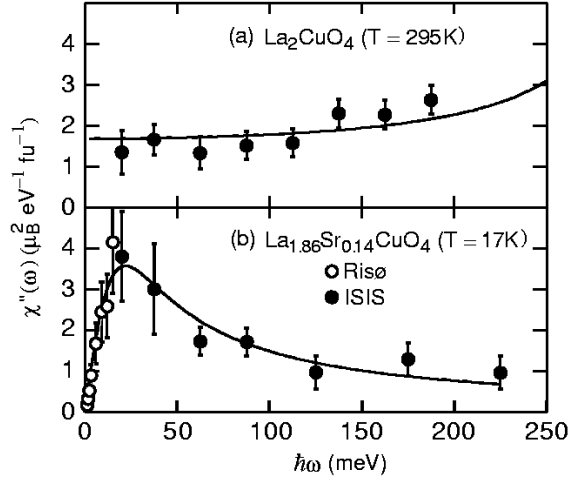}{8cm}
\caption{
Magnetic neutron scattering in \LaSr. Top panel shows the scattering by 
spin waves in the antiferromagnetic parent compound. The lower panel shows 
an underdoped sample where the scattering is peaked at $\approx$ 25 meV.
}\label{hayden2}
\bigskip
\end{figure}

The spin wave spectrum in the undoped materials, as measured around the 
antiferromagnetic $Q$, form a continuum from zero frequency up to the maximum of 
the spin wave spectrum $\approx 2J \approx 250 $ meV in both \YBax\ and 
\LaSr\  
(Hayden \etal (1997), Bourges \etal (1997)). Hole doping has several effects 
on this spectrum. The magnetic Bragg spots disappear, showing the 
destruction of long range coherence but {\it inelastic} magnetic scattering 
in the $(1/2,1/2,l)$ region remains at all doping levels including in the 
superconducting state.  The inelastic spectrum is broadened in $Q$ and the 
spectral weight is redistributed towards a broad peak at lower 
frequency. This is illustrated in Fig. \ref{hayden2} for \LaSr\ and in Figs. 
\ref{bourges3} and  \ref{hayden7} for \YBax.  

In fully oxygenated \YBax\ Rossat--Mignod \etal (1991) reported a narrow 
resonance in $\chi({\bf q},\omega)$ in the superconducting state at 41 meV 
in the  odd spectrum at {\bf Q}=$(1/2,1/2,1.7)$. 
This was subsequently verified by Mook \etal (1993) with polarized 
neutrons. 

\begin{figure}
\epsfig{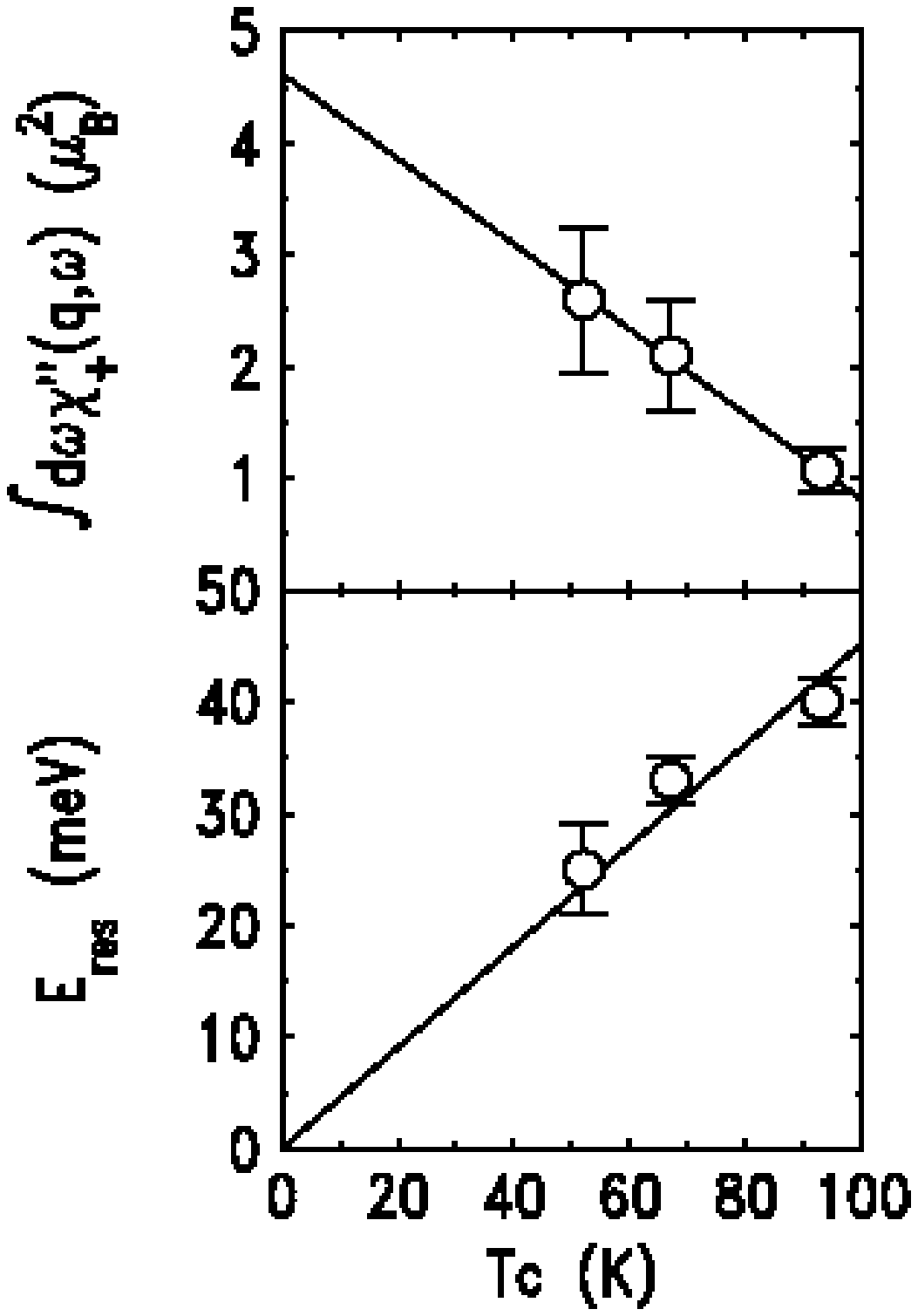}{6cm}
\caption{
The position and frequency of the magnetic scattering peak as function of 
doping. Unlike the pseudogap frequency which is doping independent, the 
magnetic scattering peak,or resonance frequency, varies linearly with $T_c$.
}\label{fong4}
\bigskip
\end{figure}

\begin{figure}[t]
\epsfig{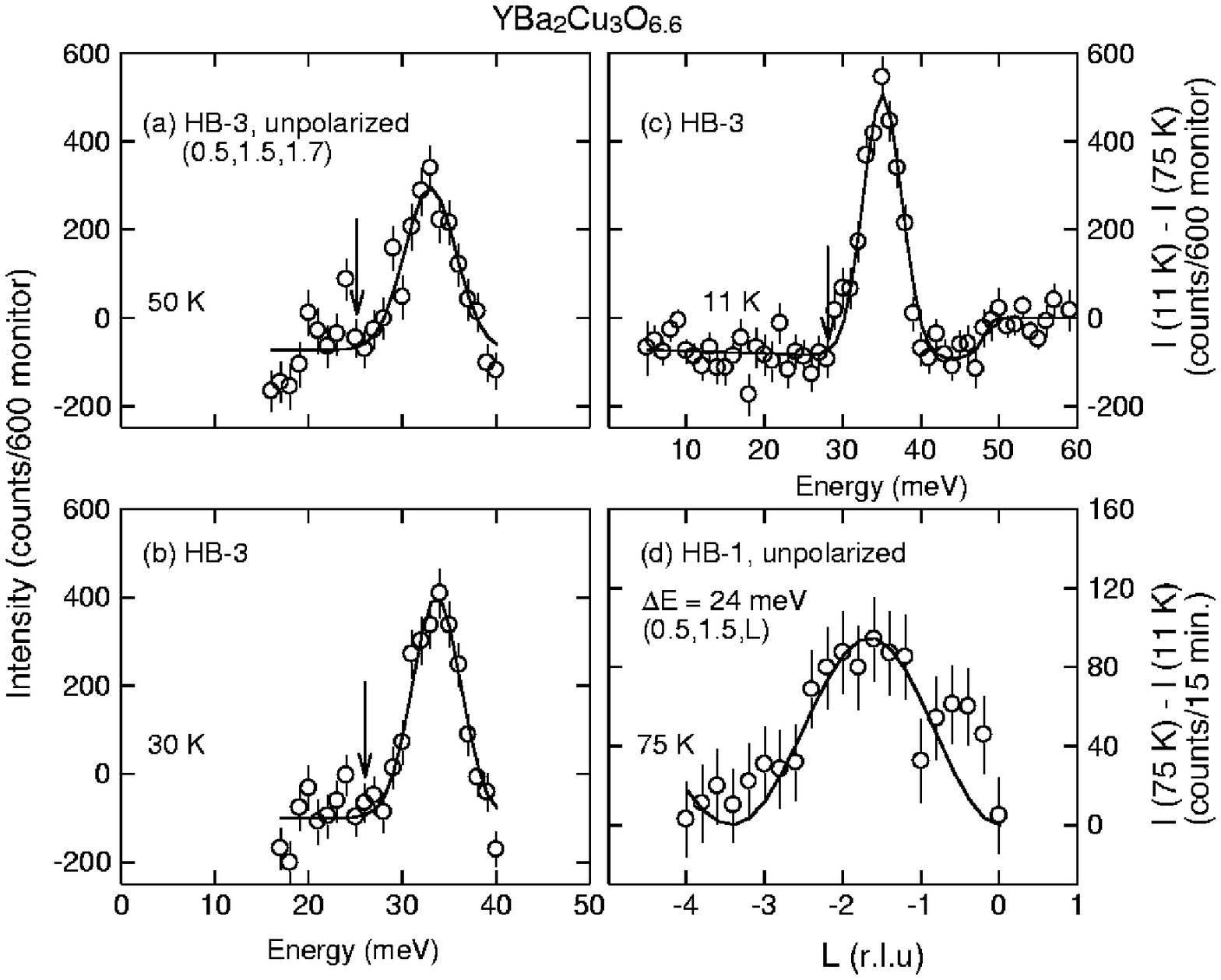}{13cm}
\caption{
The resonance peak in underdoped YBCO 6.6. Magnetic neutron scattering is 
concentrated in the ($\pi,\pi$) region of reciprocal space forming a sharp 
peak at 35 meV in the odd channel: (a) to (c). Panel (d) shows that the 
variation of scattering with ${\bf q}$ normal to the planes is sinusoidally 
modulated with a spatial frequency corresponding to the spacing between the 
bilayers. 
}\label{dai3}
\bigskip
\end{figure}

\begin{figure}[!h]
\epsfig{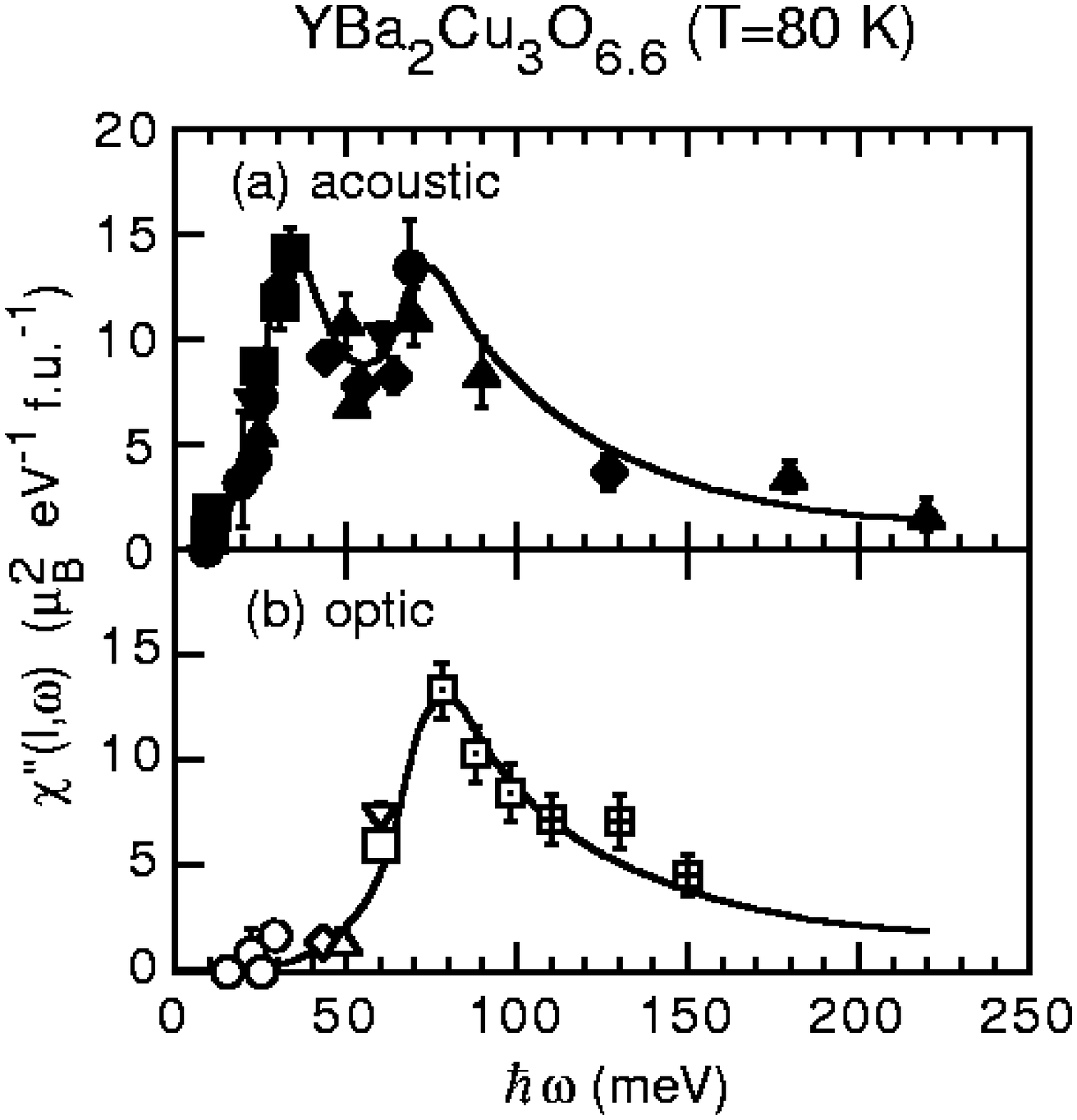}{8cm}
\caption{
Acoustic (odd) and  optic (even) spectra in underdoped YBCO 6.6. 
}\label{hayden7}
\bigskip
\end{figure}

The resonance peak can also be seen in underdoped \YBax. The frequency 
position of the peak varies with doping, decreasing as $T_c$ is reduced in 
proportion to $T_c$ with a linear dependence (Bourges \etal (1995), Dai 
\etal (1996), Fong \etal (1997)). Fig. \ref{fong4} shows the doping 
dependence of the resonance energy from the work of Fong \etal (1997).
Fig. \ref{dai3} (top right panel) shows the peak in underdoped YBCO 6.6 from 
the work of Dai \etal (1996). The bottom right panel shows intensity of the 
resonance as a function of $L$, the reciprocal lattice vector normal to the 
layers. It is clear that the resonance is only seen in the odd spectrum 
since it has zero intensity at $L=0$ where the even channel is 
expected to peak. The left panels show the resonance at higher temperatures. 

We now turn to the temperature dependence of the resonance and the question 
of its possible presence in the normal state. 

To start, we note that the odd channel susceptibility of underdoped YBCO in 
the normal state is 
dominated by a broad peak in the 25--40 meV region . This peak develops from 
the spin wave spectrum of the undoped cuprates through a shift of spectral 
weight from higher (and lower) frequencies to the peak.                                     

%47

As the temperature is lowered the low frequency tail of this peak develops a 
gap--like depression which has been termed a pseudogap (Rossat--Mignod \etal 
(1991), Tranquada \etal (1992), Sternlieb \etal (1994)). This depression is 
seen in the normal state but as the peak sharpens in the superconducting 
state a true gap develops below the peak (Bourges \etal (1996) and Dai \etal 
(1996)). 

Bourges \etal (1996) study the magnetic scattering of overdoped YBCO 6.97. 
In the overdoped region, the magnetic contribution in the normal 
state has almost completely vanished. Bourges \etal  find that the magnetic response 
is restricted to the 33--46 meV region with a very narrow (in $\bf q$) resonant 
contribution at 39 meV.  Since there is no scattering below 33 meV they 
assign this figure to the spin gap and note that it is higher than what is 
seen in underdoped samples. Very similar results are reported by Dai \etal 
(1996) who use polarized neutrons and find in underdoped samples of 
YBCO 6.6 the resonance at 35 meV and the gap at 28 meV, Fig. \ref{hayden7}. 

The ($\pi,\pi$) resonance and the associated gap are only seen in the odd 
spectra with $l=1.83$. The even spectra recorded with $l=0$ show no 
scattering at low energy but above a threshold of 50 
-- 60 meV strong scattering sets in (Bourges \etal (1997), Hayden \etal 
(1997)) as  shown in Figs. \ref{bourges3} and \ref{hayden7}. Another way of 
interpreting the odd--even channel difference is to say that above 50 meV the 
interlayer coherence is destroyed since the sinusoidal modulation in the $l$ 
direction is replaced by uniform rods.

The magnetic scattering in \LaSr\ has many parallels with that in \YBax\ 
(Hayden \etal (1997)). The main difference is the appearance of four 
incommensurate peaks symmetrically placed around ($\pi,\pi$). Their location 
can be understood in terms of Fermi surface nesting. In a Fermi liquid the 
excitations correspond to spin flip transitions across the Fermi surface 
that are enhanced for ($\pi,\pi$) scattering because this vector, at least 
for the simple tight binding Fermi surface, nests regions of the Fermi 
surface with high density of states. In good agreement with this picture in 
\LaSr\ there are four incommensurate peaks shifted away from the commensurate 
($\pi,\pi$) position by an amount $\delta$ that matches the shrinking Fermi 
surface as  hole doping proceeds (Cheong \etal (1991), Mason \etal 
(1993,1996) and Hayden \etal (1996)) 

Mason \etal (1993) looked for a gap in the magnetic excitations associated 
with the four incommensurate peaks found a suppression of excitations below 
$3.5k_BT_c$ but no clear gap. They find no evidence of the node structure 
associated with clean d--wave superconductivity. Instead there is isotropic 
residual scattering at the lowest frequencies. The authors suggest this may 
be due to localized magnetic impurities.

% 48

%49

Yamada \etal (1995) report the observation of a low lying gap in 
the optimally doped \LaSr\ with a $T_c=37.3$ K in the form of a clear 
depression of low frequency scattering below the transition temperature, below 
a frequency of 3.5 meV which they call a magnetic superconducting gap.  It 
should be noted that within the Fermi liquid picture of a d--wave 
superconductor, the gap measured in a magnetic neutron scattering experiment 
is not the maximum gap.  In terms of detailed calculations within d--wave 
models (Tanamoto \etal (1994), Bulut and Scalapino (1994), Zha \etal (1993)) 
these measurement are consistent with $d_{x^2-y^2}$ models with a maximum 
gap of $2\Delta_0=19.3$ meV (155 \cm-1).

\section{Theories of the Pseudogap} 
 
Just as the experimental evidence of the pseudogap does not 
yet provide a single consistent view, neither do the available theoretical 
models. A few selected models are described below. 

%Patrick Lee

        Several of the models described below involve preformed 
pairs. Early references to these ideas include the work of 
Uemura \etal (1989) and Randeria \etal (1989). 
One scenario which has received considerable attention involves
spin--charge separation. This idea was first proposed by P.W. Anderson
(1987) as the resonating valence bond, RVB, model with a mean--field
treatment developed by Baskaran, Zou and Anderson (1997). 
The possibility of separate transition temperatures for the RVB state
and the BEC state was discussed by Kotliar and Lin (1988).
Nagaosa and Lee (1992) produced a Ginzburg--Landau theory of the 
spin--charge separated system calculating various transport
properties in the pseudogap state. Tanamoto \etal (1991) have
calculated NMR properties using the Nagaosa and Lee model.

        Spin--charge separation creates holons with zero spin
and spinons which are zero charge, spin--1/2 fermions. 
In the mean--field description the spinons pair to form a gap
in the spin excitations, identified as the pseudogap. The holons
Bose--condense at $T_c$ to form the superconducting state. 
At present it is believed that even though it is the holons
which Bose--condense, gauge field fluctuations lead to a strong 
coupling between the spinons and holons. A gauge theory of
the normal state, including the pseudogap, has been developed
by Lee and Nagaosa (1992) and Lee and Wen (1997).

% 
% Emery and Kivelson 

Emery and Kivelson (1997) have developed a preformed pair model 
of the pseudogap based on microstripes. Phase separation takes 
place on a microscopic scale generating dynamical charged stripes 
separated by insulating antiferromagnetic stripes. These microstripes 
form at the upper crossover temperature $T^{\circ}$. 
Above this temperature the charge is uniformly distributed. Below 
$T^{\circ}$ charge is confined 
to the metallic stripes forming a one--dimensional electron gas (1DEG). 
Spin and charge are separated as spin resides in the AF stripes. 
As the temperature is lowered, AF correlations build up. At the 
lower crossover temperature $T^*$ pairing behaviour emerges. 
 
Pairing is a result of a spin gap in the AF stripes. This is 
manifested in the 1DEG via pair hopping between the 1DEG and 
AF stripes.  Emery and Kivelson describe this as a magnetic 
proximity effect. At this point there are only 1D superconducting 
correlations. The pseudogap is associated with this spin gap.
At $T_c$ Josephson coupling between the metallic stripes 
becomes large enough to yield global phase coherence. Note 
that the pairing correlations below $T^*$ are not giving 
rise to real space pairing; the pairing correlations are dynamical. 

A natural consequence of this model is that the symmetry
of the order parameter as measured by local probes will
hold in the pseudogap state, as will other superconducting
properties.

For underdoped systems the transition into the superconducting state  is 
determined by the stiffness against classical phase fluctuations (Emery (1995)), 
proportional to $n_s(T)$. In this context the observation by Uemura \etal of 
a universal relation between $T_c$ and $\lambda^{-2}(0) \propto n_s$ can be 
thought of as an upper bound on $T_c$ given by the ordering temperature for 
classical phase fluctuations.

%Pines and co.
             
Pines and co--workers have developed the Nearly Antiferromagnetic
Fermi Liquid (NAFL) model to describe the physics of the cuprate
superconductors, reviewed in Pines (1997a) and Pines (1997b). 
As described in the original MMP paper (Millis \etal (1990)),
the dominant interaction between
quasiparticles arises from spin fluctuations, as characterized in
the dynamical spin susceptibility $\chi({\bf q}, \omega)$. 
The strong AF correlations cause $\chi({\bf q}, \omega)$ to
peak at {\bf Q} = ($\pi, \pi$).  This has led the authors to
distinguish two classes of quasiparticles: {\it hot} quasiparticles
which are located near ($0,\pi$), connected to each other
by {\bf Q}, and {\it cold} quasiparticles which are not strongly
connected by AF fluctuations. The cold quasiparticles behave
like a strongly coupled Landau Fermi liquid whereas the hot
quasiparticles are anomalous and non--Landau Fermi liquid like.

Chubukov \etal (1996) have developed a scenario to 
describe the crossover temperatures and pseudogap. At high 
temperatures,  mean field behaviour is observed with weak AF 
correlations. As the temperature is lowered, in the underdoped 
region, a crossover occurs at $T^{\circ}$ where the correlation 
length $\xi \approx 2a$. Below this temperature non--universal 
pseudoscaling is observed where, for example,  $^{63}T_1T$, 
$^{63}T_{2G}$ and the Knight shift are linear in $T$. This 
behaviour is a result of the hot quasiparticle spectrum being 
modified by the strengthening AF correlations with decreasing 
temperature; {\bf i.e.} $1/\xi \approx a + bT$. 

Upon reaching $T^*$ a gap opens in the hot quasiparticle spectrum,
thus establishing the pseudogap.
Below $T^*$ the spin--relaxation  rates
and Knight shift decrease more quickly than above $T^*$. 
Superconductivity occurs when the cold quasiparticles become gapped,
independent of the hot quasiparticles which are already gapped at $T_c$.
Thus the magnitude of the pseudogap and $T^*$ are unrelated to $T_c$.
As it is only the cold quasiparticles which participate in the superconducting
transition, the superfluid density $n_s$ will be considerably less than the
total number of quasiparticles. 

Millis and Monien (1993) proposed that the pseudogap in bilayer compounds
was due to interlayer exchange coupling. Further work by Altshuler, Ioffe
and Millis (1996) have produced a model with interlayer pairing
of holons producing a spin pseudogap. The authors claim that the pseudogap
crossover temperature $T^*$ for single layer materials, namely \LaSr, is 
just slightly above $T_c$. Interlayer coupling enhances $T^*$ to temperatures
well above $T_c$. Recent data on other single layer compounds throws this
scenario into doubt. The single layer mercury compound shows a crossover
temperature well above $T_c$ (Bobroff \etal (1997)). 

%<--- Put this into NMR table.

Maly \etal (1998) introduce a description of the pseudogap state in terms of 
resonant pairs, somewhere between the preformed pairs that Bose condense at 
$T_c$ and free Fermions. Their paper offers a nice review 
of the currently competing models.
                                     
Castellani \etal (1997) find a quantum critical point at 
optimal doping associated  with a charge density wave instability 
related to the observations of Boebinger  \etal in high pulsed magnetic 
fields of an apparent metal--insulator transition at $T=0$ below optimal doping.
They find an order 
parameter that has d--wave symmetry and a superconducting $T_c$ 
that strongly depends on doping in the overdoped region but is flat 
at optimal doping in agreement with experiments. 

\section{Summary and Conclusions}

%Settled issues.

The experiments reviewed here have given evidence of the presence of a 
pseudogap in the normal state of all the cuprates. The pseudogap seems to be 
related to the superconducting gap in the sense that it evolves smoothly 
into the superconducting gap and has the same d--wave symmetry. Unlike a 
conventional  superconducting gap though, the pseudogap magnitude is 
temperature independent. Tunneling, ARPES and optical data indicate that the 
magnitude of the gap does not change significantly on going through $T_c$. 
Thus in the case of the cuprates, in contrast to a conventional 
superconducting gap, the gap starts as the pseudogap in the normal state and 
evolves into the superconducting gap below $T_c$ ({\it i.e.}  $\Delta(T)$ 
does not go to zero at $T_c$). 

%50 
\begin{figure}[t]
\epsfig{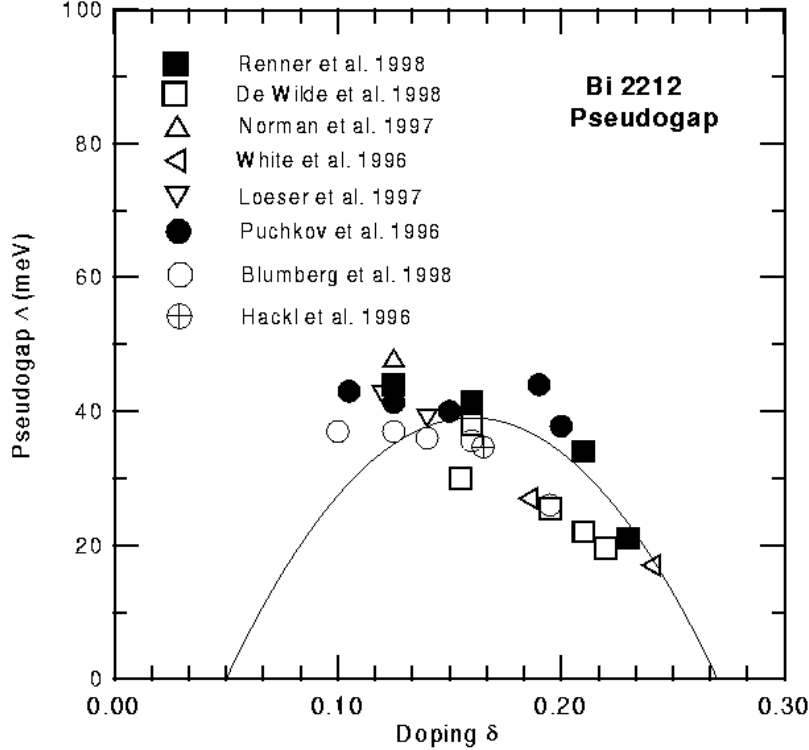}{11cm}
\caption{
The width $\Delta$ of the pseudogap in Bi 2212 measured with various 
experimental techniques as function of doping $\delta$. The dashed line is 
an empirical formula for $T_c$ assuming $2\Delta=9.5 k_BT_c$. 
The triangles  
are photoemission data, the squares tunneling, the filled 
circles $ab$--plane optical conductivity, and the open circles Raman scattering.
}\label{gaps}
\bigskip
\end{figure}

The pseudogap appears in the underdoped side of the phase diagram and 
weakens as optimal doping is approached. A weak pseudogap is still present at 
optimal doping but disappears not too far into the overdoped region. This is 
illustrated in the phase diagram where $T^*$ approaches $T_c$ just into the 
overdoped region. $T^*$ is strongly temperature dependent in the underdoped 
region in contrast to the magnitude of the pseudogap which weakly decreases 
as the doping level is increased. Thus $T^*$ is not  a direct measure of the 
pseudogap energy scale. Well into  the overdoped region, the pseudogap merges 
with the superconducting gap, i.e. $T^*$ has merged into $T_c$ and there are 
no remaining signatures of the pseudogap in the normal state. 
 
As an example, Fig. \ref{gaps} shows a summary of pseudogap energy scale 
from various laboratories on Bi 2212 as a function of doping (according to 
an empirical law $T_c/T_{c,max}=1-82.6(\delta-0.16)^2$ (Presland \etal (1991) 
and Groen \etal (1990)).  The general 
trend for any given technique is a  rough tracking of $T_c$ in the overdoped 
ranges keeping $2\Delta/k_BT_c$ roughly constant at 9.5 (shown as the dashed 
curve) and an approximately constant, or even slightly increasing gap value, 
as oxygen is removed, in the underdoped region. This graph should be treated 
with some caution.  First, all techniques measure slightly different 
properties of the gap and in the absence of a microscopic theory such 
comparisons are very approximate. Secondly, there are built in 
sample--to--sample variations and there is some uncertainty in the case of 
Bi 2212 whether a given sample is overdoped or underdoped. 
 
%Outstanding issues 
The above discussion largely is concerned with frequency dependent 
information. ARPES, INS, NMR and Raman spectroscopy 
provide {\bf q}--dependent information. 
INS, in principle, probes all of reciprocal space but due to background 
subtraction is predominantly probing the region near ($\pi,\pi$). 
The Knight shift is proportional to $\chi(q=0)$ and the spin--lattice 
relaxation rate  $1/^{63}T_1$ is dominated by fluctuations near ($\pi,\pi$). 
ARPES more or less uniformly covers the whole zone. NMR measurements 
reveal that the crossover temperature $T^*$ is somewhat {\bf q} dependent. 
The temperature evolution of the pseudogap provided by ARPES is 
also {\bf q} dependent. Hence it is insufficient to describe the pseudogap 
evolution with only a temperature dependent suppression factor. 
 
The pseudogap in the excitation spectrum opens up in the region near 
($\pi,0$), largely influenced by the susceptibility at ($\pi,\pi$). 
Excitations on the Fermi surface near the diagonal remain ungapped. The 
susceptibility near $q$=0 averages over both of these regions. Given the 
nature of the ARPES results it is not surprising that NMR yields somewhat 
different results for $q$=0 and {\bf q} = ($\pi,\pi$). 

Measurements taken with a wide variety of techniques demonstrate that the 
pseudogap is present in both the spin and charge channel. For example, ARPES 
measures the charge channel and magnetic susceptibility the spin channel. 
Loram's work has shown quantitatively, through Wilson's ratio, that the 
number of excitations in both channels is approximately the same. Thus 
retention of the term pseudogap seems justified as this is not just a spin 
gap as some would suggest. Near equal occupation of excitations in both 
channels also serves to rule out extreme models which have all the 
excitations in one channel only. For example, a charge separation model in 
which the holons condense out with no consequence to the spinon excitations 
is inconsistent with this observation. 

Many measurements of the pseudogap depict a smooth evolution of a physical 
observable as a function of temperature. Typical examples are the electronic 
specific heat  and magnetic susceptibility of underdoped \YBax. On the other 
hand, some techniques reveal a distinct crossover temperature, namely 
resistivity and NMR. Magnetic resonance is unique in that it reveals two 
crossover temperatures in several compounds. Comparison with other 
techniques suggest that the lower crossover temperature $T^*$ is associated 
with the pseudogap and that the origin of the upper crossover $T^{\circ}$ 
may be unrelated to the pseudogap. Indeed, the linear temperature evolution 
between $T^*$ and $T^{\circ}$ of the Knight shift, $^{63}T_1T$ and 
$^{63}T_{2G}$ are more indicative of an physical evolution than a gap. In 
particular, if the curves are extrapolated to zero temperature they have 
positive intercepts, whereas below $T^*$, the curves extrapolate to zero at 
a positive temperature -- consistent with a gap. 

Crossovers observed in some compounds and not others raises the 
question as to whether a crossover at $T^*$ is intrinsic to the pseudogap. 
At this point it is unclear as to whether the crossover is unique to some 
compounds and not others. One possibility is that the crossover is smeared 
out by inhomogeneities in only some compounds.

Special mention must be made of the \LaSr\ compound. Although many 
measurements are indicative of a pseudogap, even in the strongly overdoped 
region, there are several contradictory observations. Susceptibility and 
resistivity data in the ``pseudogap'' region extrapolate to a positive intercept 
rather than intersecting zero at a positive temperature as is the case with 
other compounds. A pseudogap is completely absent in the spin--lattice 
relaxation rate in the underdoped region. One possible cause of these 
differences is the presence of paramagnetic centres. These moments have 
been observed in specific heat measurements of Mason \etal (1993) 
and are intrinsic to \LaSr, a random alloy. Similar to the case of 
Zn doped YBCO 124, local moments may be responsible for the absence 
of the pseudogap in the spin--lattice relaxation rate. In any case, 
it seems that \LaSr\ must be treated as a special case; it is not 
consistent with the general properties of the pseudogap observed 
in the other cuprate families. Finally, last but not least, \LaSr\  is unique 
in having a maximum $T_c$ of only 40 K in contrast to the other compounds, 
both single layer and double layer, with maximum $T_c$'s of 90 K and above.

Theoretical views of the pseudogap are just as diverse as the 
views on the mechanism of high--$T_c$ superconductivity. One point 
of view which stands out in the context of experimental evidence 
is the idea of preformed pairs. Several relevant observations are: 
\smallskip
\begin{itemize}
\item[(1)] The pseudogap evolves into the superconducting gap. There is no 
evidence that the magnitude of the superconducting gap goes to zero at $T_c$. 
 
\item[(2)] Both the pseudogap and superconducting gap have d--wave symmetry. 
 
\item[(3)] The crossover temperature $T^*$ merges with $T_c$ in the overdoped 
region of the phase diagram. 
\end{itemize}
\smallskip 

These factors are consistent with the existence of preformed pairs, of some 
description. They need not be real space pairs and could just be dynamic 
correlations. Although the evidence so far is consistent with preformed 
pairs, it does not prove their existence.

\section*{Acknowledgments} 
 
We would like to acknowledge the support of this work by the Natural 
Science and Engineering Research Council of Canada (NSERC) and the Canadian 
Institute of Advanced Research (CIAR). We have also gained valuable insight 
from discussions with the following people: P.W. Anderson, D.N. Basov, 
A.J.~Berlinski, D.A.~Bonn, G.~Blumberg, J.P. Carbotte, J.C.~Cooper, V.J.~Emery, 
W.N.~Hardy,  C.C.~Homes, J.C.~Irwin, B.~Keimer, C.~Kallin, P.A.~Lee, 
J.W.~Loram, F.~Marsiglio, T.E.~Mason, H.A. Mook, A.J.~Millis, M.~Norman, 
J.~Preston, A.V.~Puchkov, D.~Pines, M.~Reedyk, C.~Renner, T.~R\~o\~om, Z.-X.~Shen, 
T.~Startseva, R.~Stern, D.B.~Tanner, S.~Tajima,  D. van der Marel, 
Y.~Uemura, M.B. Walker and J. F.~Zasadzinski. 

\section*{References}

\newenvironment{Alist}% Define ROP author list style
{\parindent=-0.25in \advance\leftskip by0.25in}
{}
\begin{Alist}

Abrikosov A A and Fal'kovskii L A 1961 Sov. Phys. JETP {\bf 13} 179 

Allen J W and Mikkelsen J C 1976 Phys. Rev. B {\bf 15} 2953 

Allen P B 1971 Phys. Rev. B {\bf 3} 305 

Alloul H, Ohno T and Mendels D 1989 Phys. Rev. Lett. {\bf 63} 1700 

Alloul H, Mendels P, Casalta H, Marucco J F and Arabski J
1991 Phys. Rev. Lett. {\bf 67} 3140

Altshuler B L, Ioffe L B, and Millis A J  1996 \prb {\bf 53}, 415

Anderson R O, Claessen R, Allen J W, Olson C G, Janowitz C, 
Liu L Z, Park J-H, Maple M B, Dalichaouch Y, de Andrade M C, 
Jardim R F, Early E A, Oh S-J and Ellis W P 1993 Phys. Rev. 
Lett. {\bf 70} 3163 

Anderson P W 1987 Science {\bf 235} 1196

Arnold G B, Mueller F M and Swihart J C 1991 Phys. Rev. B {\bf 
78} 1771

%Atkinson W A, Wu W C and Carbotte J P preprint: cond--mat/9701029

Ashcroft N W and Mermin N D 1996 {\it Solid State Physics} Holt, 
Rinehart and Winston, New York p. 252

Bankay M, Mali M, Roos J and Brinkmann D 1994 Phys. Rev. B {\bf 50} 6416 

Barzykin V and Pines D 1995 Phys. Rev. B {\bf 52} 13585

Baskaran G, Zou Z and Anderson P W 1987 Solid State Comm. {\bf 63} 973

Basov D N, Liang R, Dabrowski B, Bonn D A, Hardy W N and Timusk T
1996 Phys. Rev. Lett. {\bf 77} 4090
\comment{ \YBa Pseudogap and charge dynamics in CuO$_2$ planes in
YBCO.  28 c-axis \YBa \LaSr compared, low frequency mode.}

Basov D N, Timusk T, Dabrowski B and Jorgensen J D 1994a Phys. Rev. B
{\bf 50} 3511
\comment{C-axis Response of YBa$_2$Cu$_4$O$_8$}

Basov D N, Mook H A, Dabrowski B and Timusk T 1995b Phys. Rev. B
{\bf 52} R13141
\comment{28 c-axis \YBa LaSr compared, low frequency mode.}

Basov D N, Liang R, Bonn D A, Hardy W N, Dabrowski B, Quijada M,
Tanner D B, Rice J P, Ginsberg D M and Timusk T 1995a Phys. Rev.
Lett. {\bf 74} 595

Basov D N and Timusk T 1998 
{ \it Infrared Properties of High-$T_c$ Superconductors: an
Experimental Overview} in Handbook on the Physics and Chemistry
of Rare Earths edited by K.A.Dschneidner, LeRoy Eyring and M.B.Maple,  
North-Holland 1999.

Bednorz J G and M\"uller K A 1996 Zeitschrift f\"ur Physik {\bf 64} 
189 (1986)

Batlogg B, Hwang H Y, Takagi H, Cava R J, Kao H L and Kwo J 1994 Physica C
{\bf 235-240} 130

Blumberg G, Klein M V, Kadowaki K, Kendziora C, Guptasarma 
P and Hinks D 1998 {\it preprint} cond-mat/9711073

Bobroff J, Alloul H, Mendels P, Viallet V, Marucco H F and 
Colson D 1997 Phys. Rev. Lett. {\bf 78} 3757

Boebinger G S, Ando Y, Passner A, Kimura T, Okuya M, Shimoyama 
J, Kishio K, Tamasaku K, Ichikawa N, and Uchida S 1996 Phys. Rev. 
Lett. {\bf 77}  5417

Bonn D A, Dosanjh P, Liang R and Hardy W N 1992 Phys. Rev. Lett. {\bf 68}
2390
\comment{quasiparticle lifetime loss at Tc \YBa microwave surface impedance}

Bourges P, Fong H F, Regnault L P, Bossy J, Vettier C, Milius D L, Aksay I 
A and Keimer B 1997 Phys. Rev. B {\bf 56} R11439

Bourges P, Regnault L P, Henry J Y, Vettier C, Sidis Y and Burlet P 1995
Physica B {\bf 215} 30

Bourges P, Regnault L P, Sidis Y and Vettier C 1996 Phys. Rev. B {\bf 53} 
876
\comment{overdoped neutron scattering 6.97 \YBa}

Branch D 1996  {\it Optical Properties of Strongly-Coupled d-wave 
Superconductors with an Anisotropic Momentum Dependent 
Interaction}, Thesis, McMaster University.
                                        
Bucher B, Steiner P, Karpinski J, Kaldis E, and Wachter P 1993 
Phys. Rev. Lett. {\bf 70} 2012

Butaud P, Horvati M, Berthier Y, Segransan P, Kitaoka Y,
Berthier C and Katayama-Yoshida H 1990 Physica C {\bf 166} 301

Campuzano J C, Jennings G, Faiz M, Beaulaigue L, Veal B W, Liu 
J Z, Paulikas A P, Vadervoort K, Claus H, List R S, Arko A J, and 
Bartlett R J 1990 Phys. Rev. Lett. {\bf 64} 2308

Carbotte J P 1990 Rev. Modern Phys. {\bf 62} 1027

Carrington A, Walker D J C, Mackenzie A P and Cooper J R 1993 \prb {\bf 48} 
130351 

Castellani C, Di Castro C and Grilli M 1997 Z. f\"ur Physik 
{\bf 130} 137

Chakravarty S, Halperin B I and Nelson D R 1989 Phys. Rev. B {\bf 39} 2344 

Chen X K, Naeni J G, Hewitt K C, Irwin J C, Linag R and Hardy W N
1994 Phys. Rev. Letters {\bf 73} 3290 

Cheong S W, Aeppli G, Mason T E, Mook H, Hayden S M, Canfield P C, Fisk Z, 
Clausen K N and Martinez J L 1991 Phys. Rev. Lett. {\bf 67} 1791

Chubukov A V, Pines D and Stojkovi\'c B P 1996 J. Phys. Cond.
Matt. {\bf 8} 10017

Coffey L and Coffey D 1993 \prb {\bf 48} 4184 

Cooper S L, Reznik D, Kotz A L, Karlow M A, Liu R, Klein M V,
Lee W C, Giapintzakis J, Ginsberg D M, Veal B W and Paulikas A P 1993a
Phys. Rev. B {\bf 47} 8233

Cooper S L, Nyhus P, Reznik D, Klein M V, Lee W C, Ginsberg D M,
Veal B W, Paulikas A P, Dabrowski B 1993b Phys. Rev. Lett. {\bf 70} 
1533

Corey R L, Curro N J, O'Hara K, Imai T, Slichter C P,
Yoshimura K, Katoh M and Kosuge K 1996 Phys. Rev. B {\bf 53} 5907

Dai P, Mook H A and Do\^gan F 1998 Phys Rev. Lett. {\bf 80} 1738

Dai P, Yethiraj M, Mook H A, Lindmer T B and Do\^gan F 1996 Phys. Rev. Lett. 
{\bf 77} 5425         

D\"aumling M 1991 Physica (Amsterdam) {\bf 183c} 293

Dessau D S, Shen Z-X, King D M, Marshall D S, Lombardo L W, Dickinson P H, 
Loeser A G, DiCarlo J, Park C-H, Kapitulnik A, and Spicer W E 1993 Phys. 
Rev. Lett. {\bf 71} 2781 

Dessau D S, Wells B O, Shen Z-X, Spicer W E, List R S, Arko A J, Mitzi D 
B, Kapitulnik A, and  1991 Phys. Rev. Lett. {\bf 66} 2160 
\comment{first obsevation of dip in ARPES}

Deveraux T P and Einzel D 1995 Phys. Rev. B {\bf 51} 16336

DeWilde Y, Miyakawa N, Guptasarma P, Iavarone M, Ozyuzer L, Zasadzinski J 
F, Romano P, Hinks D G, Kendziora C, Crabtree G W and Gray K E 1998 Phys. 
Rev. Lett. {\bf 80} 153

Ding H, Norman M R, Yokaya T, Takeuchi T, Randeria M, Campuzano J C, 
Takahashi T, Mochiku T and Kadowaki K 1997 Phys. Rev. Lett. {\bf 78} 
2628

Ding H, Yokaya T, Campuzano J C, Takahashi Randeria M, T, Norman M R, 
Mochiku T, Kadowaki K and Giapinzakis J 1996 Nature {\bf 382} 51
\comment{Psedogap in the normal state by ARPES}

Einzel D and Hackl R 1996 J. of Raman Spectroscopy {\bf 27} 307

Emery V J and Kivelson S A 1995 Nature {\bf 374} 4347

Emery V J, Kivelson S A and Zachar O 1997 Phys. Rev. B {\bf 56} 6120

Farnworth B and Timusk T 1974 Phys. Rev. B {\bf 10} 2799

Fong H F, Keimer B, Reznik D, Milius D L and Aksay I A 1996 
Phys. Rev. B {\bf 54} 6708

Fong H F, Keimer B, Milius D L and Aksay I A 1997 
Phys. Rev. Lett. {\bf 78} 713
\comment{Keimer's group doping variation of the resonance frequeny \YBa 
magnetic neutron scattering underdoped}

Friedberg R and Lee T D 1989 Phys. Rev. B {\bf 40} 6745

Fujimori A, Ino A, Mizokawa T, Kim C and Shen Z.-X 1998 J. Phys. Chem. of Solids 
(Proceedings of SNS'97,  to be published)

Fukuoka A, Tokiwa-Yamamoto A, Itoh M, Usami R, Adachi S and Tanabe K 1997
Phys. Rev. B {\bf 55} 6612

Giaver I 1960 Phys. Rev. Lett. {\bf 5} 147

Groen W A, de Leeuw D M and Feiner L F 1990 Physica C {\bf 165} 55

Gurvich M, and Fiory A T 1987 Phys. Rev. Lett. {\bf 59} 1337

G\"otze W and W\"olfe P 1972 Phys. Rev. B {\bf 6} 1226 

Hackl R, Kaiser R and Schicktanz 1983 J. Phys. C {\bf 16} 1729

Hackl R, Krug G, Nemetshcek R, Opel M, and Stadlober B 1996 in 
{\it Spectroscopic Studies of Supercondcutors V}, edited by Ivan Bozovic  and 
Dirk van der Marel, Proc. SPIE {\bf 2696} 194     

Harris J M, Loeser A G, Marshall D S, Schabel M C and Shen Z.-X 1996 
Phys. Rev. B {\bf 54} R15665

Harris J M, White P J, Shen Z.-X, Ikeda H, Yoshizaki R, Eisaki H, Uchida S, 
Si W D, Xiong J W, Zhao Z.-X Dessau D S 1997 Phys. Rev. Lett. 
{\bf 79} 143

Hayden S M, Aeppli G, Dai P, Mook H A, Perring T G, Cheong S-W, Fisk Z, 
Do\^gan F and Mason T E 1998 Physica B  {\bf 241-243} 765 

Homes C C, Timusk T, Liang R, Bonn D A, and Hardy W N 1993
Phys. Rev. Lett. {\bf 71}, 1645

Homes C C, Timusk T, Bonn D A, Liang R and Hardy W N 1995 Physica C 
{\bf 265-280} 265

Homes C C, Timusk T, Bonn D A, Liang R and Hardy W N 1995b Can. J. Phys. {\bf 
73} 663

Horvatic M, Auler T, Berthier Y, Butaud P, Clark W G, Gillet J A, and 
Segransan P 1993 Phys. Rev. B {\bf 47} 3461

Hosseini A, Kamal S, Bonn D A, Liang Ruxiang and W.N. Hardy W N 1998 
Phys. Rev. Lett. {\bf 81} 1298

Hsueh Ya--Wei, Statt B W, Reedyk M, Xue J S, Greedan J E 
1997 Phys. Rev. B {\bf 56} 8511 \comment{correct}

Hwang H Y, Batlogg B, Takagi H, Kao H L, Kwo J Cava R J, 
Krajewski J J, and Peck W F Jr. 1994 Phys. Rev. Lett. {\bf 72} 
2636

Hwu Y, Lozzi L, Marsi M, La Rosa S, Winkour M, Davis P, Onellion M, Berger 
H, Gozzo F, L\'evy F, and Margaritondo G 1991 \prl {\bf 67} 2573

Imai T, Yasuoka H, Shimizu T, Ueda Y, Yoshimura K and Kosuge K
1989 Physica C {\bf 162--164} 169

Imer J.-M, Patthey F, Dardel B, Schneider W.-D, Baer Y, Petroff Y, and Zettl 
A 1989 Phys. Rev. Lett. {\bf 62} 336 

Ino A, Kim C, Mizokawa T, Shen Z.-X, Fujimori A, Takaba M, Tamasaku K, 
Eisaki H, and Uchida S 1997 {\it Preprint} cond-mat/9809311 
\comment{LaSr ARPES high energy pseudogap}

Ito T, Takenaka K and Uchida S 1993 Phys. Rev. Lett. {\bf 70} 3995

Itoh Y, Machi T, Fukuoka A, Tanabe K and Yasuoka H 1996
J. Phys. Soc. Jpn. {\bf 65} 3751

Julien M-H, Carretta P, Horvati\'c M, Berthier C, Berthier Y,
S\'egransan P, Carington A and Colson D 1996 Phys. Rev. Lett.
{\bf 76} 4238 

Junod A 1989 in {\it Physical Properties of High Temperature 
Superconductors} edited by D.M.~Ginsberg (World Scientific, 
Singapore), {\bf 2} 13 

Kendziora C A, Kelley R J and Onellion M 1996 in 
{\it Spectroscopic Studies of Supercondcutors V}, edited by Ivan Bozovic  and 
Dirk van der Marel, Proc. SPIE {\bf 2696} 223 

Kendziora C A, Kelley R J and Onellion M 1996 
Phys. Rev. Lett. {\bf 77} 727

King D M, Shen Z-X, Dessau D S, Wells B O, Spicer W E, Arko A 
J, Marshall D S, DiCarlo J, Loeser A G, Park C-H, Ratner E R, 
Peng J L, Li Z Y and Greene R L 1993 Phys. Rev. Lett. {\bf 70} 
3159 

Kitaoka Y \etal 1989 in {\it Strong Correlation and Superconductivity} 
edited by by H. Fukuyama, S. Maekawa and A.P. Malozemoff (Springer, Berlin) 

Kitezawa 1996 Science {\bf 271} 313

Klein M V and Dierker S B 1984 Phys. Rev. B {\bf 29} 4976 

Kostur V N and Eliashberg G M 1991 JETP Lett. {\bf 53} 391

Kotliar G and Liu J 1988 Phys. Rev. B {\bf 38} 5142

LaRosa S, Vobronik I, Zwick F, Berger H, Grioni M, 
Margaritondo G, Kelley R J, Onellion M and Chubukov A 1988 
Phys. Rev. B {\bf 56} R525  

Lee P A and Nagaosa N 1992 Phys. Rev. B {\bf 46} 5621

Lee P A and Wen X.-G  1997 Phys. Rev. Lett. {\bf 78} 4111

Leggett A J 1994 Braz. J. Phys. B {\bf 50} 496 

Liu R, Veal B W, Paulikas A P, Downey J W, Shi H, Olson C G, Gu 
C, Arko A J and Joyce R R 1991 Phys. Rev {\bf 45} 5614 

Loeser A G, Shen Z.-X, Dessau D S, Marshall D S, Park C H, 
Fournier P and Kapitulnik A 1996 Science {\bf 273} 325

Loeser A G, Shen Z.-X, Schabel M C, Kim C, Zhang M, Kapitulnik A and 
Fournier P 1997 Phys. Rev. B {\bf 56} 14185

Loram J W , Mirza K A, Cooper J R, Liang W Y and Wade J M 
1994a J. of Superconductivity {\bf 7} 243 

Loram J W, Mirza K A, Wade J M, Cooper J R and Liang W Y 
1994b Physica C {\bf 235} 134

Loram J W, Mirza K A, Cooper J R and Liang W Y 
1993 Phys. Rev. Lett. {\bf 71} 1740

Loram J W, Mirza K A, Cooper J R and Tallon J L 1997 Physica C 
{\bf 282-287} 1405

Loram J W, Mirza K A, Cooper J R, Athanassopoulou N and Liang W Y
1996 {\it Proc. of the 10th HTS Anniversary Workshop on
Physics, Materials and Applications} edited by B. Batlogg, C.W. Chu, W.K.
Chu, D.U. Gubser and K.A. M\"uller (World Scientific,
Singapore) 341

Mahajan A V, Alloul H, Collin G and Marucco J F
1994 Phys. Rev. Lett. {\bf 72} 3100

Maly J, Boldizar J and Levin K 1998 {\it preprint} cond--mat/9805018 

Mandrus D, Forro L, Koller D, and Mihaly L 1991 Nature {\bf 351} 460

Manzke R, Buslaps T, Claessen R, and Fink J 1989 Europhys. Lett. {\bf 9} 477

Mandrus D, Hartge J, Kendziora C, Mihaly L and Forro L 1993 Europhys. 
Lett. {\bf 22} 199

Martin S, Fiory A T, Fleming R M, Schneemeyer L F and Waszczak J V 1990 
Phys. Rev. Lett. {\bf 41} 846

Marshall D S, Dessau D S, Loeser A G, Park C.-H, Matsuura A Y, Eckstein J N, 
Bozovic I, Fournier P, Kapitulnik A, Spicer W E  and Shen Z.-X 1996 
Phys. Rev. Lett. {\bf 76} 4841
\comment{pocket cause by pseudogap}

Marsiglio F, Startseva T and Carbotte J P 1998 Physics Letters A 
{\bf 245} 172

Martindale J A and Hammel P C 1996 Philosophical Magazine B
{\bf 74} 573 

Mason T E, Aeppli G, Hayden S M, Ramirez A P and Mook H 1993 \prl {\bf 71} 
919

Mason T E, Schr\"oder A, Aeppli G, Mook H and Hayden S M 1996 \prl {\bf 77} 
1604

Mattis D C and Bardeen J 1958 Phys. Rev. {\bf 111} 412

Millis A J, Monien H and Pines D, 1990 Phys. Rev B {\bf 42} 167

Millis A J and  Monien H 1993 \prl {\bf 70} 2810 

Miyatake T, Yamauchi K, Takata T, Koshizuka N and
Tanaka S 1991 Phys. Rev. B {\bf 44} 10139 

Momono M, Ido M, Najano T, Oda M, Okajima and Yamaya K 
1994 Physica C {\bf 233} 395

Monien H, Monthoux P and Pines D 1991 Phys. Rev. B {\bf 43} 275 

Mook H A, Yethiraj M, Aeppli G, Mason T E and Armstrong T 1993
Phys. Rev. Lett. {\bf 70} 109

Mori H 1965 Prog. Theor. Phys. {\bf 34} 399

Morr D K and Pines D 1998 Phys. Rev. Lett. {\bf 81} 1086

Naeini J G, Chen X K, Hewitt K C, Irwin J C, Devereaux T P, Okuya M, 
Kimura T and Kishio K 1998 Phys. Rev. B {\bf 57} R11077

Nagaosa N and Lee P A 1992 Phys. Rev. B {\bf 45} 966

Nemetschek R, Opel M, Hoffman C. M\"uller P F, Hackl R, Berger H, Forr\'o 
L, Erb A and Walker E 1997 Phys. Rev. Lett. {\bf 78} 4837

Norman M R, Ding H, Randeria M, Campuzano J C, Yokaya T, Takeuchi T, 
Takahashi T, Mochiku T, and Kadowaki K Guptasamrma P and Hinks D G 
1997a {\it preprint} cond-mat/9710163

Norman M R, Ding H, Campuzano J C, Takeuchi T, Randeria M, Yokaya T, 
Takahashi T, Mochiku T and Kadowaki K 1997b Phys. Rev. Lett. {\bf 79}
3506

Ohsugi S, Kitaika Y, Ishida K and Asayama K 1991 J. Phys. Soc. Jpn.
{\bf 60} 2351

Olson C G, Liu R, Yang A.-B, Lynch D W, Arko A J, List R S, Veal B W, 
Chang Y C, Jiang P Z and Paulikas A P 1989 Science {\bf 245} 731 

Orenstein J, Thomas G A, Millis A J, Cooper S L, Rapkine D H, 
Timusk T, Schneemeyer L F and Waszczak J V 1990 Phys. Rev {\bf B 
42} 6342

Pines D 1997b Z. f\"ur Physik B {\bf 103} 129 

Pines D 1997a Physica C {\bf 282-287} 273 

Plakida N M 1997 Z. Phys. {\bf 103} 383 

Presand M R, Tallon J L, Buckley R G, Liu L S and Flower N F 
1991 Physica C {\bf 176} 95      

Puchkov A V, Basov D N and Timusk T 1996a J.Phys. Cond. Mat. {\bf 8}
10049

Puchkov A V, Fournier P, Basov D N, Timusk T, Kapitulnik A and
Kolesnikov N N 1996c Phys. Rev. Lett. {\bf 77} 3212

Puchkov A V,  Fournier P, Timusk T and Kolesnikov N N 1996b Phys.
Rev. Lett. {\bf 77} 1853

Randeria M and Campuzano J.-C 1997 {\it Varenna Lectures}, {\it preprint} 
cond-mat/9709107 

Randeria M, Duan J-M and Shieh L-Y 1989 Phys. Rev. Lett. {\bf 62} 981

Reedyk M, Timusk T, Xue J S and Greedan J E 1997 
Phys. Rev B  {\bf 56} 9134 

Regnault L P, Bourges P, Burlet P, Henry J Y, Rossat-Mignod J, Sidis Y
and Vettier C 1994 Physica C {\bf 235-240} 59

Renner C, Revaz B, Genoud J-Y and Fischer O 1996 J. Low Temp. Phys. 
{\bf 105} 1083

Renner Ch, Revaz B, Genoud J-Y, Kadowaki K and Fischer O 1998 Phys. Rev. 
Lett. {\bf 80 } 149

Reznik D, Bourges P, Regnault L P, Bossy J, Vettier C, Milius D 
L, Aksay I A, and Keimer B 1996 Phys. Rev B {\bf 53} R14741 

Reyes AP, MacLaughlin DE, Takigawa M, Hammel PC, Heffner RH, Thompson JD
and Crow JE 1991 \prb {\bf 43} 2989

Rozenberg M I, Kotliar G, Kajueter H, Thomas G A, Rapkine D H, 
Honig J M and Metcalf P 1995 Phys. Rev. Lett. {\bf 75} 105

Rossat-Mignod J, Regnault L P, Bourges P, Burlet P, Vettier C and 
Henry J Y 1993 in {\it Selected Topics Superconductivity} (Frontiers in 
Solid State Physics, {1}) edited by L.C. Gupta and M.S. Multani, (World Scientific,  
Singapore) 

Rossat-Mignod J, Regnault L P, Vettier C, Bourges P, Burlet P, Bossy J, 
Henry J Y and Lapertot G 1991 Physica C {\bf 185-189} 86
\comment{first report of 41 meV resonance \YBa neutron magnetic scattering}

Rotter L D, Schlesinger Z, Collins R T, Holtzberg F, Feild C, Welp U,
Crabtree G W, Liu J Z, Fang Y, Vandervoort K G and Fleshler S 1991
Phys. Rev. Lett. {\bf 67} 2741

Scalapino D J 1969 in {\it Superconductivity} edited by R, D. Parks (Marcel 
Dekker, New York, 1969) {\bf 1} 449 

Schulz H J 1991 Progress in High Temperature Superconductivity {\bf 29} 
57

Shastry B S and Shraiman B I 1990 Phys. Rev. Lett. {\bf 65} 1068

Shen Z.-X and Dessau D S 1995 Physica Reports {\bf 253} 1

Shen Z.-X and Schrieffer J R 1997 Phys. Rev. Lett. {\bf 78} 1771

Shen Z.-X, Dessau D S, Wells B O, King W E, Spicer W E, Arko A J, 
Marshall D S, Lombardo L W, Kapitulnik A, Dickinson P, Doniach S, DiCarlo 
H, Loeser T and Park C.-H 1993 Phys. Rev. Lett. {\bf 70} 3999
\comment{d-wave ARPES}

Shulga S V, Dolgov O V and Maksimov E G 1991 Physica C {\bf 178} 266

Slakey F, Klein M V, Rice J P and Ginsberg D 1990 Phys. Rev. B 
{\bf 42} 2643

Sooryakumar R and Klein M V  1980 Phys. Rev. Lett. {\bf 54} 660

Startseva T, Timusk T, Okuya M, Kimura T and Kishio K 1998a (unpublished)

Startseva T, TimuskT,  Puchkov A, Basov D, Okuya M, Kimura T and 
Kishio K 1998b (unpublished)

Statt B W and Griffin A 1992 Phys. Rev. B {\bf 46} 3199

Statt B W and Griffin A 1993 Phys. Rev. B {\bf 48} 619

Stern R, Mali M, Mangelschots I, Roos J, Brinkmann D, Genoud J.-Y, 
Graf T and J. Muller 1994 Phys. Rev. B {\bf 50} 426

Stern R, Mali R, Roos J and Brinkmann D 1995 Phys. Rev. B {\bf 51}
15478  

Sternlieb B J, Tranquada J M Shirane G, Sato M and Shamato S 
1994
Phys. Rev. B {\bf 50} 12915

Stojkovi\'c B P and Pines D 1997 Phys. Rev. B {\bf 56} 11931

Tajima S, Gu G D, Miyamoto S, Odagawa A and Koshizuka N 1993 Phys. Rev. B
{\bf 48} 16164

Tajima S, Sch\"utzmann J and Miyamoto S 1995 Solid State Comm. 
{\bf 95} 759

Takagi H, Batlogg B, Kao H L, Kwo J, Cava R J, Krajewski J J, and Peck Jr. W F,
1992 Phys. Rev. Lett. {\bf 69} 2975

Takenaka K Mizuhashi K, Takagi H and Uchida S 1994 Phys. Rev. B {\bf 50} 6534
\comment{first correlation of ab plane resitivity to c-axis pseudogap}

Takenaka K, Fukuzumi Y, Mitzuhasi K, Uchida S, Asaoka H and Takei H 
1997 Phys. Rev. B {\bf 56} 5654
\comment{no pseudogap in electronic thermal conductivity no T* seen}

Takigawa M, Reyes A P, Hammel P C, Thompson J D, Heffner R H, Fisk Z 
and Ott K C 1991 Phys. Rev. B {\bf 43} 247 

Tanamoto T, Kuboki K and Fukuyama H 1991 J. Phys. Soc. Jpn. {\bf 60} 3072

Tanner D B and Timusk T 1992 in {\it The physical properties of 
high temperature superconductors} edited by Donald M. Ginsberg 
(World Scientific, Singapore) {\bf 3} 363

Tao H J, Lu F and Wolf E J 1997 Physica C {\bf 282-287} 1507
\comment{first tunneling pseudogap }

Thomas G A, Orenstein J, Rapkine D H, Capizzi M, Millis A J,
Bhatt R N, Schneemeyer L F and Waszczak J V 1988 Phys. Rev. Lett.
{\bf 61} 1313

Timusk T, Cao N, Basov D N and Homes C C 1996 in
{\it Spectroscopic Studies of Superconductors,} edited by Ivan Bozovic and 
Dirk van der Marel, Proc. SPIE {\bf 2696} 2

Tinkham M 1975 {\it Introduction to Superconductivity} Robert E. Krieger 
Publishing Co (Malabar, Florida) 

Tranquada J M, Gehring P M, Shirane G, Shamato S and Sato M 1992 Phys. Rev. B 
{\bf 46} 5561

Uemura Y J,  Luke G M , Sternlieb B  J, Brewer J H, Carolan J F , Hardy W N, Kadano R, Kempton J R, Kiefl R F, Kreitzman S R, Mulhern P, Riseman T M, Williams D Ll, Yang B X, Uchida S, Takagi H, Gopalakrishnan J, Sleight A W, Subramanian M A, Chien C L,
Zieplak M Z, Xiao Gang, Lee V Y, Statt B W, Stornach C E, Kossler W J and Yu X H 
1989 Phys. Rev. Letters {\bf 62,} 2317

Uchida S, Ido T, Takagi H, Arima T, Tokura Y, and Tajima S 1991 
Phys. Rev {\bf B 43} 7942

Uchida S, Tamasaki K and Tajima S 1996 Phys. Rev. B {\bf 53} 14558

Varma C M, Littlewood P B, Schmitt-Rink S, Abrahams E and 
Ruckenstein A E 1989 Phys. Rev. Lett. {\bf 63} 1996
\comment{ marginal Fermi liquid theory} 
    
Walstedt R E, Warren W W Jr., Bell R F, Cava R J, Espinosa G P,
Schneemeyer L F and Waszczak J V 1990 Phys. Rev. B {\bf 41} 9574

Walstedt R E, Bell R F and Mitzi D B 1991 Phys. Rev. B {\bf 44} 7760 

Warren W W Jr., Walstedt R E, Brennert J F, Cava R J, Tycko R,
Bell R F and Dabbagh G 1989 Phys. Rev. Lett. {\bf 62} 1193 

Wells B O, Shen Z.-X, Dessau D S, Spicer W E, Mitzi D B, 
Lombardo L, Kapitulnik A and Arko 
A J 1992 Phys. Rev. B {\bf 46} 11830

White P J, Shen Z.-X, Kim C, Harris J M, Loeser A G, Fournier P and 
Kapitulnik A 1996 Phys. Rev. B {\bf 54} R15 669
\comment{gap in overdoped matl's by ARPES}

Williams J V M, Tallon J L, Haines E M, Michalak R
and Dupree R 1997 Phys. Rev. Lett. {\bf 78} 721

Winzek N, Mattausch Hj, Eriksson S-G, Str\"om C, Kremer R K, 
Simon A and Mehrig M 1993 Physica C {\bf 205} 45 

Yamada K, Wakimoto S, Shirane G, Lee C H, Kastner M A, Hosoya 
S, Greven M, Endoh Y, Biergenau R J 1995 Phys. Rev. Lett. {\bf 75} 1626

Zawadowski A and Cardona M 1990 Phys. Rev. B {\bf 42} 10732

Zheng G.-q, Odaguchi T, Mito T, Kitaoka Y, Asayama K and Kodama Y
1993 J. Phys. Soc. Jpn. {\bf 62} 2591

Zheng G.-q, Odaguchi T, Kitaoka Y, Asayama K, Kodama Y,
Mizuhashi K and Uchida S 1996 Physica C {\bf 263} 367

Zimmermann H, Mali M, Brinkmann D, Karpinski J, Kaldis E 
and Rusiecki 1989 Physica C {\bf 190} 681

Zimmermann H, Mali M, Mangelshots I, Roos J, Pauli L, Brinkmann D, 
Karpinski J, Kaldis E and Rusiecki 1990 J. Less Common Met. {\bf 164-165} 138

\end{Alist}

\end{document}